\newif\iffullver\fullvertrue

\iffullver

\documentclass[11pt]{llncs}
\usepackage[hmargin=1.1in,vmargin=1.1in]{geometry}

\else

\documentclass{llncs}

\fi

\usepackage{tikz}
\usetikzlibrary{arrows,calc}

\usepackage{tikz}
\usetikzlibrary{cd}
\usepackage{xcolor}

\tikzcdset{scale cd/.style={every label/.append style={scale=#1},
    cells={nodes={scale=#1}}}}

\definecolor{light-gray}{gray}{0.83}

\usetikzlibrary{backgrounds,calc,plotmarks, arrows, matrix,shapes,shapes.symbols, fit,tikzmark,
  decorations.pathmorphing, decorations.pathreplacing, positioning}

\usepackage{spverbatim}
\usepackage{multirow}
\usepackage{subfigure}
\usepackage{adjustbox}
\usepackage{graphics}
\usepackage[noend]{algpseudocode}
\usepackage{algorithm}
\usepackage{microtype}
\usepackage{todonotes}

\usepackage{url}

\usepackage{amsmath}   
\usepackage{bbm}
\usepackage{epsfig}
\usepackage{amsfonts}

\usepackage{hyperref}
\usepackage{amssymb}
\usepackage{pythonhighlight}

\usepackage{environ}
\RenewEnviron{python}{}{}

\usepackage{xspace}

\newcommand{\vect}[1]{\boldsymbol{#1}}
\renewcommand{\vec}[1]{\boldsymbol{#1}}

\newcommand{\ie}{\textsl{i.e.}\xspace}

\newcommand{\ignore}[1]{}

\newcommand{\abs}[1]{\left|#1\right|}

\newcommand{\near}[1]{\left\lfloor#1 \right\rceil}
\newcommand{\infnorm}[1]{\left\|#1\right\|_\infty}

\newcommand{\scp}[2]{\left\langle#1,#2 \right\rangle}

\newcommand{\rlwe}{\ensuremath{\mathsf{RLWE}}\xspace}

\newcommand{\ckks}{\ensuremath{\mathsf{CKKS}}\xspace}

\newcommand{\Ecd}{{\sf Ecd}}

\newcommand{\stc}{\ensuremath{\mathsf{SlotToCoeff}}} % as in ckks18 paper
\newcommand{\cts}{\ensuremath{\mathsf{CoeffToSlot}}}

\newcommand{\pr}[2]{{\sf Pr}_{#1 \rightarrow #2}} % product
\newcommand{\tr}[2]{{\sf Tr}_{#1 \rightarrow #2}} % trace
\newcommand{\ct}{{\sf ct}}
\newcommand{\sk}{{\sf sk}}
\newcommand{\pk}{{\sf pk}}
\newcommand{\evk}{{\sf evk}}
\newcommand{\cpp}{C\texttt{++}}

\newcommand{\rep}[2]{(\underbrace{#1,\ldots,#1}_{#2 \textrm{~rep.}})}
\newcommand{\repl}[2]{(\underbrace{#1,\ldots\ldots,#1}_{#2 \textrm{~rep.}})}

\begin{document}

\title{Low-Latency Bootstrapping for CKKS using Roots of Unity}

\author{Jean-S\'ebastien Coron \and Robin K\"ostler}

\institute{University of Luxembourg}

\iffullver
\else
\author{~}
\institute{~}

\fi

\maketitle

\iffullver
\else
\vspace{-1.5cm}
\fi

\begin{abstract}
We introduce Sparse Roots of Unity (SPRU) bootstrapping, a new bootstrapping algorithm for the \ckks homomorphic encryption scheme for approximate arithmetic. The original \ckks bootstrapping method relies on homomorphically evaluating a polynomial that approximates modular reduction modulo $q$. In contrast, SPRU bootstrapping directly embeds the additive group modulo $q$ into the complex roots of unity, which can be evaluated natively 
in the \ckks scheme.
 This approach significantly reduces the multiplicative depth required for bootstrapping, enabling the use of a smaller ring dimension and improving efficiency. In practice, using the OpenFHE \cpp~library, SPRU bootstrapping achieves up to a 5× reduction in latency when applied to ciphertexts with a small number of slots.
 \end{abstract}

\iffullver
\else
\fi

\section{Introduction}

\subsubsection{Fully Homomorphic Encryption.}
Homomorphic encryption (HE) enables to perform operations on encrypted data without knowing the
decryption key. \textit{Fully homomorphic encryption} (FHE) extends this capability, enabling the 
evaluation of arbitrary circuits on encrypted data.
FHE has a wide range of applications, including secure cloud computing, 
 multi-party computation, and secure machine learning.

% SV14 (published in 2011 did SIMD operation for Add and Mult. [GHSx12] does  slots permutations
%A security reduction to lattice-based public-key encryption (PKE) has been proposed in \cite{BV14}, putting FHE on firm grounds.

Since Gentry's invention of FHE in 2009 \cite{GEN09}, numerous homomorphic 
encryption schemes have been developed, leading to significant improvements in FHE performance. Gentry's 
key innovation was the introduction of \textit{bootstrapping}, a ciphertext refreshing procedure based
 on homomorphically evaluating the decryption circuit. Building on this breakthrough, schemes such as BGV 
 \cite{BGV11} and BFV \cite{Bra12,fv12} further enhanced FHE by basing their security on the hardness of 
 the ring learning with errors (RLWE) problem \cite{SSTX09,LPR10}. These schemes can achieve high 
 computational throughput by utilizing Single Instruction, Multiple Data (SIMD) operations 
 \cite{GHSx12,SV14}.

The FHEW scheme, introduced in \cite{DM15}, reduced the time for bootstrapping a single-bit encryption to 
under one second, building on the approach from \cite{AP14} and the matrix-based scheme of \cite{GSW13}. 
This was further improved in \cite{CGGI16} with the torus-FHE (TFHE) scheme, achieving a bootstrapping time
of less than 0.1 second. 
%However, these schemes do not natively support ciphertext packing or SIMD 
%operations. 
%Additionally, the size of the plaintext input is constrained by the ring dimension of the 
%intermediate Ring GSW scheme used in bootstrapping.

\ignore{
These schemes support exact arithmetic operation over some discrete spaces, for example a finite field.
More precisely, bitwise encryption schemes such as FHEW  can evaluate a
boolean gate with bootstrapping very efficiently. Word encryption schemes such as BGV 
or BFV  can encrypt multiple large field elements in a ciphertext, but the bootstrapping
operation is more costly. 
}

\subsubsection{The \ckks scheme.}
In \cite{CKKS17}, the authors presented a construction for homomorphic encryption supporting approximate arithmetic, based on the RLWE problem. The core idea 
involves adding noise to the plaintext to ensure security, with this noise being treated
as part of the error inherent in approximate computations. % as part is more correct (chatGPT)
Given a ciphertext 
$\ct$, for a secret key $\sk$, the decryption equation
$\langle \ct, \sk \rangle \pmod{q}$ outputs an approximation $m+e$ of the original message $m$, for a small error $e$.
The authors introduced a rescaling procedure to control the magnitude of plaintexts, enabling the 
construction of a leveled homomorphic encryption scheme, where the ciphertext modulus grows linearly with 
the depth of the evaluated circuit. % I reverted to the previous formulation (the authors control is a bit strange)
Through rescaling, the scheme can emulate fixed-point addition and multiplication on encrypted messages.
 Additionally, they described a packing method that allows the encryption of up to \( N/2 \) complex numbers in a single ciphertext to perform SIMD operations.
 % modulo the cyclotomic polynomial \( X^N + 1 \).

%Gentry's bootstrapping consists in homomorphically evaluating the decryption function
%to refresh a ciphertext to perform more computations.

\subsubsection{\ckks bootstrapping.}
In \cite{CHKKS18}, the authors introduced a novel ciphertext
refreshing procedure for CKKS, extending the leveled homomorphic
encryption scheme to  fully homomorphic encryption based on
Gentry’s bootstrapping technique. Specifically, the initial leveled scheme
from \cite{CKKS17} can only evaluate circuits of a fixed depth, as
each homomorphic multiplication reduces the ciphertext modulus until
it becomes too small to support further computation.

\begin{figure}[h]
\centering
\includegraphics[scale=.3]{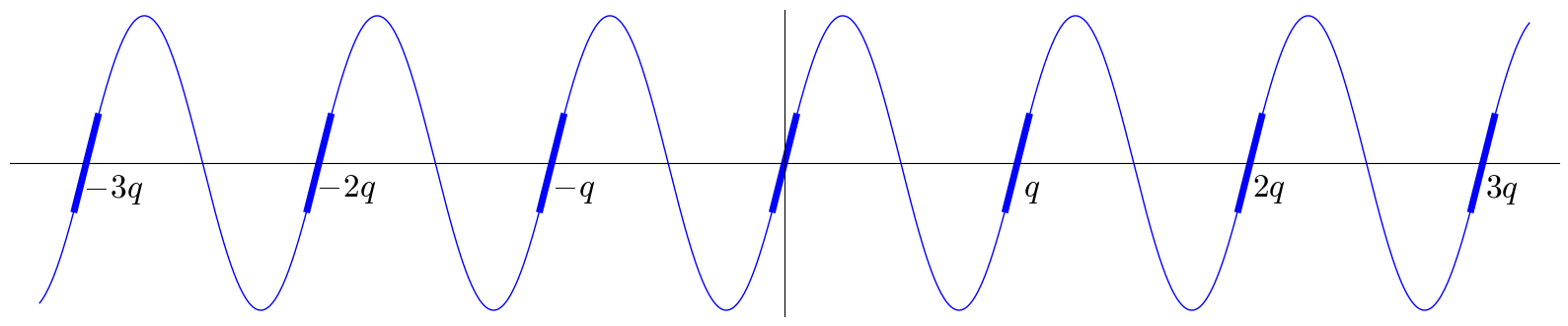} % .35
\caption{Approximating modular reduction via a scaled sine function.}
\label{f:sine}
\end{figure}

\label{s:CKKSboot}

As illustrated in Figure \ref{f:sine},   \ckks bootstrapping
relies on approximating the modular reduction modulo $q$ using a scaled sine function. Since the sine 
function is $2\pi$-periodic, the decryption function can be expressed as:
$$ [\langle \ct,\sk \rangle]_q \approx \frac{q}{2\pi} \cdot \sin \left( \frac{2\pi}{q} \cdot 
\langle \ct,\sk \rangle \right).$$
which is periodic modulo $q$ and closely approximates  $\langle \ct,\sk \rangle$ near $0$.
Given that $m \ll q$, the polynomial approximation only
has to be good for small values modulo $q$.
The bootstrapping procedure homomorphically evaluates this polynomial approximation, 
producing a refreshed ciphertext that encrypts the same message but under a larger modulus, 
thereby enabling further computation. 
 The error resulting from bootstrapping is kept sufficiently small to preserve plaintext
 precision.

The authors also demonstrated how to refresh packed ciphertexts containing $n$ slots with ${\cal O}(n)$ complexity.
In particular, they use a linear transformation that moves the polynomial coefficients into the plaintext slots via the 
\textit{coefficient-to-slot} (\cts) procedure, and reverses the latter using the \textit{slot-to-coefficient} 
(\stc) operation. The complexity of these operations was further reduced to ${\cal O}(\log n)$ in 
\cite{CHH18,CCS19}. 
We recall  the original \ckks bootstrapping
procedure in more details in Appendix \ref{app:CKKSboot}, first for a single
slot, and then for multiple slots, using the $\cts$ and $\stc$ operations.

The main drawback of \ckks bootstrapping is its relatively high runtime, primarily due to its substantial multiplicative depth,
 which necessitates a large modulus.
  Reducing the multiplicative depth would free up more levels for useful computations after bootstrapping,
  thereby improving overall throughput.
   Additionally, it would allow for the selection of a smaller ring dimension, leading to reduced latency.

% see https://eprint.iacr.org/2024/1637.pdf (Asiacrypt 2024)

% see [BTPH22] J.-P. Bossuat, J. Troncoso-Pastoriza, and J.-P. Hubaux. Bootstrapping for approximate homomorphic encryption with negligible failure-probability by using sparse-secret encapsulation. In ACNS, 2022.

% What is the value of Delta in OpenFHE ?

\subsubsection{Sparse Roots of Unity (SPRU) bootstrapping.}

Our main contribution is SPRU bootstrapping, a new bootstrapping algorithm for \ckks that achieves reduced multiplicative depth. The key idea is to embed the additive group $\mathbb{Z}_q$ into the complex roots of unity via the homomorphism:
\[
f: \mathbb{Z}_q \longrightarrow \mathbb{C}, \qquad x \longmapsto \exp(2i\pi \cdot x/q).
\]
This embedding enables modular reduction to be evaluated natively within the \ckks scheme, avoiding expensive polynomial approximations and enabling more efficient ciphertext refreshing, especially for a small number of slots.

Considering for simplicity an LWE ciphertext $\vec{c}$,
we can  compute the decryption equation 
$m = \langle \vec{s}, \vec{c} \rangle = \sum_{k=0}^{N-1} s_k c_k \pmod q$ 
over this multiplicative group, using  $s_k \in \{0,1\}$  as a selector:
$$
\exp(2i\pi \cdot  \langle \vec{s}, \vec{c} \rangle/q) 
 =\prod\limits_{k=0}^{N-1} \exp(2i\pi \cdot c_{k} s_k/q) \\
 =\prod\limits_{k=0}^{N-1}  \left(1 + \left( \exp(2i\pi \cdot c_{k} /q) - 1 \right) \cdot s_k  \right).
$$
% I removed the \abs{} here, it is not necessary at this point.
Moreover, considering a message $m \in {\mathbb Z}$ such that $m = [\langle \vec c, \vec s \rangle]_q \ll q$, we can use the approximation  $\exp(x) \simeq 1+x$ for small $x$, which yields:
\begin{equation}
  \label{eq1:1mpiq}
1+ m \cdot \frac{2i\pi}{q} \simeq \prod\limits_{k=0}^{N-1}  \left(1 + \left( \exp(2i\pi \cdot c_{k} /q) - 1 \right) \cdot s_k  \right) 
\end{equation}

We claim that this provides a new bootstrapping equation for \ckks,
as it can be homomorphically evaluated to provide a refreshed ciphertext.
Specifically, it suffices to encrypt each $s_k$, and following
(\ref{eq1:1mpiq}), homomorphically multiply it by
$\exp(2i\pi \cdot c_{k} /q) - 1$.
The full product is then computed homomorphically using \ckks multiplications.
Eventually, by extracting the imaginary part and applying appropriate
scaling, we obtain a new ciphertext encrypting the same $m \in {\mathbb Z}$, but
 under a larger modulus $q \cdot p$ than the original modulus $q$, enabling further computation.

This bootstrapping equation differs significantly from the original \ckks bootstrapping algorithm, as we 
avoid using a polynomial approximation of the sine function.
Instead, the modular reduction modulo \(q\) is achieved by directly 
embedding the ciphertext coefficients \(c_k \in \mathbb{Z}\) into the complex circle group, where 
computations are performed using \ckks' homomorphic operations. 
 Notably, when bootstrapping multiple slots, the \textit{coefficient-to-slot}  procedure from 
% Notably, when bootstrapping across multiple slots, the \textit{coefficient-to-slot} procedure from 
 the \ckks bootstrapping is dispensable, as the ciphertext is already embedded within the complex slots. 
 However, the \textit{slot-to-coefficient} (STC) operation is still needed to revert to the coefficient 
 representation. 
 
 % From reviewer A.
Our approach shares conceptual similarities with the blind rotation algorithms used in FHEW/TFHE bootstrapping 
\cite{DM15,CGGI16}, where the additive group ${\mathbb Z}_q$ is
embedded into the set $Y^i$ of roots of unity
of  $X^N+1$, for $N=k \cdot q$ and $Y=X^{2k}$. In both cases, the computation involves a chain of products with factors selected homomorphically over the encrypted coefficients of the secret key.
However, we stress that the underlying mathematics of FHEW/TFHE and \ckks are fundamentally different.
FHEW/TFHE bootstrapping is optimized for low-latency processing of single-slot messages with limited precision (typically Boolean to 5-bit integers).
In contrast, \ckks supports the encryption and bootstrapping of up to $N/2$ high-precision complex values in a single ciphertext, albeit at the cost of significantly higher latency.

The goal of this paper is therefore to explore this new bootstrapping approach, prove a corresponding bootstrapping 
theorem, and introduce several optimizations that make it practical. Finally, we present concrete parameter 
sets and benchmark our results based on the OpenFHE \cpp ~library \cite{OpenFHE}, comparing the performance of our bootstrapping with the original 
\ckks bootstrapping.

\subsubsection{First optimization: packing the secret-key bits.}

%In Section \ref{sec:packing_sk_bits},  
We introduce a first optimization
for the homomorphic evaluation of our bootstrapping equation (\ref{eq1:1mpiq}),
 where the $N$ secret-key bits $s_i$ are packed into the
$N/2$ slots of \ckks ciphertexts.
As a consequence, the bootstrapping key size decreases from $N$ to only two \ckks ciphertexts. 
The product of  the slots is then computed using a product operator (corresponding to the standard field norm, see Sec. \ref{s:trace:product}),
most notably reducing the number of ciphertext-ciphertext multiplications from ${\cal O}(N)$ to ${\cal O}(\log N)$.
 When homomorphically evaluating 
 (\ref{eq1:1mpiq}), we further take advantage of \ckks natively supporting complex numbers in the slots.

\subsubsection{Second optimization: a sparse block secret-key.}
%In Section \ref{s:boot-trace}, 
As a second optimization, we assume that the secret key is
structured into
 $h$ blocks, each of size $B=N/h$, with exactly one non-zero bit per block. 
This technique was previously explored in the context of TFHE bootstrapping \cite{LMSS23}.
Leveraging  this structure, we can rewrite the bootstrapping equation (\ref{eq1:1mpiq}) using partial sums over 
each block instead of products. 
For a packed ciphertext, these sums are efficiently computed using the trace operator, while the remaining 
products are still handled by the product operator. 
\ignore{To apply the trace operator, we re-index the secret-key 
coordinates such that each block corresponds to a congruence class of indices, over which the trace 
operator computes the sum.}Since most homomorphic multiplications are replaced by additions, 
this approach reduces the multiplicative depth from $\log_2 N +1$ to $\log_2 h+1$.
Additionally,  the number of homomorphic multiplications decreases 
from ${\cal O}(\log N)$ to ${\cal O}(\log h)$, further improving efficiency.
We formally describe our bootstrapping algorithm for a single slot by
combining the two previous optimizations in Section \ref{s:boot_single}.

\subsubsection{Bootstrapping any number of slots.}
Equation (\ref{eq1:1mpiq}) applies to bootstrapping a single slot $m \in \mathbb{Z}$, but it can be naturally extended to handle multiple slots.
 We first generalize the approach to $n$ slots for $n \leq B/2$, where $B$ is the block size in the secret key. 
To achieve this, we treat each coefficient of the polynomial message as the decryption of an independent 
LWE ciphertext. The key bits and the corresponding embeddings of the LWE ciphertext coefficients are then 
packed into the $N/2$ slots of $2n$ \ckks ciphertexts.
As before, the sum is computed in two steps: first, by summing across $2n$ distinct ciphertexts, and 
then by applying the trace operator to the result. The product is again evaluated using the product operator.
 Finally, we apply the \stc ~operation to move the slots into the polynomial coefficients.
 Note that, as opposed to the original \ckks bootstrapping, we do not need the \cts ~operation.
 To support ciphertexts with up to $n \leq N/2$ slots, we apply the above bootstrapping procedure sequentially over ciphertexts containing $B/2$ slots.

Asymptotically, our method has a complexity of ${\cal O}(n+ \log N)$ homomorphic operations, whereas the original \ckks 
bootstrapping scales as ${\cal O}(\log n+ \log N)$. Consequently, the original \ckks bootstrapping is more 
efficient for large $n$. However, for a small number of slots, our approach is significantly faster in practice 
due to its 
lower multiplicative depth. Additionally, our bootstrapping method is highly parallelizable; with  $n$ 
processors, it achieves the same ${\cal O}(\log N)$ complexity as the original \ckks bootstrapping.

\subsubsection{Open-source implementation based on OpenFHE.}

We present our open-source implementation of the new bootstrapping algorithm, developed using the 
OpenFHE \cpp~library. This enables a direct comparison with the existing implementation of the 
original CKKS bootstrapping provided in OpenFHE. We recall the parameter set used for the 
original bootstrapping and introduce a concrete parameter set for our method, both ensuring a  
128-bit security level. Our results demonstrate that, in practice, the reduced multiplicative depth of 
our approach yields up to an 5× speedup when bootstrapping ciphertexts with a small number of slots.
\iffullver 
The source code is provided in 
\begin{center}
\url{https://github.com/coron/spru_boot}
\end{center}
\else
We provide our \cpp ~source code  in the auxiliary files. \fi

\subsubsection{Efficient implementations of \ckks.}

Efficient implementations of \ckks  use a \textit{residue number system} (RNS), which decomposes the 
ciphertext modulus into several smaller moduli.
This allows operations to be performed on native 64-bit word-sized integers, providing significant 
computational advantages. The first {\sf RNS-CKKS} scheme was introduced in \cite{CHKKS18SAC}, 
achieving an order of magnitude improvement in efficiency,
by adapting the double-CRT representation in the BGV scheme \cite{GHSa12}.
Since the RNS technique is implemented in the OpenFHE \cpp ~library, 
 we are able to directly compare our method against the state-of-the-art implementation of the original CKKS bootstrapping available in OpenFHE.

A significant portion of \ckks\!’ runtime is attributed to its linear transformations, 
specifically \(\stc\) and \(\cts\), which perform homomorphic encoding and decoding. 
These transformations can be executed using matrix multiplication, requiring ${\cal O}(n)$
 homomorphic operations \cite{CHKKS18}. This was later optimized to ${\cal O}(\log n)$ by employing a 
 homomorphic Discrete Fourier Transform (DFT) \cite{CHH18,CCS19}. Both techniques are implemented in the OpenFHE \cpp~library for the original CKKS bootstrapping.
  Our bootstrapping implementation reuses the existing \(\stc\) implementation without modification.

Another optimization involves GPU-accelerated implementations, as described
 in \cite{SYL+23} and \cite{JKA+21}. Additionally, the authors of \cite{CCKS23} proposed a new technique for performing a
ciphertext multiplication that consumes fewer bits of the ciphertext
modulus. All these optimizations are fully 
compatible with our new bootstrapping algorithm.

\subsubsection{Related work.}

The \ckks approximation of the modular reduction function, specifically the 
\({\sf EvalMod}\) algorithm, has been optimized in \cite{KPK+23} and \cite{LLK+22} to consume fewer bits of the ciphertext modulus. The approximation  of the sine function has been improved  
in \cite{LLL+21} and \cite{JM22},
improving the bootstrapping precision. 

% The following section is a bit too detailed, why is this important for our paper?
In \cite{KDE+21}, the authors introduced a new bootstrapping approach for BGV, BFV and \ckks, 
based on the blind rotation technique used in FHEW/TFHE bootstrapping. 
For \ckks, their approach enables high-precision bootstrapping with a relatively small 
ring dimension of \(N = 2^{13}\).
Since FHEW/TFHE can only bootstrap a single slot, 
the ciphertext is split according to the number of slots.
Then, each part is bootstrapped independently, and 
the results are recombined into a single ciphertext. However, this approach 
requires \({\cal O}(N)\) homomorphic operations for a single slot, which 
could render it impractical for \ckks.
The authors do not provide timing results for their implementation.

In \cite{BCC+22}, the authors introduced a method for achieving high-precision bootstrapping by combining multiple low-precision bootstrapping steps. As a result, it is possible to keep
 the same parameters (in particular, the same modulus size) for bootstrapping while achieving higher precision, albeit with the trade-off of requiring more iterations. Finally, it was recently shown
 that \ckks can be used in a black-box manner to perform computations on binary data \cite{DMPS24}.
 This was further improved in \cite{BCKS24} with  a
 \ckks bootstrapping algorithm designed specifically for ciphertexts encoding binary data.

\subsubsection{Independent work.}
Recently, Cheon, Hanrot, Kim, and Stehlé introduced a new bootstrapping algorithm for \ckks at Eurocrypt 2025 \cite{CHKS25}.
 Like our approach, their method is based on complex roots of unity and achieves a reduction in multiplicative depth compared to the original \ckks bootstrapping.

The main difference is that they focus  on a full $N/2$-slots bootstrapping using a large number of processors, while we focus on efficient bootstrapping for a small number of slots, with a single or few  
processors. Additionally, the underlying techniques are fundamentally different. Their method 
homomorphically evaluates the polynomial shift arising in the decryption process through a blind rotation procedure 
— inspired by the DM/CGGI bootstrapping technique. This involves encrypting 
the index $j$ of each of the $h$ non-zero secret-key elements (where $h$ is the Hamming weight), and 
then performing a blind rotation by $j$ positions on a vector of complex roots of unity.
In contrast, our method does not encode the secret-key coefficients by their indices. Instead, the 
secret-key coefficients are packed directly into the slots of a \ckks ciphertext, and bootstrapping is  
performed by homomorphically evaluating the decryption equation for each slot over the complex roots 
of unity.

\section{Preliminaries}

\subsubsection{Basic notations.}
For a real number $r$, we denote by $\near{r}$ the nearest integer to $r$, rounding upwards. We denote
vectors in bold, and $\langle \vec{a},\vec{b} \rangle$ denotes the scalar product of the 
vectors $\vec{a}$ and $\vec{b}$. For an integer $q$, we use ${\mathbb Z} \cap (-q/2,q/2]$
as a representative of ${\mathbb Z}_q$, and use $[z]_q$ or $z \bmod q$ to denote the central
 reduction of
the integer $z$ modulo $q$ into that interval. % this is really \bmod, \pmod is for mod equivalence.
For a finite set $S$, we denote by $v \leftarrow S$ the sampling of $v$ uniformly at random from $S$.
By $v \leftarrow \mathcal D$ we also denote sampling $v$ from a distribution $\mathcal D$.
We utilize the same notations for sampling coefficients of vectors/polynomials independently and identically from a set/distribution. 
For an integer $\kappa >0$, we let $v \leftarrow \chi_\kappa$ be sampled from the \textit{centered binomial distribution} $\chi_\kappa$.
It is defined as outputting $v := \sum_{j=1}^\kappa (a_i - b_i)$, where $\vec a, \vec b \leftarrow \mathbb \{0,1\}^\kappa$. In this case, it holds that $\abs v \le \kappa$.
We denote by $\lambda$ the security parameter; all known attacks
should take $\Omega(2^\lambda)$ bit operations. 

\subsubsection{Cyclotomic rings.}
For a power-of-two $N$, we set $\mathcal R := \mathbb{Z}[X]/( X^N + 1)$ as the $2N$-th cyclotomic ring.
%, a quotient ring with  $X^N + 1$ as the quotient polynomial.
%Thus, we have the property $X^{2N} = 1$ in $\mathcal R$.
We also write $\mathcal R_q := \mathbb{Z}_q[X]/(X^N + 1)$ for the
residue ring of ${\cal R}$ modulo $q$.
As previously, coefficients of elements of $\mathcal R_q$ are represented as integers in $(-q/2, q/2]$.
We represent an arbitrary element of the cyclotomic ring ${\cal S}={\mathbb R} [X]/ ( X^N+1 )$
as a polynomial $a(X)=\sum_{j=0}^{N-1} a_j X^j$ of degree strictly  less than $N$, and
identify it with its coefficient vector $(a_0,\ldots,a_{N-1}) \in {\mathbb R}^N$. 
We consider the norms $\|a\|_{\infty} = \max_{i \in \{0,\dots, N-1\}} \abs{a_i}$ and $\|a\|_1 = \sum_{i = 0}^{N-1} \abs{a_i}$ on the coefficient vector of $a$.

\ignore{
\newcommand{\cnorm}[1]{\|#1\|_{\infty}^{\sf can}}

For $a \in {\mathbb R}[X] / (X^N+1)$, we denote by 
$\sigma(a)=(a(\zeta^j))_{j \in {\mathbb Z}_{2N}}$
for $\zeta =\exp(i\pi/N)$ the canonical embedding into ${\mathbb C}^N$. Its 
$\ell_{\infty}$ norm is called the {\sl canonical embedding norm}, 
denoted by $\cnorm{a}=\|\sigma(a)\|_{\infty}$.
It satisfies the following properties: for all $a,b \in {\cal S}$,
we have $\cnorm{a \cdot b} \leq \cnorm{a} \cdot \cnorm{b}$,
and $\cnorm{a} \leq \|a\|_1 \leq N \cdot \|a \|_{\infty}$.
Moreover, for a power-of-two $N$, we have 
$\|a\|_{\infty} \leq \cnorm{a}$, see \cite{DPSZ12}.
}

\section{The \ckks scheme}

\subsection{Basic \rlwe-based encryption}
 
\label{s:basic_CKKS}

In \cite{CKKS17}, the authors described a fully homomorphic encryption scheme designed for approximate 
arithmetic. Given ciphertexts encrypting \(m_1\) and \(m_2\), the scheme enables secure computations of 
encrypted approximate values of \(m_1 + m_2\) and \(m_1 \cdot m_2\), with a predefined level of precision.
We recall the concrete instantiation based on the BGV scheme \cite{BGV11}, using the multiplication
from \cite{GHSa12} based on raising the ciphertext modulus.
Note that the \ckks scheme differs from {\sf BGV} in the sense that the plaintext space
is the polynomial ring ${\cal R}={\mathbb Z}[X]/(X^N+1)$, 
instead of 
${\mathbb Z}_t[X] / (X^N+1)$ for a small $t$. %; see Table \ref{t:HEtype}.

%\the\intextsep

\ignore{

\begin{table}[h]
\centering
\renewcommand{\arraystretch}{1.2}
\resizebox{0.72\columnwidth}{!}{%
\begin{tabular}{|l|c|c|} \hline
~Scheme type~ & ~Plaintext space~ & ~Decryption structure~ \\ \hline
~{\sf BGV}~ & ~${\mathbb Z}_t[X] / (X^N+1)$~  & ~$\langle \ct,\sk \rangle = m + t e \pmod q$~ \\ \hline
~{\sf FV}~ & ~${\mathbb Z}_t[X] / (X^N+1)$~  & ~$\langle \ct,\sk \rangle = (q/t)m + e \pmod q$~ \\ \hline 
~\ckks~ & ~${\mathbb Z}[X] / (X^N+1)$~  & ~$\langle \ct,\sk \rangle = m + e \pmod q$~ \\ \hline
\end{tabular}
}
\medskip
\caption{Plaintext space and decryption structure for various HE families.}
\label{t:HEtype}
\end{table}

}

Let $q$ be an integer and let $s$ be a 
secret key with small components in $\cal R$.
A \ckks ciphertext $\ct=(c_0,c_1)$ for a plaintext 
$m \in \cal R$ satisfies:
$$
\langle \ct,\sk \rangle=c_0 + c_1 \cdot s = m + e \pmod{(q,X^N+1)}
$$
for some small error $e \in \cal R$.
This implies that during decryption, the message $m$ is not recovered exactly, as the low-order bits of the coefficients of $m(X)$
are masked by the error $e(X)$.
Such an error $e$ is inserted to guarantee the security of the hardness assumption
 of the ring learning with error (\rlwe) problem. During homomorphic operations, 
 the coefficients of the plaintext polynomial $m(X)$ should remain sufficiently small compared to the ciphertext
 modulus $q$.

\begin{definition}[\rlwe]
For an integer $q$, a ring dimension $N$ and  distributions ${\cal D}_s$ and ${\cal D}_e$,
the decisional \rlwe problem consists, for $s \leftarrow {\cal D}_s$, in distinguishing polynomially 
many samples $(a,a \cdot s + e) \in {\cal R}_q^2$ from uniform  in ${\cal R}_q^2$, where $a \leftarrow {\cal R}_q$
and $e \leftarrow {\cal D}_e$. 
\end{definition}

We now formally describe the scheme, following \cite{CHKKS18}, beginning with the construction of a \textit{leveled} homomorphic encryption scheme for approximate arithmetic.
The plaintext space is ${\cal R}=\mathbb{Z}[X]/( X^N + 1)$, while % I think that it is OK to recall R here
the ciphertext space is ${\cal R}_q^2$.
The scheme defines various layers of moduli, \ie
$q_\ell=q_0 \cdot \Delta^\ell$ for
$1 \leq \ell \leq L$ and a base $\Delta$.
For the error sampling, for simplicity we use the centered binomial distribution
$\chi_\kappa$ with parameter $\kappa$, rather than a discrete Gaussian
distribution.  
For $h \in \mathbb N$, we let ${\cal HW}(h)$ be the set of ternary
vectors in $\{0,\pm 1\}^N$ whose Hamming weight is exactly $h$. 
For a real \( 0 \leq \rho \leq 1 \), let \({\cal ZO}(\rho)\) denote the distribution that outputs a vector in \(\{0, \pm 1\}^N\), where each entry is sampled independently, with probability \(\rho/2\) for both \(-1\) and \(1\), and probability \(1 - \rho\) for \(0\).

% We may include {\sf KSGen_sk} in {\sf KeyGen}
% Then pk will include evk
%
\begin{itemize}

\item ${\sf KeyGen}(1^\lambda)$: 

\begin{itemize}

\item For a base $\Delta$, a base modulus $q_0$ and an integer $L$, let $q_\ell=q_0 \cdot \Delta^\ell$ for 
$\ell=1,\ldots,L$. 
Given the security parameter $\lambda$, choose a power-of-two $N$, an integer
$P$, and an integer $\kappa>0$ for an \rlwe problem that achieves a security level of $\lambda$-bits, 
for a modulus $P \cdot q_L$ and a ring degree $N$.

\item Sample $s \leftarrow {\cal HW}(h)$, $a \leftarrow {\cal R}_{q_L}$
and $e \leftarrow \chi_\kappa$. Set the secret key as $\sk \leftarrow (1,s)$ and the
public key as $\pk \leftarrow (b,a) \in {\mathcal R}_{q_ L}^2$
where $b=-as+ e \pmod{q_L}$.
\end{itemize}

\smallskip

\item 
${\sf KSGen_\sk}(s')$: for $s' \in {\cal R}$, sample
$a \leftarrow {\cal R}_{P \cdot q_L}$ and $e' \leftarrow \chi_\kappa$.
Return the switching key as ${\sf swk} \leftarrow (b',a') \in {\cal R}_{P \cdot q_L}$
where $b' =-a's + e' + Ps' \pmod{P \cdot q_L}$.
\begin{itemize}
\item Set the evaluation key as $\evk \leftarrow {\sf KSGen}_\sk(s^2)$. 
\end{itemize}

\smallskip

\item ${\sf Enc}_{\pk}(m)$: for $m \in {\cal R}$, sample
$v \leftarrow {\cal ZO}(1/2)$ and $e_0,e_1 \leftarrow \chi_\kappa$. Output
$\ct=v \cdot \pk + (m+e_0,e_1) \pmod {q_L}$.

\smallskip

\item ${\sf Dec}_\sk(\ct)$: for $\ct=(c_0,c_1) \in {\cal R}_{q_\ell}^2$, output
$m=c_0 + c_1 \cdot s \pmod{	q_\ell}$.

\end{itemize}

We now define the homomorphic operations ${\sf Add}(\ct_1,\ct_2)$ and ${\sf Mult}_\evk(\ct_1,\ct_2)$.
We also define the more lightweight external multiplication ${\sf ExtMult}$, which multiplies a ciphertext 
by a plaintext $v \in {\cal R}$.
 
\begin{itemize}

\smallskip
\item 
${\sf Add}(\ct_1,\ct_2)$: for $\ct_1,\ct_2 \in {\cal R}^2_{q_\ell}$,
output $\ct_{\sf add}=\ct_1+\ct_2 \pmod{q_\ell}$.

\smallskip
\item 
${\sf Mult}_\evk(\ct_1,\ct_2)$: for $\ct_1=(b_1,a_1)$, $\ct_2=(b_2,a_2) 
\in {\cal R}_{q_\ell}^2$, let $(d_0,d_1,d_2)=(b_1b_2,a_1b_2+a_2b_1,a_1a_2)
\pmod{q_\ell}$. Output $\ct_{\sf mult} \leftarrow (d_0,d_1) + \lfloor
P^{-1} \cdot d_2 \cdot \evk \rceil \pmod{q_\ell}$.

\smallskip
\item 
${\sf ExtMult}(\ct, v)$: for $\ct = (b,a) \in {\cal R}^2_{q_\ell}$ and $v \in {\cal R}$,
output $\ct_{\sf extmult}=(b \cdot v, a \cdot v) \pmod{q_\ell}$.

\end{itemize}

The rescaling procedure below transforms an encryption of $m$ modulo $q$
into an encryption of $m/\Delta$, while also scaling its inherent noise $e$ to around $e/\Delta$. 
As will be seen in
Section \ref{s:reals}, the composition of homomorphic multiplication and rescaling 
enables to mimic fixed-point arithmetic, while
managing the error growth. 
\ignore{
Namely rounding enables to discard some inaccurate
LSBs of the message while keeping the error at an almost constant relative size compared to the message.
Without such rounding, the bit size of the message
would grow exponentially with the multiplicative depth $\ell$. Thanks to rounding, 
the magnitude of the message remains about the same during homomorphic evaluation, and therefore 
 the required bit size of the largest modulus
  grows only linearly with the depth of the circuit to be  evaluated. }

\begin{itemize}
\item ${\sf RS}_{\ell \rightarrow \ell'}(\ct)$:
for a ciphertext $\ct \in {\cal R}_{q_\ell}^2$ at level $\ell$, output
$\ct'=\near{\Delta^{\ell'-\ell} \cdot \ct} \pmod{q_{\ell'}}$.
We omit the subscript $\ell \rightarrow \ell'$
when $\ell'=\ell-1$.
\end{itemize}

\subsubsection{RNS arithmetic.}
In practice, efficient implementations of \ckks use the so-called RNS-CKKS variant \cite{CHKKS18SAC}, which relies on the Residue Number System (RNS) representation. In this setting, the ciphertext moduli
have the form 
 $Q_r=\prod_{0 \leq i \leq r} q_i$, where each $q_i$ is a small (word-size) prime and 
 $0 \leq r \leq \ell_m$. 

% I think that the sentences below are not useful at this point.
 %The largest
%modulus used is then $Q=Q_{\ell_m}$, where $\ell_m$ is the maximum depth.
% This RNS-CKKS variant is implemented in the OpenFHE library \cite{OpenFHE}, which we use as the basis for our implementation of the new bootstrapping algorithm. 

% I reverted to the previous formulation. The rigorous proof is not
% mainly due to the binomial distribution, it could be obtained with Gaussian.

\subsubsection{Noise growth.}
In \cite{CKKS17}, the authors state some lemmata for estimating the noise, 
although their noise estimation is only heuristic. Our lemmata below are with
rigorous proofs, though with larger upper bounds; we refer to Appendix \ref{a:error_analysis_proofs}
for the proofs.

\begin{lemma}
\label{lemma:noise_encryption}
Let $\ct \leftarrow {\sf Enc}_{\sf pk}(m)$ be an encryption of $m \in {\cal R}$. 
Then $\langle \ct,\sk \rangle=m+e_{\sf clean} \pmod{q_L}$ for some clean encryption error $e_{\sf clean} \in {\cal R}$ 
satisfying $\infnorm{e_{\sf clean}} \leq B_{\sf clean} :=3N\kappa$.
\end{lemma}

\begin{lemma}
\label{lemma:noise_rescaling}
Let $\ct' \leftarrow {\sf RS}_{\ell \rightarrow \ell'}(\ct)$
for a ciphertext $\ct \in {\cal R}_{q_\ell}^2$. Then
$\langle \ct',\sk \rangle= \near{\langle \ct,\sk \rangle \cdot q_{\ell'}/q_\ell} 
+ e_{\sf rs} \pmod{q_{\ell'}}$ for some rescaling error $e_{\sf rs} \in {\cal R}$
satisfying $\infnorm{e_{\sf rs}} \leq B_{\sf rs} := 2N$.
\end{lemma}

\ignore{
Adding two ciphertexts is straightforward, and thus their noise bounds sum, too.
Multiplying two ciphertexts does multiply their noises, therefore we will rescale afterward.
Since we only use one evaluation key's $P$ for all levels $\ell$, we need to set $P \ge N q_L \kappa$ for the following lemma.
}

\begin{lemma}
\label{lemma:noise_multiplication}
Let $\ct_{\sf mult} \leftarrow {\sf Mult}_\evk(\ct_1,\ct_2)$ for two ciphertexts $\ct_1,\ct_2 \in {\cal R}_{q_\ell}^2$.
If $P \ge Nq_{\ell}\kappa $, then $\langle \ct_{\sf mult},\sk \rangle=\langle \ct_1,\sk \rangle \cdot 
\langle \ct_2,\sk \rangle + e_{\sf mult} \pmod{q_\ell}$ for some $e_{\sf mult}$ satisfying 
$\infnorm{e_{\sf mult}} \leq B_{\sf mult} := 2N$.
\end{lemma}

Writing $\langle \ct_1,\sk \rangle=m_1+e_1 \pmod{q_\ell}$ and 
$\langle \ct_2,\sk \rangle=m_2+e_2 \pmod{q_\ell}$, 
the previous lemma shows that only the high-order bits of $m_1m_2$ can be recovered, 
as we get $\langle \ct_{\sf mult},\sk \rangle=(m_1+e_1) \cdot (m_2+e_2)+e_{\sf mult}=
m_1m_2+e^\star \pmod{q_{\ell}}$, and the 
 low-order bits 
of $m_1 m_2$ are masked by the quite large error $e^\star=e_1m_2 + e_2 m_1 + e_1 e_2+e_{\sf mult}$.
Applying Lemma \ref{lemma:noise_rescaling}, one can then perform a modulus switching to rescale the product, which then gives:
$$
\langle \ct'_{\sf mult},\sk \rangle = \near{ m_1m_2 /\Delta} + e'_{\sf mult} \pmod{q_{\ell-1}}.
$$
Therefore, one obtains an encryption of the scaled product $\near{ m_1 m_2/\Delta }$.
\ignore{
 see Figure \ref{figure:scaled_mult} for an illustration.
As shown in the next section, 
this enables to emulate fixed-point arithmetic.

\begin{figure}
\centering
\adjustbox{max width=\textwidth}{
\input{mult_CKKS.tikz}
}
\caption{The scaled multiplication in \ckks: first multiplication,
then rescaling.}
\label{figure:scaled_mult}
\end{figure}
}

\subsection{Encrypting single real numbers}

\label{s:reals}

For simplicity, we first recall the CKKS encryption of single real
numbers. We then recall the encryption of $N/2$ complex numbers in
parallel in Section \ref{s:CKKS-packing}.

Specifically, 
 the \ckks scheme can encrypt a real value \( x \in \mathbb{R} \) with a specified precision, 
by rounding to an integer after scaling.
This enables to 
 define (approximate) homomorphic 
operations directly over ${\mathbb R}$.
More precisely, to encode a real value $x\in {\mathbb R}$ 
with precision $\Delta$, we use
\begin{equation}
\label{eq:Ecdreals}
 {\sf Ecd}(x) = \lfloor x \cdot \Delta \rceil.
 \end{equation}
\ignore{
We have the approximate homomorphic properties:
\begin{align*}
{\sf Ecd}(x+y)  & \simeq {\sf Ecd}(x) + {\sf Ecd}(y) \\
{\sf Ecd}(x \cdot y) & \simeq {\sf Ecd}(x) \cdot {\sf Ecd}(y) /p
\end{align*}
} 
Therefore, to encrypt $x \in {\mathbb R}$, we  first encode $x$ 
into $m={\sf Ecd}(x)$, and then encrypt $m$ into
$\ct={\sf Enc}_{\pk}(m)$, which gives  
$\langle \ct,\sk \rangle = {\sf Ecd}(x) + e \pmod{q_L}$.
Since  ${\sf Ecd}(x) \in {\mathbb Z}$,
 we are using only a single coefficient of the plaintext space ${\cal R}={\mathbb Z}[X]/(X^N+1)$.
Formally, we define
the following encryption procedure for real numbers, and the homomorphic multiplication of two ciphertexts of real numbers.

\begin{itemize}
\item ${\sf EncR}_{\pk}(x)$: for $x \in {\mathbb R}$, return $\ct \leftarrow  {\sf Enc}_{\pk}({\sf Ecd}(x))$.

\smallskip

\item ${\sf MultR}_{\evk}(\ct_1,\ct_2)$: return ${\sf RS}({\sf Mult}_\evk(\ct_1,\ct_2))$

\smallskip

\item 
${\sf ExtMultR}(\ct, v)$: return ${\sf RS}({\sf ExtMult}(\ct,v))$

\end{itemize}

\begin{figure}
\centering
\adjustbox{max width=\textwidth}{
\newcommand{\civv}[8]{
\node[#1,draw,thick,minimum width=#2,minimum height=.5cm] (#3) {};
\node[left=0cm of #3.east,draw,thick,fill=blue!10,inner sep=0pt,minimum height=.5cm,minimum width=#4] (m) {};
\node[left=0cm of #3.east,draw,thick,fill=blue!25,inner sep=0pt,minimum height=.5cm,minimum width=#5] (no) {\normalsize #6};
\node[left=0cm of #3.west] () {\normalsize #7};
\draw[decorate,decoration={brace,raise=.07cm}]
(m.north west) -- (#3.north east) node[above=.1cm,pos=.5] { \normalsize #8};
}

\begin{tikzpicture}[auto, node distance=1cm, >=latex']
\begin{scope}[scale=.95, transform shape]

\civv{}{5cm}{ci1}{2cm}{.5cm}{$e_1$}{$q$}{${\sf Ecd}(x_1)$}
\civv{below=1cm of ci1}{5cm}{ci2}{2cm}{.5cm}{$e_2$}{$q$}{${\sf Ecd}(x_2)$}

\node [right=.8cm of ci1,yshift=-.7cm,draw,thick,circle,minimum size=.2,inner sep=0pt] (prod) {$\times$};

\draw[->] ([xshift=.1cm,yshift=-.1cm] ci1.south east) -- (prod);
\draw[->] ([xshift=.1cm,yshift=.1cm] ci2.north east) -- (prod);

\civv{right=1cm of prod}{5cm}{ci3}{4cm}{2.5cm}{$e^\star$}{$q$}{${\sf Ecd}(x_1) \cdot {\sf Ecd}(x_2)$}

\draw[->] (prod) -- ([xshift=-.5cm] ci3.west);

\civv{right=3cm of ci3,xshift=-1cm}{3cm}{ci4}{2cm}{.5cm}{$e'$}{$q/\Delta$}{${\sf Ecd}(x_1 \cdot x_2)$}

\draw[->] ([xshift=.3cm]ci3.east) -- ([xshift=-.9cm] ci4.west);

\end{scope}
\end{tikzpicture}
}
\caption{Homomorphic multiplication of real numbers in \ckks: given two 
encryptions of $x_1, x_2 \in {\mathbb R}$, we
obtain an encryption of $x_1 \cdot x_2 \in {\mathbb R}$, albeit for
a smaller modulus $q/\Delta$.}
\label{figure:scaled_mult_real}
\end{figure}
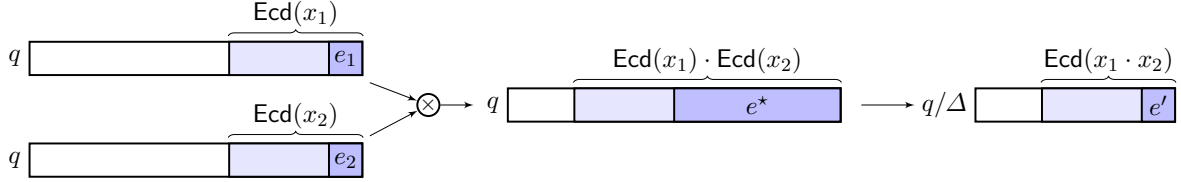

Thanks to the rescaling by a factor $\Delta$, given as input two ciphertexts
$\ct_1$, $\ct_2$ encrypting $x_1,x_2 \in {\mathbb R}$ under a modulus $q_\ell$, we can
obtain a new ciphertext $\ct={\sf MultR}_{\evk}(\ct_1,\ct_2)$
encrypting ${\sf Ecd}(x_1) \cdot {\sf Ecd}(x_2)/\Delta \simeq {\sf Ecd}(x_1 \cdot x_2)$, therefore an encryption of 
the product $x_1 \cdot x_2 \in {\mathbb R}$.
However, the new encryption is under a smaller modulus $q_{\ell-1}=q_\ell / \Delta$.
 This is captured in the following lemma;  
see Fig. \ref{figure:scaled_mult_real} for an illustration.

\begin{lemma}
\label{lemma:noise_multiplication_encoding}
Let $\ct_{\sf multR} \leftarrow {\sf MultR}_{\evk}(\ct_1,\ct_2)$
for two ciphertexts $\ct_1,\ct_2 \in {\cal R}_{q_\ell}^2$
such that 
$\langle \ct_i,\sk \rangle={\sf Ecd}(x_i)+e_i \pmod{q_\ell}$ for $i=1,2$,
for $x_i \in {\mathbb R}$ with $|x_i | \leq \nu<N$, and  $\infnorm{e_i}  \leq E$, where $E^2 \leq \Delta$. 
Then $\langle \ct_{\sf mult},\sk \rangle= {\sf Ecd}(x_1 \cdot x_2) +e_{\sf multR} \pmod{q_{\ell-1}}$
for some $e_{\sf multR}$ satisfying $\|e_{\sf multR}\|_{\infty} \leq 
 2\nu E+8N$.
\end{lemma}

\ignore{
In particular, given a degree $d$ polynomial $P(x)$ and a ciphertext $\ct$ encrypting $x$ modulo $q$, we can obtain
a ciphertext $\ct'$ encrypting $P(x)$ modulo $q/\Delta^d$. 
}

\subsection{Packing method for parallel encryption of complex numbers}

\label{s:CKKS-packing}

In \cite{CKKS17,CHKKS18}, the authors described a packing method to encrypt
multiple messages in a single ciphertext. Through a ring isomorphism, 
the set ${\cal S}={\mathbb  R}[X] / (X^N+1)$ 
can be identified with the complex coordinate space ${\mathbb C}^{N/2}$. 
This enables to encrypt $N/2$ 
complex numbers is a single ciphertext, and perform
parallel operations in a SIMD manner.

More precisely, the authors considered a variant of the Fourier transform
$$
{\sf DFT} : {\mathbb R}[X] / (X^N+1) \rightarrow {\mathbb C}^{N/2}, \qquad p \mapsto \left(p\left(\zeta^{5^0}\right), p\left(\zeta^{5^1}\right), \dots, p\left(\zeta^{5^{N/2-1}}\right) \right)
$$
where $\zeta$ is a primitive $(2N)$-th root of unity. We denote by
${\sf iDFT} : {\mathbb C}^{N/2} \rightarrow {\cal S}$ the inverse transform.
% Sentence below is redundant/can be merged with the one before.
The {\sf DFT} function is  a ring
isomorphism from the ring ${\cal S}$ (the coefficients space) to
the ring of complex vectors in ${\mathbb C}^{N/2}$ (the slots space).
Its purpose is therefore to map polynomial addition/multiplication in the ring
${\cal S}$ to component-wise addition/multiplication in ${\mathbb C}^{N/2}$.

\subsubsection{Encoding.}
\begin{figure}
\centering
\tikzset{mymatr/.style={every outer matrix/.append style={draw=black, inner xsep=4pt , inner ysep=6pt, rounded corners, thick}}}
\adjustbox{max width=\textwidth}{
\begin{tikzcd}[mymatr,column sep=large,row sep=small]
  \vec{z}=(z_1,\cdots,z_{N/2})  \arrow[r, "{\sf \Large Encode}"] 
&  m(X)={\sf Ecd}(\vec{z})   \arrow[r, "{\sf \large Encrypt}"] 
&  c={\sf Enc}(m) \\[-.3cm] 
\text{Component-wise}  &   {\rm Polynomial} & {\rm Homomorphic} \\[-.5cm] 
{\rm ~add.~and~mult.} & {\rm ~add.~and~mult.} & {\rm ~add.~and~mult.} \\[-.3cm]
{\rm Slots~space}={\mathbb C}^{N/2}		 & {\rm Coefficient~space}={\cal R}
& {\rm Ciphertext~space} = {\cal R}_q
\end{tikzcd}
}
\caption{Packing method for \ckks.}
\label{f:CKKS-packing}
\end{figure}
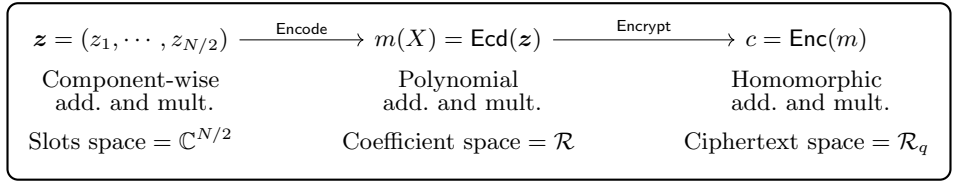

For \rlwe-type schemes, we must encode our $N/2$ complex values into polynomials with integer
coefficients in the plaintext ring ${\cal R}={\mathbb Z}[X]/(X^N+1)$, 
instead of ${\mathbb R}[X] / (X^N+1)$. 
Therefore, the encoding map ${\sf Ecd} : {\mathbb C}^{N/2} \rightarrow
{\cal R}$ uses an integer approximation of ${\sf iDFT}$, with a
scaling factor $\Delta$:
$$ \forall {\vec z} \in {\mathbb C}^{N/2} : {\sf Ecd}(\vec{z}) =
\lfloor \Delta \cdot {\sf iDFT}(\vec{z}) \rceil$$
and the corresponding decoding map ${\sf Dcd} : {\cal R} \rightarrow {\mathbb C}^{N/2}$
is defined as
$ \forall p \in {\cal R} : {\sf Dcd}(p)=\frac{1}{\Delta} \cdot {\sf
  DFT}(p)$.
 We denote by $\times$ the scaled multiplication over ${\cal R}$, \ie
 $p_1 \times p_2=\lfloor p_1 \cdot p_2 /\Delta \rceil$
 for any  $p_1,p_2 \in {\cal R}$.
We then have the approximate homomorphic properties:
\begin{align*}
    {\sf Ecd}( \vect z_1 + \vect z_2) & \simeq {\sf Ecd}(\vect z_1)
   + {\sf Ecd}(\vect z_2) \\
   {\sf Ecd}( \vect z_1 \odot \vect z_2) & \simeq {\sf Ecd}(\vect z_1)
  \times {\sf Ecd}(\vect z_2)
\end{align*}
where the vectors $\vect z_1$  and $\vect z_2$ are added and multiplied component-wise.
Therefore,  to perform parallel
addition/multiplication on $N/2$ complex 
values, we can first 
encode them as a polynomial $p(X)$
using ${\sf Ecd}$, and then perform ring addition / multiplication in ${\cal R}$.
Eventually, we can decode the polynomial into $N/2$ complex values using ${\sf Dcd}$;
see Fig.
\ref{f:CKKS-packing} for an illustration.

\subsubsection{Sparse packing.}
% Not only for efficiency reasons...
For efficiency reasons, it can be advantageous to encode fewer than $N/2$ components. 
We consider a power-of-two $n$ dividing $N/2$. 
The sub-ring ${\mathbb R}[X^{N/(2n)}]/(X^{N}+1) \simeq {\mathbb R}[Y]/(Y^{2n}+1)$
is isomorphic to ${\mathbb C}^{n}$.
In particular, if $p \in {\mathbb R}[X^{N/(2n)}]/(X^{N}+1)$, then
${\sf DFT}(p)$  contains $N/(2n)$ repetitions of the same $n$ components.  We can
therefore extend the encoding map ${\sf Ecd}$ to encode vectors of $n$
complex values instead of $N/2$,  as follows:
\begin{equation}
\label{eq:Ecdz}
 \forall {\vec z} \in {\mathbb C}^{n} : {\sf Ecd}(\vec{z}) =
\lfloor \Delta \cdot {\sf iDFT}\repl{\vec{z}}{N/(2n)} \rceil \in {\mathbb Z}[X^{N/(2n)}]/(X^{N}+1).
\end{equation}
In particular, for $n=1$, we have 
${\sf Ecd}(z) = \near{\Delta \cdot {\sf Re}(z)} + X^{N/2} \cdot  \near{\Delta \cdot {\sf Im}(z)} $
for any $z \in {\mathbb C}$.  
Therefore, the encoding map ${\sf Ecd}$ defined in (\ref{eq:Ecdz}) is a generalization of the 
encoding function defined in (\ref{eq:Ecdreals})
 for encoding $x \in {\mathbb R}$.

\subsubsection{Message packing into a ciphertext.}
Thanks to this generalized encoding function for any $\vec{z} \in {\mathbb  C}^n$, 
we can extend the
encryption procedure ${\sf EncR}_\pk(x)$ from Section \ref{s:reals}
to any $\vec{z} \in {\mathbb  C}^n$, for any power-of-two $1 \leq n \leq N/2$:

\begin{itemize}
\item ${\sf EncR}_\pk(\vec{z})$: for $\vec{z} \in {\mathbb C}^n$, return $\ct \leftarrow  {\sf Enc}_\pk({\sf Ecd}(\vec z))$.
\end{itemize}

To perform a homomorphic multiplication, we employ the same rescaled multiplication procedure
${\sf MultR}_\evk(\ct_1,\ct_2)$ as in the previous section. 
Given ciphertexts $\ct_1 \leftarrow {\sf EncR}_\pk( \vec{z}_1)$
and $\ct_2 \leftarrow {\sf EncR}_\pk( \vec{z}_2)$,
when applying ${\sf Add}(\ct_1,\ct_2)$
and ${\sf MultR}_\evk(\ct_1,\ct_2)$,
the vectors $\vec{z}_1,\vec{z}_2 \in {\mathbb C}^n$
are homomorphically added and  multiplied component-wise; see Figure \ref{f:CKKS-packing}
for an illustration.

\subsection{Rotation and conjugation of the slots}

\label{s:rot-conj}

As demonstrated in \cite{CKKS17}, rotation and conjugation of slots can be performed homomorphically by 
applying an automorphism \(\Psi\) to both components of a ciphertext. However, \(\Psi\) is also applied to 
the secret key, meaning the new ciphertext only decrypts correctly with the modified key \(s' = \Psi(s)\). 
Therefore, a key-switching procedure is required to revert to the original key \(s\).

\subsubsection{Key switching.}
The goal of key switching is to convert a ciphertext under
a secret key $\sk'=(1,s')$ into a ciphertext for the same message with respect to another
secret key $\sk=(1,s)$. The switching key is generated by the
procedure ${\sf swk} \leftarrow {\sf KSGen}_\sk(s')$.
Formally, the key switching proceeds as follows:
\begin{itemize}
\item 
${\sf KS_{swk}}(\ct)$: for $\ct=(c_0,c_1)$, return $\ct' \leftarrow (c_0,0) +
 \lfloor P^{-1} \cdot c_1 \cdot {\sf swk} \rceil \pmod{q_{\ell}}$.
 \end{itemize}

% Lemma out of context
%While a rotation itself does not increase the noise, a key switching
%does; this is captured in the following lemma.
% saying that rotation does not increase noise can be confusing.
\noindent
The noise growth from key switching is captured in the following lemma.

\begin{lemma}[Key switching]
  \label{lemma:noise_key_switching}
Let $\ct=(c_0,c_1) \in {\cal R}_{q_\ell}^2$ be a ciphertext with respect
to a secret key $\sk'=(1,s')$ and let ${\sf swk} \leftarrow
{\sf KSGen}_{\sf sk}(s')$. If $P \ge Nq_{\ell}\kappa $, then 
${\sf ct}' \leftarrow {\sf KS_{swk}}(\ct)$ 
satisfies $\langle \ct',\sk \rangle = \langle \ct,\sk' \rangle + e_{\sf ks} \pmod {q_\ell}$
for some $e_{\sf ks} \in {\cal R}$ with $\infnorm{e_{\sf ks}}  \leq 2N.$
\end{lemma}

\subsubsection{Rotation.}
For ${\cal S}={\mathbb R} [X]/ ( X^N+1 )$, the automorphism
$\Psi_r: {\cal S} \rightarrow {\cal S}$, given $m(X) \in {\cal S}$, returns $m(X^{5^r}) \pmod{X^N+1}$. 
It induces a rotation by $r$ positions to the left on the vector of encoded values.
Namely for any $m(X) \in {\cal S}$,  
letting ${\sf DFT}(m)=(z_j)_{0 \leq j <N/2}$, we get:
\begin{align*}
{\sf DFT}(\Psi_r(m)(X)) & =(\Psi_r(m)(\zeta^{5^j}))_{0 \leq j < N/2}=
(m((\zeta^{5^j})^{5^r}))_{0 \leq j < N/2} \\
& =(m(\zeta^{5^{r+j}}))_{0 \leq j < N/2}= ( z_{r+j \bmod N/2})_{0 \leq j < N/2}. % = (z_r,\ldots,z_{N/2-1},z_0,\ldots,z_{r-1})
\end{align*}
\ignore{
Since $X^{2N} = 1$ in $\cal R$ and $5^{N/2} = 1 \pmod{2N}$, we have that $\Psi_{N/2} = {\sf id}$.
}

\subsubsection{Conjugation.}
We also consider the automorphism ${\sf conj}: \cal S \rightarrow \cal S$
defined as ${\sf conj}(m)(X)=m(X^{-1}) \pmod{X^N+1}$. It enables 
to compute the conjugate of the values encoded into a polynomial, namely, 
we have that ${\sf DFT}({\sf conj}(m))=\overline{{\sf DFT}(m)}$. 
% Finally, this  enables to extract the imaginary part. 
We also denote by ${\sf Im2}:
{\cal S} \rightarrow {\cal S}$ the operator
$${\sf Im2}(m)=-X^{N/2} \cdot (m-{\sf conj}(m)).$$ 
Since $-\Delta \cdot X^{N/2}={\sf Ecd}(1/i)$,
given any $\vec{z} \in {\mathbb C}^{N/2}$, 
we have ${\sf DFT}({\sf Im2}({\sf iDFT}(\vec{z})))=2 \cdot {\sf Im}
(\vec{z})$. Therefore, the ${\sf Im2}$ operator enables to extract 
the imaginary part of the slots, up to a factor $2$.
Similarly, we define the ${\sf Re2}$ operator with ${\sf Re2}(m)= {\sf conj}(m)+m$,
so that ${\sf DFT}({\sf Re2}({\sf iDFT}(\vec{z})))=2 \cdot {\sf Re}
(\vec{z})$.

\subsubsection{Homomorphic rotation/conjugation of the slots.}
We summarize below the procedures for homomorphic rotation/conjugation
of the slots for a ciphertext $\ct = (c_0, c_1)$: \\[-.6cm] 

\begin{itemize}
\item ${\sf GenRot}_\sk(r)$: generate the rotation key ${\sf rk}_r \leftarrow
{\sf KSGen}_\sk(\Psi_r(s))$.

\smallskip
\item 
${\sf Rot}(\ct,r)$: return ${\sf KS}_{{\sf rk}_r}((\Psi_r(c_0), \Psi_r(c_1)))$.

\smallskip
\item
${\sf GenConj}_\sk$: generate the conjugation key ${\sf ck} \leftarrow
{\sf KSGen}_\sk({\sf conj}(s))$.

\smallskip
\item
${\sf Conj}(\ct)$: return ${\sf KS}_{\sf ck}(({\sf conj}(c_0), {\sf conj}(c_1)))$.

\smallskip
\item
${\sf Im2}(\ct)$: return the ciphertext ${\sf ExtMult}({\sf Conj}(\ct)-\ct, X^{N/2})$.
\end{itemize}

\ignore{
In practice, we do not need to store a rotation key ${\sf rk}_r$ 
for all possible values $1 \leq r< N/2$.
Instead, we store the
rotation key only for $r=2^i$ where $1 \leq 2^i \leq N/4$. 
}

\subsection{The trace and product operators}

\label{s:trace:product}

\subsubsection{Summing the slots.}

We recall how to homomorphically compute  partial or full summation of the slots. 
The sum of all slots could be obtained by repeatedly rotating 
the slots by one position and then accumulating the result by summing.
However, this would require $N/2-1$ rotations for $N/2$ slots, which would be impractical.
In the following, we recall the technique from \cite[Alg. 2]{CHKKS18},
 which uses only ${\cal O}(\log N)$ homomorphic 
operations. Summing all slots is algebraically known as the field trace operator 
\cite{CDKS20}.

Formally, for a power-of-two integer 
$a$ dividing $N/2$, using the $\Psi_r$ automorphism from the
previous section, we define the operator $\tr{N/2}{a}$ as 
% : {\mathbb Z}[X] / (X^N+1)  \rightarrow {\mathbb Z}[X^{N/(2a)}] / (X^N+1)$:
$$
\tr{N/2}{a} = ({\sf id} + \Psi_a) \circ 
({\sf id} + \Psi_{2a}) \circ ({\sf id} + \Psi_{4a}) \circ \dots \circ ({\sf id} + \Psi_{N/4}).
$$
We show in Appendix \ref{a:trace} that 
the $\tr{N/2}{a}$ operator computes partial sums of the slots
sharing the same index modulo $a$:
$$\tr{N/2}{a}({\sf iDFT}(z_0,\ldots,z_{N/2-1})) =
{\sf iDFT}(\underbrace{\vec{t},\ldots\ldots,\vec{t}}_{N/(2a)~\textrm{rep.}})$$
for $\vec{t}=(t_k)_{0 \leq k < a}$,
with  $ t_k=z_k + z_{k+a} + \cdots + z_{k+(N/(2a)-1) \cdot a}$. 
Therefore, the  $\tr{N/2}{a}$ operator outputs an encoding of the $a$ slots $t_0,\ldots,t_{a-1}$,
which are repeated $N/(2a)$ times to get $N/2$ slots.
The $\tr{N/2}{a}$ operator can be homomorphically evaluated, 
requiring $\log_2 (N/(2a))$ automorphisms and key switchings, 
while consuming no multiplicative levels; see Appendix \ref{a:trace}.

\subsubsection{Multiplying the slots.}

Similarly to the trace operator, we can define the product operator to compute the
partial or full product of all slots, by replacing sums by scaled 
products in the definition. For powers-of-two $b < a \leq N/2$, we define the $\pr{a}{b}$
operator as
$$
\pr{a}{b} := ({\sf id} \times \Psi_b) \circ 
({\sf id} \times \Psi_{2b}) \circ
({\sf id} \times \Psi_{4b}) \circ
\cdots \circ ({\sf id} \times \Psi_{a/2})
$$
where as in Section \ref{s:CKKS-packing} 
we denote by $\times$ the scaled multiplication over ${\cal R}$, with 
 $p_1 \times p_2=\lfloor p_1 \cdot p_2 /\Delta \rceil$
 for any  $p_1,p_2 \in {\cal R}$.
Letting $\vec{z} \in {\mathbb C}^a$, the $\pr{a}{b}$ operator computes partial products of the slots
sharing the same index modulo $b$:
$$\pr{a}{b}({\sf Ecd}(\vec{z})) \simeq {\sf Ecd}(\vec{t})$$
where $\vec{t}=(t_k)_{0 \leq k < b}$, with 
$ t_k=z_k \cdot z_{k+b}  \cdots  z_{k+(a/b-1) \cdot b}$ for $0 \leq k<b$.
\ignore{Recall that with the extended definition of the ${\sf Ecd}$ map
from (\ref{eq:Ecdz}), for $z \in {\mathbb C}^a$ the encoding ${\sf Ecd}(\vec{z})$
contains $N/(2a)$ repetitions of the same slots $\vec{z}$.
Similarly,
${\sf Ecd}(\vec{t})$ will contain $N/(2b)$ repetitions of $\vec{t} \in {\mathbb C}^b$.}
The $\pr{a}{b}$ operator can be homomorphically evaluated using 
 $\log_2(a/b) $ automorphisms and multiplications;
thus, it consumes $\log_2(a/b) $ levels; see Appendix \ref{a:product}.
Note that the product operator $\pr{a}{b}$ corresponds to the standard field norm
  of $\mathbb Q(\zeta_{a})$ over $\mathbb Q(\zeta_{b})$, where 
  $\zeta_k$ denotes a primitive $k$-th root of unity.

\subsection{From slots back to polynomial coefficients}

\label{s:slottocoeff}

Our bootstrapping algorithm for multiple slots requires putting the slots back to the
coefficients of a polynomial; this is the $\stc$ operation. The original \ckks bootstrapping
requires both $\cts$ and $\stc$ operations
\cite{CHKKS18}, while our technique
only requires $\stc$, so we only describe the latter; the two operations
are the inverse of each other.

% avoid operation overload below
Formally, $\stc_n$ takes as input a polynomial $v(X)$
encoding $n$ slots $\vec{t}=(t_0,\ldots,t_{n-1}) \in {\mathbb R}^n$:
$$ v(X)={\sf Ecd}(t_0,\ldots,t_{n-1})  \in {\mathbb Z}[X^{N/(2n)}] / (X^N+1)$$
and outputs a polynomial $m(X) \in {\mathbb Z}[X^{N/n}] / (X^N+1)$
whose $n$ coefficients are the $t_j$'s, scaled by a factor $\Delta \in {\mathbb Z}$:
\begin{equation}
\label{eq:mXStC}
m(X)=\stc_n(v)=\sum\limits_{j=0}^{n-1} \lfloor \Delta \cdot t_j  \rceil\cdot X^{j \cdot N/n}
\end{equation}

%The \(\stc\) operation can be homomorphically evaluated on ciphertexts. 
Since \(\stc\) is a linear map, it can be homomorphically evaluated  as a matrix-vector multiplication, 
requiring \({\cal O}(n)\) homomorphic operations \cite{CHKKS18}. However, 
the homomorphic Discrete Fourier Transform (DFT), which is the primary computation within \(\stc\), 
can be performed more efficiently with only \({\cal O}(\log n)\) homomorphic operations \cite{CHH18,CCS19}. 
In this case, the polynomial coefficients in (\ref{eq:mXStC}) are recovered in a bit-reversed order.
We refer to Appendix \ref{a:slottocoeff} for a detailed description.
\ignore{
In this paper, we employ the fast DFT-based algorithm \cite{CHH18,CCS19}
with complexity %for both \(\cts\) and \(\stc\), which has a
 \({\cal O}(\log n)\). 
To further reduce its multiplicative depth from \(\log_2 n\), we use a radix-4 variant, resulting 
in a lower depth of \(\ell_{\sf CtS} = \ell_{\sf StC} = \lfloor (\log_2 n)/2 \rfloor + 1\). 
We refer to Appendix \ref{a:slottocoeff} for a detailed description.
}

\section{New bootstrapping equation for \ckks: single slot}
% \ckks notation!

We now introduce our new bootstrapping algorithm,
which is quite different from the original \ckks bootstrapping
recalled in the previous section. 
 For simplicity, we first consider a \ckks ciphertext $\ct$ 
encrypting a single slot, that is
$\langle \ct, \sk \rangle = m +e \pmod{q}$ for
$m \in {\mathbb Z}$ and $e \in \cal R$.
We will describe the bootstrapping of $n \geq 2$ slots in parallel
in Section \ref{s:boot:mult}.

An RLWE-format ciphertext can be viewed as (multiple) LWE-format 
ciphertexts.
Therefore, we first convert $\ct=(b,a)$ into an LWE ciphertext
$\vec{c} \in {\mathbb Z}_q^{N}$ for the same $m \in {\mathbb Z}$, \ie we set $\vec c = (a_0 + b_0, -a_{N-1}, -a_{N-2}, \dots, -a_1)$ and $\vec s = (s_0, \dots, s_{N-1})$, assuming that $s_0=1$. 
LWE decryption now comprises computing the dot product 
$\langle \vec c, \vec s \rangle = m +e_0 \pmod q$ for $e_0 \in {\mathbb Z}$. 
Unlike the original \ckks bootstrapping, which uses sparse ternary secrets, 
we first assume  that \(\vec{s}\) is a random binary vector,  not necessarily 
constrained to a low Hamming weight.

\subsection{Our new bootstrapping equation}

Our new bootstrapping equation consists in embedding the additive
group ${\mathbb Z}_q$ into the multiplicative circle group of complex numbers, using the homomorphism:
$$
f:  {\mathbb Z}_q   \rightarrow {\mathbb C}, \qquad x \longmapsto \exp(2i\pi \cdot x/q).
$$
We can therefore compute the decryption equation 
$m = \langle \vec{s}, \vec{c} \rangle = \sum_{k=0}^{N-1} s_k c_k \pmod q$ 
over the multiplicative circle group in $\mathbb{C}$:
$$
\exp(2i\pi \cdot  m/q) 
=\prod\limits_{k=0}^{N-1} \exp(2i\pi \cdot c_{k}  s_k/q) 
$$
Moreover, since $s_k \in \{0,1\}$ works as a selector, we can write:
\begin{equation}
\label{eq:2ipisc}
\exp(2i\pi \cdot  m/q) 
=\prod\limits_{k=0}^{N-1}  \left(1 + \left( \exp(2i\pi \cdot c_{k} /q) - 1 \right) \cdot s_k  \right).
\end{equation}
By isolating the imaginary part and assuming $m \ll q$, we obtain:
\begin{equation}
\label{eq:2pimqsin}
\frac{2\pi \cdot m}{q} \simeq  \sin \left( \frac{2\pi \cdot m}{q} \right)=
{\sf Im} \left(
\prod\limits_{k=0}^{N-1}  \left(1 + \left( \exp(2i\pi \cdot c_{k} /q) - 1 \right) \cdot s_k  \right)
\right).
\end{equation}
We want to obtain an encryption of $m \in {\mathbb Z}$, which corresponds
to an encoding of $m/\Delta \in {\mathbb R}$. Therefore, we scale the above equation
by a factor $q/(2 \pi \Delta)$, and we use the approximation:
\begin{equation}
\label{eq:approxmDelta}
\frac{q}{2\pi \Delta} \cdot
\sin \left( \frac{2 \pi m}{q} \right)= 
\frac{q}{2\pi \Delta} \cdot 
\left( \frac{2 \pi m}{q} + {\cal O}\left( \frac{m^3}{q^3} \right) \right)
= \frac{m}{\Delta} + {\cal O}\left( \frac{m^3}{q^2 \Delta} \right). 
\end{equation}
Eventually, we obtain from (\ref{eq:2pimqsin}):
\begin{equation}
\label{eq:boot_basic}
\frac{m}{\Delta} \simeq \frac{q}{2\pi \Delta} \cdot  {\sf Im}   \left( \prod\limits_{k=0}^{N-1}  \left(1 + \left( \exp(2i\pi \cdot c_{k} /q) - 1 \right) \cdot s_k  \right) \right).
\end{equation}
From (\ref{eq:approxmDelta}), the above approximation requires 
$m={\cal O}(q^{2/3})$, the same condition as in the original \ckks bootstrapping.

\subsubsection{Bootstrapping.}
We claim that (\ref{eq:boot_basic}) provides a new bootstrapping equation for \ckks,
as it can be homomorphically evaluated to provide a new
ciphertext for the same $m \in {\mathbb Z}$, but
under a larger modulus $q \cdot p$ than the original modulus $q$.
For this, 
we stress that in the above
equation, we consider all variables over ${\mathbb R}$,
and since $\exp(2i\pi \cdot c_{k} /q) \in \mathbb C$, we treat 
its real and imaginary part separately. Each complex multiplication 
decomposes into $4$ multiplications and $2$ additions over $\mathbb R$.
Since a complex addition also decomposes into two real additions, we can work entirely over ${\mathbb R}$.
Similarly, we view the final scaling as a multiplication by $q/(2\pi \Delta) \in {\mathbb R}$, and the final output $m /\Delta \in {\mathbb R}$.

Correspondingly, for the homomorphic evaluation of (\ref{eq:boot_basic}), we 
consider all ciphertexts as encryptions of real numbers, 
using the encryption function ${\sf EncR}(x)$ for $x \in {\mathbb R}$ introduced in Section 
\ref{s:reals}. Consequently, all homomorphic operations are performed
on encryptions of reals.
%In particular, we use ${\sf MultR}_\evk$ for multiplying ciphertexts (see Section \ref{s:reals}), such that the underlying 
%real number plaintexts get multiplied. 
Eventually, since $m/\Delta \in {\mathbb R}$ is encoded as $m={\sf Ecd}(m/\Delta) \in {\mathbb Z}$, the encryption of $m/\Delta$ will correspond to an 
encryption of the original plaintext $m \in {\mathbb Z}$, 
but under a larger modulus $q \cdot p$.

Therefore, for bootstrapping, we assume that we are given encryptions 
 $S_k \leftarrow {\sf EncR}_\pk(s_k)$ of each 
bit $s_k \in \{0,1\}$ of the secret key $\vec s$. Here, although
$s_k \in \{0,1\}$, we again view $s_k \in {\mathbb R}$.
\ignore{
Similarly, we consider encryptions of the variables $\exp(2i\pi \cdot c_{k} /q) - 1$
for each $c_k \in {\mathbb Z}_q$; 
since $\exp(2i\pi \cdot c_{k} /q) \in \mathbb C$, we 
encrypt the real and imaginary parts separately.
}
Then, by homomorphic multiplications
(using {\sf ExtMultR}) and homomorphic additions, we obtain encryptions of
$1+\left( \exp(2i\pi \cdot c_{k} /q) - 1 \right) \cdot s_k$ for each $0 \leq k<N$,
while still encrypting the real/imaginary parts separately.
The product of the $N$ terms can then be homomorphically computed with a tree of multiplicative depth $\log_2 N$, such that, according to (\ref{eq:2ipisc}), we obtain an encryption
of $\exp(2i\pi \cdot m/q)$. 
Eventually, we only keep the encryption of its imaginary part and scale by $q/(2\pi\Delta)$. According to (\ref{eq:boot_basic}), this provides an encryption of the real $m/\Delta$, or
equivalently, an encryption of the integer $m={\sf Ecd}(m/\Delta)$.

The circuit corresponding to (\ref{eq:boot_basic}) has a total multiplicative depth of $\ell=\log_2 (N) +2$.
Thus, starting with a modulus $Q=q  \cdot p \cdot
\Delta^{\ell}$ for the encryptions $S_k$, after $\ell$ rescalings, we end up with an encryption of $m$ under the modulus $q \cdot p > q$. This achieves bootstrapping since there remains 
a level for a single homomorphic multiplication. For this last level, 
we can use a smaller scaling factor $p \simeq q^{2/3}$, because the plaintext message $m$ 
must be encoded under the condition $m={\cal O}(q^{2/3})$ for (\ref{eq:approxmDelta}).
On the other hand, we must use
a scaling factor $\Delta={\cal O}(q)$ for the main bootstrapping evaluation.
We prove the following bootstrapping theorem summarizing the above in
Appendix \ref{a:proofboot}. 
%We remark that the bounds are not tight, and that 
%in practice one can use a much smaller value for $\Delta$. 

\begin{theorem}[Bootstrapping]
  \label{theorem:bootstrapping}
Given a ciphertext $\ct$ such that
$ \infnorm{[\langle \ct,\sk \rangle]_q} < q^{2/3}$
and $\scp{\ct}{\sk} = m + e \pmod{q}$ for $m \in {\mathbb Z}$ and $e \in \cal R$,
the above evaluation procedure with 
$\Delta \geq \max(4 N^2 \kappa q,(20N^{3}\kappa)^2)$
outputs a new ciphertext $\ct'$
such that $ \langle \ct',\sk \rangle =m+e(0) +e_{\sf bt} \pmod {p \cdot q}$,
where $e(0)$ is the constant coefficient of $e$, 
and $\|e_{\sf bt}\| \leq B_{\sf bt}$ with $B_{\sf bt}:=10N$.
\end{theorem}

As opposed to the original \ckks bootstrapping,
our new bootstrapping equation (\ref{eq:boot_basic})  does not require
a small Hamming weight secret key; however, we describe an
optimization based on such a secret key in Section \ref{s:boot-trace}. 
We note that the above bootstrapping method is fully compatible with RNS arithmetic. 
In that case, the moduli are of the form $Q_r=\prod_{0 \leq i \leq r} q_i$, for small primes $q_i$.% , instead of $Q_r=q \cdot p \cdot \Delta^{r-1}$. 

\subsection{First optimization: packing the secret-key bits}
\label{sec:packing_sk_bits}

For a ring degree $N$, the above bootstrapping algorithm requires ${\cal O}(N)$ homomorphic operations,
which would be  impractical. Recall that the \ckks scheme provides a packing method
where $N/2$ plaintext slots can be packed into a single ciphertext.
%  (see Section \ref{s:CKKS-packing}).
% We don't have to refer to Section 3 at this point, since we mention the Fig. immediately after.
 These $N/2$ slots 
can then be added and multiplied independently, by performing homomorphic
operations on ciphertexts (see Fig. \ref{f:CKKS-packing} in Section \ref{s:CKKS-packing}).
Therefore, the first natural optimization consists of packing the $N$ secret 
key bits $s_k$ from (\ref{eq:boot_basic}) 
into the $N/2$ slots of two \ckks ciphertexts, instead of encrypting each bit separately. 
Similarly, we pack the terms $\exp(2i\pi \cdot c_{k} /q) - 1$ into two polynomial encodings.
%While this results in a precomputational step for the $s_k$ packings, it does not for the $c_k$ packings.
% Actually, the encryption of the s_k becomes part of the public key, this is not really a pre-computation
The products $ \left( \exp(2i\pi \cdot c_{k} /q) - 1 \right) \cdot s_k$ are then 
computed independently and in parallel in the slots, thus requiring only two homomorphic 
external multiplications ({\sf ExtMultR}). 
Moreover, the bootstrapping key size is also reduced from $N$ to
only two ciphertexts. 

% I think that the previous formulation was more clear, so I reverted to it (for the second sentence).
Another advantage of the packing method is that 
 \ckks natively supports complex numbers in the slots, so we do not need to encrypt the real/imaginary part separately anymore. 
 To compute the product in (\ref{eq:boot_basic}),  we can homomorphically
 apply the $\pr{N/2}{1}$ operator from
Section \ref{s:trace:product}
  to compute the product of the $N/2$ slots in each ciphertext. As a consequence, in the entire procedure, the multiplicative depth
remains $\ell=\log_2 (N)+2$, but now the number of homomorphic multiplications is reduced from ${\cal O}(N)$ to ${\cal O}(\log N)$. 
We will describe this optimization more formally in Section \ref{s:boot_single},
when combined with the following second optimization.

\subsection{Second optimization: using the trace operator}

\label{s:boot-trace}

In order to be competitive with the original \ckks bootstrapping, we describe a second optimization.
 It further reduces the  
multiplicative depth from ${\cal O}(\log N)$ to ${\cal O}(\log h)$, with $h \ll N$,
by replacing most homomorphic multiplications by additions, thus boosting efficiency.

 For this, we still consider a  binary secret key $s$ 
 but with a small Hamming weight $h$, where $h$ is a power-of-two, moreover with a special block structure (as in e.g. \cite{LMSS23}).
We assume that the coefficient vector $\vec{s} \in \{0,1\}^N$ can be split
into $h$ blocks, each of size $B=N/h$, where each block contains a single $1$. 
Formally,  we write 
$ \vec{s}=(s_{b \cdot B+j})_{0 \leq b <h,\,0 \leq j<B}$
where in the $b$-th block, $s_{b \cdot B+j}=0$ for all 
$j \in \{0,\ldots,B-1\} \setminus \{j^\star_b \}$, and
 $s_{b \cdot B+j^\star_b}=1$ for some 
$0 \leq j^\star_b < B$.

As previously, we consider an LWE ciphertext
 $\vec{c} \in {\mathbb Z}_q^N$ derived 
from the original \rlwe ciphertext, with decryption equation:
$$m \simeq \langle \vec{s},\vec{c} \rangle=\sum\limits_{k=0}^{N-1} s_k c_k=
\sum\limits_{b=0}^{h-1}  \sum\limits_{j=0}^{B-1}  s_{b \cdot B+j}
\cdot c_{b \cdot B+j} \pmod{q}$$
and, again, we embed the above equation
into the multiplicative circle group of complex numbers:
$$
\exp(2i\pi \cdot  \langle \vec{s}, \vec{c} \rangle/q)=\prod\limits_{b=0}^{h-1}  \prod\limits_{j=0}^{B-1} \exp(2i\pi 
\cdot s_{b \cdot B+j} \cdot c_{b \cdot B+j}/q).
$$
Since within a given block there is only a 
single $s_{b \cdot B+j}$ equal to $1$ and the others are zero, we can replace
the products within a block by a sum:
$$\exp(2i\pi \cdot  \langle \vec{s}, \vec{c} \rangle/q)=
\prod\limits_{b=0}^{h-1}
\sum\limits^{B-1}_{j=0} s_{b \cdot B+j} \cdot  
\exp(2i\pi  \cdot c_{b \cdot B+j}/q).
$$
As previously, this enables to derive  the following bootstrapping
equation:
%which is similar  to (\ref{eq:boot_basic}):
$$
\frac{m}{\Delta} \simeq \frac{q}{2\pi \Delta} \cdot  {\sf Im}   \left( \prod\limits_{b=0}^{h-1}  \sum\limits^{B-1}_{j=0} s_{b \cdot B+j} \cdot  
\exp(2i\pi  \cdot c_{b \cdot B+j}/q) \right).
$$
We can remove the last multiplication 
by $q/(2\pi \Delta)$ by scaling
%the terms $\exp(2i\pi c_{b \cdot B+j}/q)$
by a factor $\delta \in {\mathbb R}$ such that $\delta^h=q/(2\pi \Delta)$. This yields our final bootstrapping equation in the single slot case:
\begin{equation}
\label{eq:boot_trace}
\frac{m}{\Delta} \simeq {\sf Im}   \left( \prod\limits_{b=0}^{h-1}  \sum\limits^{B-1}_{j=0} s_{b \cdot B+j} \cdot  
\exp(2i\pi  \cdot c_{b \cdot B+j}/q)  \cdot \delta \right).
\end{equation}

The advantage of the above equation is that it has a smaller
multiplicative depth than  (\ref{eq:boot_basic}), since we are only
computing the product of $h$ factors instead of $N$. More precisely,
the multiplicative depth is now $\ell=\log_2 (h) + 1$ instead of $\ell=\log_2 (N)+2$. 
\ignore{ From the SHIP paper
We can further reduce the depth to $\ell=\log_2 (h)$ because the first plaintext-ciphertext 
multiplication can be performed without consuming any depth, by encrypting the secret-key elements 
using the 
auxiliary prime which is normally used for key switching.}
By lowering the multiplicative depth, we can decrease the size of the largest
modulus \( Q = q \cdot p \cdot \Delta^{\ell} \). After bootstrapping, the refreshed 
ciphertext is encrypted 
under the larger modulus $p \cdot q$ rather than the initial 
modulus $q$, enabling a single additional
homomorphic multiplication between two bootstrapped ciphertexts. Similar to the original CKKS 
bootstrapping, further homomorphic operations after bootstrapping can be supported 
by choosing a larger $Q$.

 \ignore{
 This reduction in \( Q \) enables the use of a smaller 
 ring dimension \( N \), while preserving the same level of security against lattice attacks, further 
 enhancing the efficiency of our scheme. Concrete parameters, along with practical implementation results, 
 will be provided in Section \ref{s:implem}.
}

\subsection{Our bootstrapping algorithm for a single component}

\label{s:boot_single}

\begin{figure}
\centering
\adjustbox{max width=\textwidth}{
\begin{tikzpicture}
\tikzstyle{bucket}=[rectangle, draw,minimum width=5cm,minimum height=.7cm]

\newcommand{\enc}[3]{
\node[bucket,#1] (#2) {\large #3};
\draw (#2) ++(-2.2,-.35) -- ++ (0,.7);
\draw (#2) ++(-1.9,-.35) -- ++ (0,.7);
\draw (#2) ++(1.9,-.35) -- ++ (0,.7);
\draw (#2) ++(2.2,-.35) -- ++ (0,.7);
}

\enc{}{B1}{$s_{b \cdot B+j}$}
\enc{below=.5cm of B1}{B2}{$s_{b \cdot B+j+\frac{B}{2}}$}
\node[right=.2cm of B1] {$\times$};
\node[right=.2cm of B2] {$\times$};
\enc{right=1cm of B1}{E1}{ \normalsize $\exp(2i\pi c_{b \cdot B+j} /q) \cdot \delta$ }
\enc{right=1cm of B2}{E2}{\normalsize $\exp(2i\pi c_{b \cdot B+j+\frac{B}{2}} /q) \cdot \delta$ }
\node[right=.2cm of E1] {$=$};
\node[right=.2cm of E2] {$=$};
\enc{right=1cm of E1}{P1}{$\cdots$}
\enc{right=1cm of E2}{P2}{$\cdots$}
\node[below=0cm of P1] {$+$};

\draw (P2) ++(-2.3,-.7) -- ++(4.6,0);

\enc{below=.7cm of P2}{S}{$\cdots$}
\enc{below=1cm of S}{T}{\normalsize $\exp(2i\pi \cdot c_{b \cdot B+j^\star_b} /q)$}

\draw [decorate,decoration={brace,amplitude=5pt,raise=4ex}]
  (-2.1,0) -- (2.1,0) node[midway,yshift=3em]{\normalsize $0 \leq j < B/2$};

\draw[->] (S.south) ++(0,-.1)  --  ++(0,-.8) node[midway,xshift=2.5em]{\normalsize ${\sf Tr}_{N/2 \rightarrow h}$};

\end{tikzpicture}
}
\caption{Illustration of the computation within the half-blocks of size $B/2$.
After the application of the $\tr{N/2}{h}$ operator,
all $B/2$ slots contain the same value $\exp(2i\pi \cdot c_{b \cdot B+j^\star_b})$,
where $j^\star_b$ is the index for which $s_{b \cdot B+j^\star_b}=1$. 
Note that each polynomial encoding contains $h$ such half-blocks, for a total of $h \cdot B/2=N/2$ slots.}
\label{fig:blocks}
\end{figure}
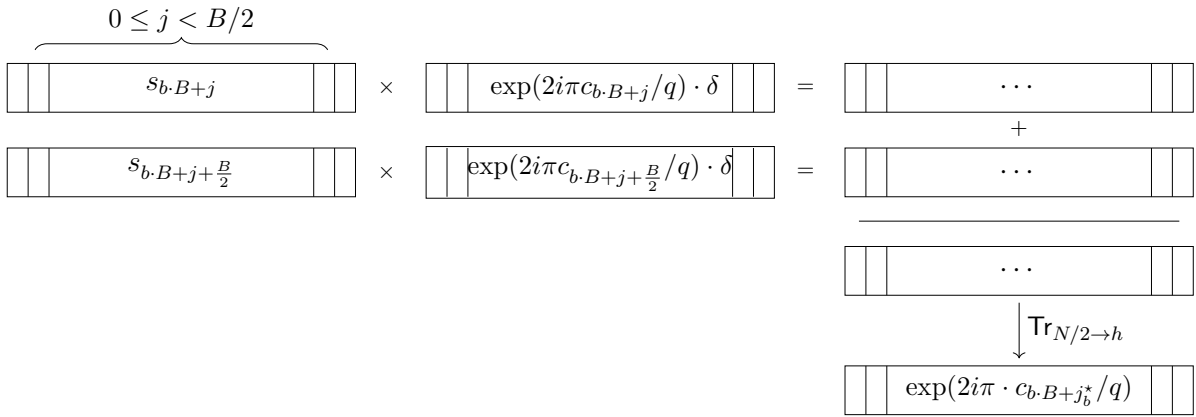

We now formally describe our bootstrapping algorithm for a single slot by
combining the two previous optimizations.
% Unnecessary, already stated in the previous sections multiple times.
% We will consider the bootstrapping of $n \geq 2$ slots in parallel
% in Section \ref{s:boot:mult}. 
We consider the sparse block secret key $\vec{s}$ as in Section \ref{s:boot-trace}, 
split into $h$ blocks, each of size $B=N/h$. 
In order to combine the two optimizations, 
we must rewrite equation (\ref{eq:boot_trace}) 
to support packed slots.
\ignore{
The variables from a given block will be stored in the slots of a polynomial encoding, 
and we  use the partial sum operator ${\sf Tr}$
recalled in Section \ref{s:trace:product} to compute the sum over the slots. 
}
Recall that we must encode $N$ secret-key bits, but a polynomial over the plaintext space
${\mathbb Z}[X]/(X^N+1)$ can only
encode $N/2$ slots, so we need two polynomial encodings.
To speed up computation, we  split each block of the secret key in two 
halves. The first halves of each block will be placed in the first polynomial encoding, and the second
halves in the second encoding; see Figure \ref{fig:blocks} for an illustration.

After the external multiplications 
by the encoding of the
$\exp(2i\pi \cdot c_{b \cdot B + j}/q)$ values, we compute the sum in (\ref{eq:boot_trace}) by  
first adding the two  polynomial encodings, while the
rest of the sum's evaluation within each block is handled by the $\tr{N/2}{h}$ operator applied to a single encoding.

However, we must be careful with the indexing of the slots when applying the $\tr{N/2}{h}$ operator. 
As illustrated in Appendix \ref{a:trace}, %Figure \ref{f:trace},
$\tr{N/2}{h}$ computes the partial sums of slots
sharing the same index modulo $h$. This implies that we cannot
 encode the elements of the blocks
contiguously in the slots, but only separated modulo $h$. Therefore, we utilize a 
``reversed"
indexing, in which we put the $j$-th element of the $b$-th
block at index $j \cdot h + b$ for all $0 \leq j <B/2$
for the first half, and similarly at index $(j-B/2) \cdot h + b$
for all $B/2 \leq j <B$ for the second half.
More precisely, we denote by $\tilde{s}^{0}_\iota$ and $\tilde{s}^{1}_\iota$
the corresponding slots, for $0 \leq \iota <N/2$.
According to the above indexing, 
we let $\tilde{s}^{0}_{j \cdot h +b}=s_{b \cdot B + j}$
for $0 \leq j <B/2$ (first block halves), and
$\tilde{s}^{1}_{(j-B/2) \cdot h +b}=s_{b \cdot B + j}$
for $B/2 < j <B$ (second block halves),
for all $0 \leq b< h$. In total, the $N$ secret-key bits
$s_k$ are encoded into the two polynomials:
$$
S_0(X)={\sf Ecd}((\tilde{s}^{0}_\iota)_{0 \leq \iota <N/2}), \qquad
S_1(X)={\sf Ecd}((\tilde{s}^{1}_\iota)_{0 \leq \iota <N/2}).
$$
 
 We encode the ciphertext's components following the same indexing. More precisely, we define  the slots $(e^0_\iota)_{0 \leq \iota <N/2}$ and $(e^1_\iota)_{0 \leq \iota <N/2}$,
 with $e^0_{j \cdot h + b}=\exp(2i\pi \cdot c_{b\cdot B+j}/q) \cdot \delta$
for $0 \leq j<B/2$, 
and $e^1_{(j-B/2) \cdot h + b}=\exp(2i\pi \cdot c_{b\cdot B+j}/q) \cdot \delta$
for $B/2 \leq j < B$, 
for all $0 \leq b < h$. This yields the following polynomial encodings:
$$ E_0(X)={\sf Ecd}((e^{0}_\iota)_{0 \leq \iota <N/2}), \qquad
 E_1(X)={\sf Ecd}((e^{1}_\iota)_{0 \leq \iota <N/2}).
 $$
Under this  indexing, the bootstrapping equation (\ref{eq:boot_trace}) can be rewritten as:
\begin{equation}
\label{eq:eqmDstilde}
\frac{m}{\Delta} \simeq {\sf Im}   \left( \prod\limits_{b=0}^{h-1} 
 \sum\limits^{B/2-1}_{j=0} \left(\tilde{s}^0_{j \cdot h + b} \cdot e^0_{j \cdot h + b}
 +\tilde{s}^1_{j \cdot h + b}  \cdot e^1_{j \cdot h + b}\right) \right).
\end{equation}

With this new indexing now compatible with the \(\tr{N/2}{h}\) operator, we can perform the same operation 
homomorphically on polynomials. Specifically, we use \(\tr{N/2}{h}\) to compute the sums over each half-block and \(\pr{h}{1}\) to calculate the product of the \(h\) factors, 
after which the $N/2$ slots contain the same value.
Finally, we extract the imaginary part using the ${\sf Im2}$ operator (see Section \ref{s:rot-conj}).
Eventually, 
from (\ref{eq:eqmDstilde}), we obtain $m/\Delta$ in all $N/2$ slots, which corresponds
to an encoding of  $m={\sf Ecd}(m/\Delta) \in {\mathbb Z}$. In total, we arrive at
our bootstrapping equation for polynomial encodings:
\begin{equation}
\label{eq:mIm2}
 m  \simeq  
 {\sf Im2} \left( 
  \pr{h}{1} \left( 
\tr{N/2}{h} \left(S_0(X) \times E_0(X)+S_1(X) \times E_1(X) 
\right) \right) \right).
\end{equation}
Note that  we must
use a scaling factor  $\delta \in {\mathbb R}$ such that $\delta^h=q/(4\pi \Delta)$
instead of $\delta^h=q/(2\pi \Delta)$, because of the factor $2$ in the ${\sf Im2}$
operator.

\subsubsection{Bootstrapping.}

As before, we claim that Equation (\ref{eq:mIm2}) provides a decryption equation that is compatible with 
bootstrapping. Namely, all the previous polynomial operations can be applied homomorphically to 
ciphertexts.

More precisely, 
let ${\sf cs}$ denote the bootstrapping key consisting of \ckks encryptions ${\sf cs}_0, {\sf cs}_1$ of $S_0(X)$ and $S_1(X)$, modulo the largest modulus $Q$. 
The products $S_i(X) \times E_i(X)$ are then performed homomorphically
on ciphertexts using ${\sf ExtMultR}({\sf cs}_i,E_i)$, including
rescaling. Similarly, the operators $\tr{N/2}{h}$, $\pr{h}{1}$ 
and ${\sf Im2}$ are applied homomorphically on ciphertexts, as explained in sections 
\ref{s:rot-conj} and \ref{s:trace:product}.
% Is the sentence below not a bit redundant? It is already stated in the previous paragraph.
% Yes, this is a bit redundant, but I think this is fine, because this is important.
Eventually,  we obtain an 
 encryption of $m/ \Delta \in {\mathbb R}$ over all slots, and
  equivalently an encryption of $m ={\sf Ecd}(m/\Delta) \in {\mathbb Z}$.
This corresponds to  a new  
  \ckks ciphertext $\ct'$ such that $\langle \ct',\sk \rangle=m+e \pmod{pq}$, for a new 
  modulus $pq > q$. 
  For this, it suffices to set the largest modulus $Q :=q \cdot p \cdot \Delta^\ell$,
  where $\ell=\log_2 (h) + 1$ is the multiplicative depth. 
We provide a pseudo-code description of this bootstrapping algorithm in Appendix \ref{a:algo_single_slot}.

The main advantage of the packing approach is that the total number 
% log (h) for the multiplications, this is maybe more accurate, indeed we should mention it, greetings from the marketing department
  of homomorphic operations  is now ${\cal O}(\log N)$ instead of ${\cal O}(N)$, which
  makes our bootstrapping equation practical.
  In terms of the costly homomorphic multiplications, which dominate the number of bit-operations,
  the complexity further decreases to ${\cal O}(\log h)$. 
  We will
  consider the bootstrapping of $n \geq 2$ slots in parallel in Section \ref{s:boot:mult},
  and concrete implementation results in Section \ref{s:implem}.

\subsection{Security analysis of block binary secret key}

\label{s:attack}

Our scheme uses a block binary secret key, divided into \(h\) blocks, each of size \(B = N/h\), with exactly one '1' in each block. This distribution slightly differs  from that used in the original \ckks
 bootstrapping: a ternary secret key with Hamming weight \(h\), but without 
 the regular block structure. To ensure security against the best-known attacks, we provide a specific 
 security analysis for this modified distribution and present corresponding parameter sets in Section 
 \ref{s:implem}.  We follow the analysis from \cite{LMSS23}, which uses  the same distribution in the 
 context of TFHE.

We first consider 
the hybrid dual attack on LWE with a sparse secret \cite{A17}. This attack, which 
combines a dual lattice attack with exhaustive search, is used in the Lattice Estimator \cite{APS15} 
to estimate the security of LWE. In the first phase, the secret key $\vec{s}$ is partitioned 
as $\vec{s}=(\vec{s}_0,\vec{s}_1)$. Using lattice reduction, the LWE problem is reduced
to a single LWE instance with a smaller dimension. The secret key \(\vec{s}_0\) can 
then be recovered using combinatorial techniques such as exhaustive search or a meet-in-the-middle 
algorithm.

\ignore{
The LWE problem can be written as \(\vec{b} = - {\bf A} \vec{s} + \vec{e} \pmod{q}\), 
where \(\vec{b} \in \mathbb{Z}_q^m\), \({\bf A} \in \mathbb{Z}_q^{m \times n}\), 
and \(\vec{s} \in \mathbb{Z}_q^n\). 
By partitioning the secret \(\vec{s}\) as \(\vec{s} = \begin{bmatrix} \vec{s}_0 \\ \vec{s}_1 \end{bmatrix}\) and the matrix \({\bf A}\) as \({\bf A} = [{\bf A}_0 \mid {\bf A}_1]\), where \(\vec{s}_0 \in \mathbb{Z}_q^{n-k}\), \(\vec{s}_1 \in \mathbb{Z}_q^k\), \({\bf A}_0 \in \mathbb{Z}_q^{m \times (n-k)}\), and \({\bf A}_1 \in \mathbb{Z}_q^{m \times k}\), we have the relation:
$$ \vec{b} + {\bf A}_0 \vec{s}_0 = -{\bf A}_1 \vec{s}_1 + \vec{e} \pmod{q}. $$
The dual lattice attack involves finding a short vector \((\vec{x}, \vec{y})\) in the lattice
$$ {\cal L} = \{ (\vec{x}, \vec{y}) \in \mathbb{Z}^m \times 
\mathbb{Z}^k \mid \vec{x}^\intercal {\bf A}_1 = \vec{y}^\intercal \pmod{q} \}, $$
which leads to the equation \(\vec{x}^\intercal(\vec{b} + {\bf A}_0 \vec{s}_0) = -\langle 
\vec{y}, \vec{s}_1 \rangle + \langle \vec{x}, \vec{e} \rangle := e'\). 
}

For the second, combinatorial phase, the authors of \cite{LMSS23} considered two possible approaches. 
The original method involves guessing the positions of the zeros in the secret key.
However, for the block binary distribution, 
the lattice dimension can be reduced by the block size \(B = N/h\) by guessing the position of a '1' among \(B\) possible cases. 
While the latter approach seemed potentially more efficient, the authors provide an 
analysis showing that the original approach of guessing the zero positions is, in fact, more efficient. 
Consequently, the block binary structure seems to offer no advantage for an attacker.
Thus, the hybrid dual attack, as modeled in 
the Lattice Estimator, remains the best attack. 
Since, for convenience, we assume \(s_0 = 1\), we must adjust the 
ring dimension to \(N' = N \cdot (1 - 1/h)\) and the Hamming weight to \(h' = h - 1\) in the estimator.

The authors of \cite{LMSS23} also considered the recently improved meet-in-the-middle
 attack algorithm from May \cite{May21}.
 It is based on
recursively splitting the secret vector $\vec{s}$, which requires at least ${\cal S}^{0.25}$ time,
where ${\cal S}$ is the size of the key space. In our case,
the search space is ${\cal S}=(N/h)^{h-1}$. Therefore, the complexity
of May's attack is at least ${\cal S}=(N/h)^{0.25(h-1)}$. 
As in \cite{CHKKS18}, we will take $h=64$ in the concrete parameters. In that case, 
for $N \geq 2^{15}$, the complexity is at least $2^{141}$ operations.
 
Finally, we observe that the complexity of our bootstrapping algorithm grows only as 
\({\cal O}(\log h)\) homomorphic multiplications, 
making it relatively insensitive to increases in \(h\). Even if the attacks exploiting the 
sparsity of the secret key were to improve significantly, increasing $h$ would not 
result in a substantial performance penalty. For example, augmenting \(h\) from 64 to 128 would only increase the number of homomorphic operations in our 
bootstrapping by about 7\%.
Conversely, the complexity of May's attack would escalate from \(2^{141}\) to 
\(2^{254}\). This originates from the double-exponential gap between 
the complexity of our bootstrapping and that of May's attack.

We conclude that, as in the TFHE case, the use of block binary secrets in \ckks
does not affect the hardness of RLWE in the usual parameter setup.

\section{Bootstrapping for multiple slots}

\label{s:boot:mult}
% do you really want to write our new \ckks bootstrapping algorithm?
In the previous section, we have described our new \ckks bootstrapping 
algorithm for a single slot only. That is, we considered a \ckks ciphertext
$\ct$ such that $\langle \ct, \sk \rangle =m +e \pmod{q}$, for a message 
$m \in {\mathbb Z}$. In this section, we generalize our bootstrapping
equation (\ref{eq:boot_trace}) to handle multiple slots in parallel.

\subsection{Bootstrapping equation for multiple slots} 
\label{sec:boot_eq_mult_slots}

For a power-of two $n \geq 2$, 
we consider the plaintext space ${\cal P}_n={\mathbb Z}[X^{N/n}]/(X^N+1) \simeq {\mathbb Z}[Y] / (Y^n+1)$ for $Y=X^{N/n}$. Therefore, the plaintext space contains only $n$ coefficients. 
We consider a ciphertext $\ct=(c_0,c_1)$ encrypting a message $m \in {\cal P}_n$.

Our first step is to extract the $n$ corresponding LWE ciphertexts. For this, we consider
 the decryption equation $\langle \ct,\sk \rangle =c_0 + s \cdot c_1=m+e \pmod q$,
where 
$$m(X)=\sum_{a=0}^{n-1} m_a X^{aN/n} \in {\cal P}_n,$$
and $e \in {\mathbb Z}[X] / (X^N+1)$. For ease of notation, in the decryption equation, we include the coefficients
of $X^{aN/n}$ in the error $e(X)$ directly in $m(X)$.
By the linearity of polynomial multiplication, we can rewrite the decryption equation as a vector-matrix multiplication, keeping only the coefficients in $X^{aN/n}$ for $0 \leq a<n$. 
By setting $s(X) = \sum_{i=0}^{N-1} s_i X^i$, the decryption equation
 can be written as:
$$
\sum\limits_{a=0}^{n-1} m_a X^{aN/n} +e = c_0+ c_1 \cdot
  \sum_{i=0}^{N-1} s_i X^i  =c_0
+  \sum_{i=0}^{N-1} s_i \cdot c_1 \cdot X^i \pmod{q}.
$$
Therefore, we consider the vector $\vec{c}_0 \in {\mathbb Z}_q^{n}$
of coefficients of powers of $X^{N/n}$ of $c_0(X)$, 
and the matrix ${\bf C}_1 \in {\mathbb Z}_q^{N \times n}$ whose
 $i$-th row contains the $n$ coefficients of powers of $X^{N/n}$ of $c_1(X) \cdot X^i$,
  for $0 \leq i < N$.
Using this notation, we arrive at the equivalent
decryption equation
$ \vec{m}=\vec{c_0} + \vec{s} \cdot {\bf C}_1   \pmod q$.
Assuming that $s_0=1$, we can add the row vector $\vec{c_0}$ to the first row of ${\bf C}_1$ and obtain the decryption equation for $\vec{m} \in {\mathbb Z}^n$, 
$\vec{s} \in {\mathbb Z}^N$ and ${\bf C} \in {\mathbb Z}_q^{N \times n}$:
\begin{equation}
\label{eq:msC}
 \vec{m}=\vec{s} \cdot {\bf C}   \pmod q
 \end{equation}
We call ${\bf C}$ the \textit{decryption matrix}, and by 
$ {\bf C} \leftarrow {\sf DecMat}(\ct)$ we denote the 
above algorithm extracting ${\bf C} \in {\mathbb Z}_q^{N \times n}$ from the ciphertext $\ct$.
% Should we write the explicit algorithm here?

As in Section \ref{s:boot-trace}, we consider a binary secret key $s$ with a 
small power-of-two Hamming weight $h$, such that when written as a 
coefficient vector $\vec{s} \in \{0,1\}^N$, it can be separated
into $h$ blocks, each of size $B=N/h$, where there is exactly a single $1$ in each block.
From (\ref{eq:msC}), for each component $m_a$ of $\vec{m} \in {\mathbb Z}^n$,
 we can write $m_a= \langle \vec{s}, {\bf C}_a \rangle \pmod q$
  for the 
 corresponding  column vector ${\bf C}_a$ of ${\bf C}$. Therefore, each $m_a$ is decrypted as an independent LWE ciphertext given by the column
 vector ${\bf C}_a$. This implies that  
 Equation (\ref{eq:boot_trace}) generalizes to the
 following bootstrapping equation for each of the $m_a$ for $0 \leq a <n$:
\begin{equation}
\label{eq:boot_trace_slots}
\frac{m_a}{\Delta} \simeq   {\sf Im}   \left( \prod\limits_{b=0}^{h-1}  \sum\limits^{B-1}_{j=0} s_{b \cdot B+j} \cdot  
\exp(2i\pi  \cdot C_{b \cdot B+j,a}/q)  \cdot \delta \right).
\end{equation}
In the next section, we show how to perform this computation efficiently in parallel
using all available $N/2$ slots of polynomials in ${\mathbb Z}[X]/(X^N+1)$, and then homomorphically
over ciphertexts to achieve bootstrapping.

\subsection{Bootstrapping algorithm}

\label{s:bootstrapp:nslots}

Our approach for evaluating (\ref{eq:boot_trace_slots}) homomorphically  is essentially the same as in 
Section \ref{s:boot_single}.  The difference is that we must compute
(\ref{eq:boot_trace_slots}) for the $n$ components $m_a$ in parallel 
instead of a single one. Since each polynomial encoding provides a maximum of $N/2$ slots,
we only have $N/(2n)$ slots per component $m_a$ at our disposal in each polynomial encoding.
 Recall that the binary secret key $\vec{s}$ is split into $h$ blocks
of size $B=N/h$, each containing a single $1$. With only $N/(2n)$ slots at our disposal and $h$ blocks, we will  consider sub-blocks of size $N/(2n)/h=B/(2n)$. In other words, 
we split each 
 block of $B$ components into $2n$ sub-blocks with $B/(2n)$
components each, and we encode each of the $2n$ sub-blocks
in a separate polynomial.

\begin{figure}
\centering
\adjustbox{max width=\textwidth}{
\begin{tikzpicture}
\tikzstyle{bucket}=[rectangle, draw,minimum width=5cm,minimum height=.7cm]

\newcommand{\enc}[3]{
\node[bucket,#1] (#2) {\large #3};
\draw (#2) ++(-2.2,-.35) -- ++ (0,.7);
\draw (#2) ++(-1.9,-.35) -- ++ (0,.7);
\draw (#2) ++(1.9,-.35) -- ++ (0,.7);
\draw (#2) ++(2.2,-.35) -- ++ (0,.7);
}

\enc{}{B1}{$s_{b \cdot B+k}$}

\node[below=.3cm of B1] {$\vdots$};

\enc{below=1.5cm of B1}{B2}{$s_{b \cdot B+k+(2n-1)\frac{B}{2n}}$}

\node[right=.2cm of B1] {$\times$};
\node[right=.2cm of B2] {$\times$};
\enc{right=1cm of B1}{E1}{\normalsize $\exp(2i\pi \cdot C_{b \cdot B+k,a} / q) $ }

\node[below=.3cm of E1] {$\vdots$};

\enc{right=1cm of B2}{E2}{\normalsize $\exp(2i\pi \cdot C_{b \cdot B+k+ \cdot,a} / q) $ }
\node[right=.2cm of E1] {$=$};
\node[right=.2cm of E2] {$=$};
\enc{right=1cm of E1}{P1}{$\cdots$}
\enc{right=1cm of E2}{P2}{$\cdots$}

\node[below=0cm of P1] {$+$};
\node[below=.3cm of P1] {$\vdots$};
\node[above=0cm of P2] {$+$};

\draw (P2) ++(-2.3,-.7) -- ++(4.6,0);

\enc{below=.7cm of P2}{S}{$\cdots$}
\enc{below=1cm of S}{T}{\normalsize $\exp(2i\pi \cdot C_{b \cdot B+j^\star_b,a} / q)$}

\draw [decorate,decoration={brace,amplitude=5pt,raise=4ex}]
  (-2.1,0) -- (2.1,0) node[midway,yshift=3em]{\normalsize $0 \leq k < B/(2n)$};

\draw[->] (S.south) ++(0,-.1)  --  ++(0,-.8) node[midway,xshift=2.5em]{\normalsize ${\sf Tr}_{N/2 \rightarrow hn}$};

\draw [decorate,decoration={brace,amplitude=5pt}]
   (-2.7,-2.3) -- (-2.7,0) node[midway,xshift=-1.5em] {\normalsize  $2n$};

\end{tikzpicture}
}
\caption{Illustration of the computation within a given block.
After application of the $\tr{N/2}{hn}$ operator,
all $B/(2n)$ slots contain the same value $\exp(2i\pi \cdot C_{b \cdot B+j^\star_b,a}/q)$,
where $j^\star_b$ is the index for which $s_{b \cdot B+j^\star_b}=1$. }
\label{fig:blocks:slots}
\end{figure}
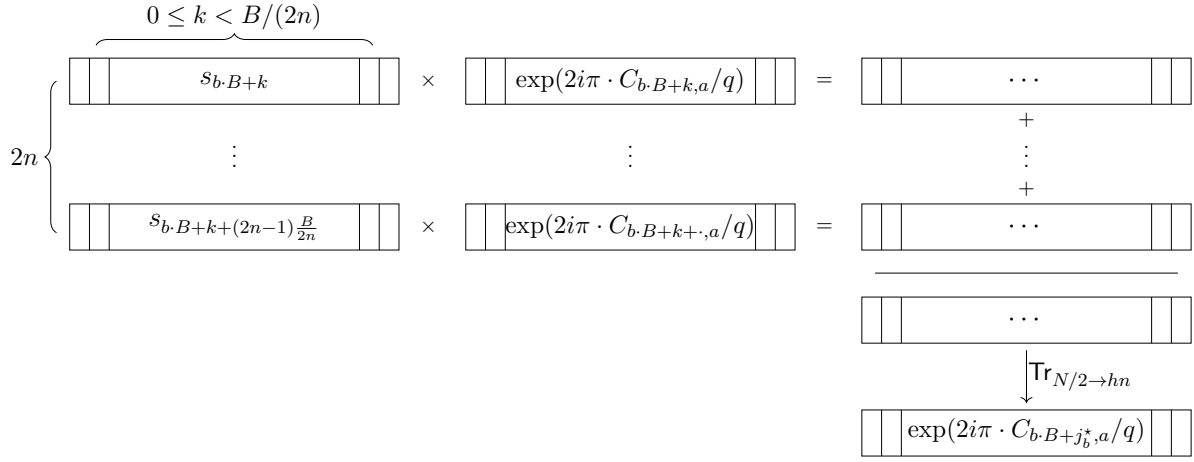

As illustrated in Figure \ref{fig:blocks:slots}, we first compute
the products $s_{b \cdot B+j} \cdot  
\exp(2i\pi  \cdot C_{b \cdot B+j,a}/q)  \cdot \delta$
from (\ref{eq:boot_trace_slots})
in parallel over each sub-block of size $B/(2n)$, for each of the $2n$ sub-blocks.
To compute the sum in (\ref{eq:boot_trace_slots}),
we first compute the sum of the $2n$ corresponding encodings, 
such that we end up with a single sub-block with $B/(2n)$ components. Then we can apply the trace operator to 
 finish the computation of the sum of each block. After the application
 of the trace operator, each of the $B/(2n)$ slots contains the same value
 $\exp(2i\pi \cdot C_{b \cdot B+j^\star_b,a} / q) \cdot \delta$. 
 This computation is performed in parallel for each of the $h$ blocks and for each of the $n$ components $m_a$ of  the plaintext.
In total, this corresponds to $h \cdot n$ independent slots; therefore, we must
 use the $\tr{N/2}{hn}$ operator. 
  Eventually we apply the $\pr{hn}{n}$ 
 operator to compute the final product of the $h$ elements, for each of the $n$ components of the message in parallel.
 
 As previously, we must
 use an indexing of the slots that is compatible with the 
 $\tr{N/2}{hn}$ operator, which computes partial sums of the slots
sharing the same index modulo $hn$ (see Appendix \ref{a:trace} for an illustration). 
 This implies that we must not
 encode the block's elements 
contiguously in the slots, but separated modulo $nh$. Therefore, we use a 
``reversed" indexing as in Section \ref{s:boot_single}. 
In this reversed indexing, the index $0 \leq j<B$
in (\ref{eq:boot_trace_slots}) is decomposed as $j=u \cdot B/(2n)+k$
for $0 \leq u<2n$ and $0 \leq k<B/(2n)$. Consequently, for 
an index $0 \leq u<2n$, 
we consider the slots $(\tilde{s}^u_\iota)_{0 \leq \iota <N/2}$,
such that for all $0 \leq k<B/(2n)$, $0 \leq b<h$ and $0 \leq a <n$, we have:
$$
\tilde{s}^{u}_{k \cdot hn +b \cdot n+a}=s_{b \cdot B + u \cdot B/(2n)+k}.
$$
We use the same reversed indexing for the ciphertext components,
with the slots $(e^u_\iota)_{0 \leq \iota <N/2}$,
such that for all $0 \leq k<B/(2n)$, $0 \leq b<h$ and $0 \leq a <n$, we have:
$$
e^{u}_{k \cdot hn +b \cdot n+a}=\exp(2i \pi \cdot C_{b \cdot B + u \cdot B/(2n)+k,a}/q) \cdot \delta.
$$
Under this new indexing, we may rewrite the bootstrapping equation (\ref{eq:boot_trace_slots}) as:
$$
\frac{m_a}{\Delta} \simeq   {\sf Im}   \left( \prod\limits_{b=0}^{h-1} 
\sum\limits_{k=0}^{B/(2n)-1} \sum\limits_{u=0}^{2n-1}
\tilde{s}^{u}_{k \cdot hn +b \cdot n+a} \cdot e^{u}_{k \cdot hn +b \cdot n+a} \right)
$$

Since the indexing is now compatible with the $\tr{N/2}{hn}$ operator, 
we can transfer the above equation to polynomial notation. Namely, we consider for an index
$0 \leq u<2n$ the polynomials $S_u(X)={\sf Ecd}((\tilde{s}^{u}_\iota)_{0 \leq \iota <N/2})$
and $E_u(X)={\sf Ecd}((e^{u}_\iota)_{0 \leq \iota <N/2})$, and we obtain the equivalent equation:
$$
\Ecd \left( \left(\frac{m_a}{\Delta} \right)_{0 \leq a<n} \right)
  \simeq 
 {\sf Im2} \left( 
  \pr{hn}{n} \left( 
\tr{N/2}{hn} \left( \sum\limits_{u=0}^{2n-1}
S_u(X) \times E_u(X)
\right) \right) \right).
$$
Note that in the above equation, we only have the decrypted values \( m_a / \Delta \) in the slots, so we need to move the \( m_a \)'s to the coefficient space,
using the same $\stc$ procedure as in the original
\ckks bootstrapping procedure (see Section \ref{s:slottocoeff}).
 This eventually gives us the decrypting equation:
 $$
 \sum\limits_{a=0}^{n-1} m_a X^{aN/n}
  \simeq {\sf StC} \left(
 {\sf Im2} \left( 
\pr{hn}{n} \left( 
\tr{N/2}{hn} \left( \sum\limits_{u=0}^{2n-1}
S_u(X) \times E_u(X)
\right) \right) \right) \right).
$$
Finally, since $\stc$ is based on a homomorphic DFT computation to achieve ${\cal O}(\log n)$ complexity,
 the coefficients $m_a$ are, in fact, recovered in bit-reversed order\footnote{In the OpenFHE library, this is also the case for the matrix-vector implementation of \stc.}. To recover the $m_a$ coefficients
 with the normal order, we must therefore encode the slots in the polynomials $S_u(X)$ and $E_u(X)$
 with bit-reversed order.

\subsubsection{Bootstrapping.}
As previously, we claim that this equation is compatible with bootstrapping, since it can be
 homomorphically evaluated over ciphertexts, such that eventually, we obtain a refreshed ciphertext
 of the same plaintext $m(X)=\sum_{a=0}^{n-1} m_a X^{aN/n}$, but under a larger modulus $q \cdot p$.
 
As in the single slot case, the bootstrapping key consists of \ckks
 encryptions of the polynomials $S_u(X)$ for $0 \leq u<2n$; see Alg. \ref{alg:bkey_multiple} below.
 The products $S_u(X) \times E_u(X)$ are evaluated homomorphically
on ciphertexts using the ${\sf ExtMultR}$ procedure.\footnote{
  As an optimization, we may compute the rescaling operation inside the ${\sf ExtMultR}$ procedure only after computing the sum over $0 \le u <2n$.
  This way, we only rescale once compared to $2n$ rescalings.
}
 Similarly, the
operators $\tr{N/2}{hn}$, $\pr{hn}{1}$ 
and ${\sf Im2}$ are applied homomorphically on ciphertexts; see Alg. \ref{alg:boot_multiple_slots}
below.

The multiplicative depth is now  $\ell=\log_2 h + 1+\ell_{\sf StC}$, % is now is the correct construction, not now is
where $\ell_{\sf StC}$ is the depth of $\stc$ (see Appendix
\ref{a:slottocoeff}).
Therefore, we set the big modulus $Q=\Delta^{\ell} \cdot q \cdot p$.
The total number of homomorphic operations is ${\cal O}(n + \log N)$.

\begin{algorithm}[H]
\caption{Bootstrapping key generation, multiple slots}
\label{alg:bkey_multiple}
\begin{algorithmic}[1]
	\Require A length $N$ secret key $\vec{s}$ with Hamming weight $h$, and $B = N/h$
	\Ensure A bootstrapping key ${\sf cs} = ({\sf cs}_u)_{0 \leq u<2n}$
  \smallskip
    \For {{\bf all} $0 \leq u<2n$}
      \For {{\bf all} $0 \leq k <B/(2n)$ {\bf and} $0 \leq b  <h$ {\bf and} $0 \leq a <n$}
        \State $\tilde{s}^{u}_{k \cdot hn +b \cdot n+a} = s_{b \cdot B + u \cdot B/(2n)+k}$ 
     \EndFor
	\State $S_u(X) \leftarrow {\sf Ecd}((\tilde{s}^{u}_\iota)_{0 \leq \iota < N/2})$
	\State ${\sf cs}_u \leftarrow {\sf Enc}_\pk(S_u(X))$
	%\State ${\sf cs}_u \leftarrow {\sf EncR}_\pk((\tilde{s}^{u}_\iota)_{0 \leq \iota < N/2})$
  \EndFor
    \State \Return $({\sf cs}_u)_{0 \leq u<2n}$
\end{algorithmic}
\end{algorithm}

\begin{algorithm}[H]
  \caption{Bootstrapping, for at most $B/2$ slots}
  \label{alg:boot_multiple_slots}
  \begin{algorithmic}[1]
    \Require A modulus $q$, a bootstrapping key $({\sf cs}_u)_{0 \le u < 2n}$, an \rlwe ciphertext $\ct$
    containing $n \le B/2$ slots, where $B = N/h$,
    and $\delta = (q/(4\pi \Delta))^{1/h}$
    \Ensure A refreshed ciphertext $\ct'$ modulo $p \cdot q$.
    \smallskip
      \State ${\bf C} \leftarrow {\sf DecMat}(\ct)$
%      \Comment{see Section \ref{sec:boot_eq_mult_slots}}
      \State ${\sf acc} \leftarrow (0, 0)$
 %     \Comment{The accumulator, an empty ciphertext}
      \For {$u = 0$ to $2n - 1$}
	 \For {{\bf all} $0 \leq k <B/(2n)$ {\bf and} $0 \leq b  <h$ {\bf and} $0 \leq a <n$}
       \State $e^{u}_{k \cdot hn +b \cdot n+a} = \exp(2 i \pi \cdot {\bf C}_{b \cdot B + u \cdot B/(2n)+k,a} / q) \cdot \delta$ 
\EndFor      
      
		\State $E_u(X) \leftarrow {\sf Ecd}((e^{u}_\iota)_{0 \leq \iota < N/2})$      
        \State $T_u  \leftarrow {\sf ExtMultR}({\sf cs}_u, E_u)$
        
        \State ${\sf acc} \leftarrow {\sf Add}({\sf acc}, T_u)$
      \EndFor
      \State \Return $\stc ({\sf Im2}(\pr{hn}{n}(\tr{N/2}{hn}({\sf acc}))))$
  \end{algorithmic}
\end{algorithm}

Note that as previously, the above bootstrapping method is fully compatible with RNS arithmetic. 
% Redundant to specify the moduli chain again!
%In that case, one would have moduli of the form $Q_r=\prod_{0 \leq i \leq r} q_i$, where each $q_i$ is a small prime, instead of $Q_r=q \cdot p \cdot \Delta^{r-1}$. 

\ignore{

\subsection{Noise estimation in bootstrapping}

\label{s:heuristicnoise_slots}

Our estimate of the noise growth is similar to 
the single slot case (see Section \ref{s:heuristicnoise}).
Before the application of the $\tr{N/2}{hn}$ operator,
we expect the noise to be ${\cal O}(n \cdot \sqrt{N})$ instead
of ${\cal O}(\sqrt{N})$, because of the summing of the $2n$ ciphertexts
in the accumulator ${\sf acc}$. As previously, each of the ${\cal O}(\log N)$
 key switchings in $\tr{N/2}{hn}$
will increase the noise by an additive term of ${\cal O}(\sqrt{N})$, 
therefore the noise becomes ${\cal O}((n+\log N) \cdot \sqrt{N})$.
After the $\pr{h}{1}$ operator and taking into account the scaling factor, the noise
becomes ${\cal O}(h \cdot \sqrt{N} (n+\log N) \cdot q/\Delta + \sqrt{N})$.
Finally, the ${\cal O}(\log n)$ linear layers in $\stc$
introduce an additive noise term of ${\cal O}(\sqrt{N} \log n)$,
so the total noise is ${\cal O}(h \cdot \sqrt{N} (n+\log N) \cdot q/\Delta + \sqrt{N} \log n)$.
Therefore, by taking $\Delta={\cal O}(h \cdot  (n+\log N) \cdot q)$, the bootstrapping error 
remains heuristically ${\cal O}(\sqrt{N} \log n)$.

}

\subsection{Bootstrapping up to $N$ slots.}

\label{s:fullboot}

The previous bootstrapping algorithm is limited to bootstrapping up to 
\( n \leq n_{\sf max} = B/2 = N/(2h) \) components. However, it can be extended to support more 
slots, specifically  $n' \leq N$, by shifting the coefficients of the input ciphertext and applying the 
previous bootstrapping procedure as a black box for each group of \( n_{\sf max} \) coefficients.
We refer to Appendix \ref{a:boot_max_slots} for the details.
The number of homomorphic operations remains ${\cal O}(n + \log N)$ for $n$ slots
and the depth also remains unchanged. Note that our bootstrapping method is highly parallelizable; with  $n$ 
processors, it achieves the same ${\cal O}(\log N)$ complexity as the original CKKS bootstrapping.
We summarize 
the asymptotic complexities of $\ckks$ bootstrapping approaches in Table \ref{t:bootcomp}, expressed in the
number of homomorphic operations.

\begin{table}[h]
\def\arraystretch{1.2}
\centering
%\resizebox{0.7\columnwidth}{!}{%
\begin{tabular}{|l|c|} \cline{2-2}
\multicolumn{1}{c|}{} & ~Complexity~ \\ \hline
 ~Original \ckks bootstrapping \cite{CHKKS18}~ & ~${\cal O}(\log n+\log N)$~ \\ \hline
 ~Blind rotation bootstrapping \cite{KDE+21}~ & ${\cal O}(n \cdot N)$ \\ \hline
 ~Our new bootstrapping~ & ${\cal O}( n+\log N )$ \\ \hline
\end{tabular}
%}
\medskip
\caption{Complexities of different bootstrapping approaches for \ckks, expressed in number of
ciphertext multiplications, for a ring dimension $N$ and $n$ slots.}
\label{t:bootcomp}
\end{table}

\section{Implementation and performance comparison}

\label{s:implem}

In this section, we describe the results of an implementation of our new bootstrapping based
on the OpenFHE library \cite{OpenFHE}. We compare it against the state-of-the-art implementation of the original \ckks bootstrapping also available in OpenFHE.
\iffullver 
The source code is provided in 
\begin{center}
\url{https://github.com/coron/spru_boot}
\end{center}
\else
We provide our \cpp ~source code  in the auxiliary files. \fi

\subsection{Parameter sets}

In the OpenFHE library \cite{OpenFHE}, the ring dimension is automatically determined based on the size of the 
largest modulus and the desired security level, which is 128-bit by default. We also independently verified the 
security level of our parameter sets using the standard Lattice Estimator \cite{APS15}. The complete parameter 
sets used for both the original and our new bootstrapping methods are summarized in Table~\ref{t:paramsCKKS}.

\newcommand{\mr}[1]{\multirow{2}{*}{#1}}
\newcommand{\mrr}[1]{\multirow{4}{*}{#1}}

\subsubsection{Fixing the precision.}
With the RNS arithmetic, the ciphertext moduli
have the form 
 $Q_r=\prod_{0 \leq i \leq r} q_i$, where each $q_i$ is a small prime and 
 $0 \leq r \leq \ell_m$, with $\ell_m$ being the maximum depth. 
 OpenFHE allows to determine the bit-size 
 of $q_0$ and the bit-size of the $q_i$'s for $i \geq 1$ separately. 
  For the  original \ckks
 bootstrapping, OpenFHE uses $\log_2 q_0 \simeq 60$ and $\log_2 q_i \simeq 59$ for $i \geq 1$
achieving roughly $14$ bits of precision,
with a scaling factor $\Delta=2^{59}$. 
For our bootstrapping, we use  $\log_2 q_0 \simeq 60$ and $\log_2 q_i \simeq 49$ for $i \geq 1$ to reach the same level of precision, with scaling factor  $\Delta=2^{49}$. In Table \ref{t:paramsCKKS}, we denote by $Q=Q_{\ell_m}$ the largest modulus in the ladder, and by $PQ$ the modulus used for key switching.
 
\subsubsection{Bootstrapping depth.}
 
We denote by $\ell_c$ the number of levels consumed by bootstrapping, excluding $\cts$ ~and $\stc$. 
Let $\ell_{op}$ be the number of levels available after bootstrapping. The total depth  $\ell_m$ is then given by 
$ \ell_{m}= \ell_{\sf CtS} +\ell_{c} + \ell_{\sf StC} + \ell_{op} $,
where 
$\ell_{\sf CtS}$ and $\ell_{\sf StC}$ are the number of levels consumed by 
$\cts$ and $\stc$, respectively.

Since our goal is to minimize bootstrapping latency, we only leave a single
level for computation after bootstrapping, that is $\ell_{op}=1$, for both schemes. The implementation of the original \ckks bootstrapping
in OpenFHE has depth $\ell_c=11$. For our bootstrapping, we use a Hamming weight $h=64$ (see our analysis in Section 
\ref{s:attack}), which gives $\ell_c=\log_2 h + 1=7$.

\subsubsection{Depth of $\cts$ and $\stc$.}

In the OpenFHE \cpp ~library, one can directly fix the number of levels
$\ell_{\sf CtS}$ and 
$\ell_{\sf StC}$ for \cts ~and \stc.
 The complexity (in number of homomorphic operations)
is then ${\cal O}( \rho  \cdot n^{1/\rho})$ where $\rho$ is the number of levels and $n$  the number of slots in sparse packing. Therefore, when bootstrapping a small number of slots (up to $n=16$), it is more advantageous to use the matrix-vector multiplication technique for both $\cts$ and $\stc$, which gives
$\ell_{\sf CtS}=\ell_{\sf StC}=1$. When increasing the number of slots,
one can allocate a larger level budget, up to $\ell_{\sf CtS}=\ell_{\sf StC}=4$ for full slot 
$(n=N/2)$ bootstrapping. Note that our new bootstrapping does not use $\cts$, therefore
$ \ell_{\sf CtS}=0$ for our bootstrapping.

In the following, we consider bootstrapping only for  a relatively
small number of slots $n$, with $1 \leq n \leq 1024$.
For the original \ckks bootstrapping, we use  $\ell_{\sf CtS}=\ell_{\sf StC}=1$ for
$1 \leq n \leq 16$, and $\ell_{\sf CtS}=\ell_{\sf StC}=2$ for $32 \leq n \leq 1024$. 
For our new bootstrapping, internally we use at most $32$ slots, following the technique
from Section \ref{s:fullboot}.
 Therefore, we can use 
$\ell_{\sf StC}=1$, except for $n=1$ for which $\ell_{\sf StC}=0$, since
no \stc ~is required in that case. We summarize the parameters in Table \ref{t:paramsCKKS} below. Note that to bootstrap $n \geq 2$ complex slots, we
must consider the plaintext space ${\cal P}_{2n}={\mathbb Z}[X^{N/(2n)}]/(X^N+1) $, 
which corresponds to polynomials with $2n$ coefficients (see Section \ref{sec:boot_eq_mult_slots}).
Due to its smaller multiplicative depth, our new bootstrapping can operate with a smaller ring dimension 
$N = 2^{15}$, compared to $N = 2^{16}$ for the original CKKS bootstrapping.

\begin{table}[H]
\centering
\renewcommand{\arraystretch}{1.15}
\resizebox{1\columnwidth}{!}{%
\begin{tabular}{|l|c|c|c|c|c|c|c|c|c|c|c|c|} \cline{2-12}
 \multicolumn{1}{c|}{} &  ~$N$~ & ~$\log q_0$~ & ~$\log q_i$~  
 & ~$\ell_c$~ & $n$ & ~$\ell_{\sf cts}~$ & ~$\ell_{\sf stc}~$ & $\ell_m$~ 
& ~$\log Q$~ & ~$\log PQ$~ & ~precision~ \\ \hline
\mr{~\ckks boot.~} &  \mr{$2^{16}$} & \mr{60} & \mr{59}    & \mr{10} & $1 \leq n \leq 16$ & 1 & 1 & 13 & 828  & 1128 & \mr{14 bits} \\  \cline{6-11}
 & & & & & ~$32 \leq n \leq 1024$~ & 2 & 2 & 15 & 946 & 1306 &\\ \hline
\mr{~New boot.~}  & \mr{$2^{15}$}  & \mr{60} & \mr{49} & \mr{7} & $n=1$ & 0 &  0  & 8 & 453 & 633 & \mr{14 bits} \\ \cline{6-11}
 &   &&&& $2 \leq n \leq 1024$ &0 & 1 & 9  & 502 & 742 &\\\hline
\end{tabular}
}
\medskip
\caption{Parameter sets for the two \ckks bootstrapping,
for $128$ bits of security.}
\label{t:paramsCKKS}
\end{table}

% mention \log_2 QP, the maximum key-swithing modulus.

\ignore{
We use the standard Lattice Estimator \cite{APS15} to fix the parameters
for security against the best known attacks, including the hybrid lattice attack taking advantage of 
the sparse key distribution (see Section \ref{s:attack}). 
 
In Table \ref{t:maxmod} we summarize the maximal size of 
the largest modulus $Q=q_L$ in the ladder,  
as a function of $\log_2 N$, for $\lambda=100$ bits of security. 
We  account for the fact that the switching keys $\evk, {\sf
  rk}_r,$ and ${\sf ck}$ are all encrypted modulo $P \cdot Q$, 
with $P=Q$.
Consequently, the effective largest modulus is $Q^2$.
}

\vspace{-.65cm}

\subsection{Running time comparison}

We benchmark our implementation using the parameters in Table~\ref{t:paramsCKKS}, on an Apple M2 CPU with 8 cores. We also run the OpenFHE implementation of the original \ckks bootstraping for comparison.
\iffullver \else
Our source code is provided in the auxiliary files. \fi

\begin{table}
  \centering
  \def\arraystretch{1.1}
  \resizebox{\columnwidth}{!}{%
  \begin{tabular}{|l||c|c|c|c|c|c|c|c|c|c|c|} \hline
  ~Number of slots~ & 1 & 2 & 4 & 8 & 16 & 32 & 64 & 128 & 256 & 512 & 1024 \\ \hline \hline
 ~\ckks boot.~ & ~1.7 s~ & ~1.7 s~ & ~1.7 s~ & ~1.7 s~ & ~1.8 s~ & ~2.0 s~ & ~2.2 s~ & ~2.8 s~ & ~3.1 s~ & ~3.6 s~ & ~3.6 s~ \\ \hline
  ~New boot.~  & ~0.3 s~ & ~0.4 s~ & ~0.4 s~ & ~0.4 s~ & ~0.5 s~ & ~0.6 s~ & ~1.3 s~ & ~2.7 s~ & ~5.6 s~ & ~11.3 s~ & ~23.6 s~ \\ \hline
  \end{tabular}
  }
  \medskip
  \caption{Running time comparison of the original \ckks bootstrapping and our new bootstrapping, for a
  variable number of slots $n$.}
  \label{t:timings}
  \end{table}

 Table~\ref{t:timings} shows that, due to its reduced multiplicative depth, 
 our bootstrapping is approximately 5× faster than the original CKKS bootstrapping for a single slot ($n = 1$). 
 It remains significantly faster up to $n = 64$ slots.
\iffullver As illustrated in Figure \ref{f:comparison}, the \else The \fi
 crossover point lies around $n=128$ slots,
after which the original \ckks bootstrapping scales as ${\cal O}(\log n)$, whereas 
our new bootstrapping scales as ${\cal O}(n)$, thus it becomes impractical for $n>1024$.
% I really don't like the "scaling after" formulations (here and earlier), as it suggests that
% the scaling doesn't apply for n < 128, which is not true, and moreover does it
% overshadow the real reason, why our bootstrapping is faster for n < 128.

\iffullver
\begin{figure}
  \centering
  \includegraphics[scale=.55]{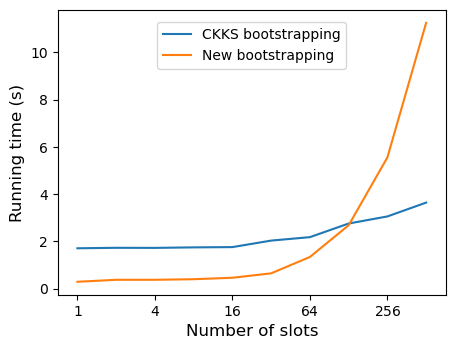} % .7
  \caption{Comparison of the running times of \ckks bootstrapping and our new bootstrapping.}
  \label{f:comparison}
\end{figure}
\fi

\section{Conclusion}

In this paper, we presented an alternative bootstrapping algorithm for the \ckks scheme, 
based on embedding the additive group modulo \( q \) into the circle group of complex numbers, 
which can be evaluated natively in \ckks. Due to its lower multiplicative depth, our new bootstrapping 
algorithm can operate with a smaller ring dimension \( N \), resulting in significant efficiency 
improvements for few slots \( n \). 
This new bootstrapping approach could be particularly useful in scenarios where circuit evaluation requires 
a dynamic number of slots, offering an alternative to scheme-switching methods.

%However, unlike the original \ckks bootstrapping, which 
%scales as \( {\cal O}(\log n) \), our method scales as \( {\cal O}(n) \), making it less efficient for 
%large \( n \).
% Given that bootstrapping is 
%a fundamental algorithm, we believe it is valuable to explore new techniques that offer different 
%mathematical foundations and performance trade-offs.

\bibliographystyle{alpha}
\bibliography{newboot}

@article{BGV11,
  author    = {Zvika Brakerski and
               Craig Gentry and
               Vinod Vaikuntanathan},
  title     = {Fully Homomorphic Encryption without Bootstrapping},
  journal   = {Electron. Colloquium Comput. Complex.},
  pages     = {111},
  year      = {2011}
}

@Article{SV14,
author={Smart, N. P.
and Vercauteren, F.},
title={Fully homomorphic {SIMD} operations},
journal={Designs, Codes and Cryptography},
year={2014},
month={Apr},
day={01},
volume={71},
number={1},
pages={57-81},
issn={1573-7586},
doi={10.1007/s10623-012-9720-4},
url={https://doi.org/10.1007/s10623-012-9720-4}
}

@misc{GHSx12,
author="Gentry, Craig
and Halevi, Shai
and Smart, Nigel P.",
editor="Pointcheval, David
and Johansson, Thomas",
title="Fully Homomorphic Encryption with Polylog Overhead",
booktitle="Advances in Cryptology -- EUROCRYPT 2012",
year="2012",
publisher="Springer Berlin Heidelberg",
address="Berlin, Heidelberg",
pages="465--482",
isbn="978-3-642-29011-4"
}

@inproceedings{GEN09,
author = {Gentry, Craig},
year = {2009},
month = {05},
pages = {169-178},
title = {Fully Homomorphic Encryption Using Ideal Lattices},
volume = {9},
journal = {Proceedings of the Annual ACM Symposium on Theory of Computing},
doi = {10.1145/1536414.1536440}
}

@inproceedings{GSW13,
  author    = {Craig Gentry and
               Amit Sahai and
               Brent Waters},
  title     = {Homomorphic Encryption from Learning with Errors: Conceptually-Simpler,
               Asymptotically-Faster, Attribute-Based},
  booktitle = {Advances in Cryptology - {CRYPTO} 2013 - 33rd Annual Cryptology Conference,
               Santa Barbara, CA, USA, August 18-22, 2013. Proceedings, Part {I}},
  pages     = {75--92},
  year      = {2013}
  }

@inproceedings{AP14,
  author    = {Jacob Alperin{-}Sheriff and
               Chris Peikert},
  editor    = {Juan A. Garay and
               Rosario Gennaro},
  title     = {Faster Bootstrapping with Polynomial Error},
  booktitle = {Advances in Cryptology - {CRYPTO} 2014 - 34th Annual Cryptology Conference,
               Santa Barbara, CA, USA, August 17-21, 2014, Proceedings, Part {I}},
  series    = {Lecture Notes in Computer Science},
  volume    = {8616},
  pages     = {297--314},
  publisher = {Springer},
  year      = {2014}
  }

@inproceedings{DM15,
  author    = {L{\'{e}}o Ducas and
               Daniele Micciancio},
  editor    = {Elisabeth Oswald and
               Marc Fischlin},
  title     = {{FHEW:} Bootstrapping Homomorphic Encryption in Less Than a Second},
  booktitle = {Advances in Cryptology - {EUROCRYPT} 2015 - 34th Annual International
               Conference on the Theory and Applications of Cryptographic Techniques,
               Sofia, Bulgaria, April 26-30, 2015, Proceedings, Part {I}},
  series    = {Lecture Notes in Computer Science},
  volume    = {9056},
  pages     = {617--640},
  publisher = {Springer},
  year      = {2015}
  }

@inproceedings{CGGI16,
  author    = {Ilaria Chillotti and
               Nicolas Gama and
               Mariya Georgieva and
               Malika Izabach{\`{e}}ne},
  editor    = {Jung Hee Cheon and
               Tsuyoshi Takagi},
  title     = {Faster Fully Homomorphic Encryption: Bootstrapping in Less Than 0.1
               Seconds},
  booktitle = {Advances in Cryptology - {ASIACRYPT} 2016 - 22nd International Conference
               on the Theory and Application of Cryptology and Information Security,
               Hanoi, Vietnam, December 4-8, 2016, Proceedings, Part {I}},
  series    = {Lecture Notes in Computer Science},
  volume    = {10031},
  pages     = {3--33},
  year      = {2016}
}

@inproceedings{CKKS17,
  author    = {Jung Hee Cheon and
               Andrey Kim and
               Miran Kim and
               Yong Soo Song},
  editor    = {Tsuyoshi Takagi and
               Thomas Peyrin},
  title     = {Homomorphic Encryption for Arithmetic of Approximate Numbers},
  booktitle = {Advances in Cryptology - {ASIACRYPT} 2017 - 23rd International Conference
               on the Theory and Applications of Cryptology and Information Security,
               Hong Kong, China, December 3-7, 2017, Proceedings, Part {I}},
  series    = {Lecture Notes in Computer Science},
  volume    = {10624},
  pages     = {409--437},
  publisher = {Springer},
  year      = {2017}
}

@inproceedings{CHKKS18,
  author    = {Jung Hee Cheon and
               Kyoohyung Han and
               Andrey Kim and
               Miran Kim and
               Yongsoo Song},
  editor    = {Jesper Buus Nielsen and
               Vincent Rijmen},
  title     = {Bootstrapping for Approximate Homomorphic Encryption},
  booktitle = {Advances in Cryptology - {EUROCRYPT} 2018 - 37th Annual International
               Conference on the Theory and Applications of Cryptographic Techniques,
               Tel Aviv, Israel, April 29 - May 3, 2018 Proceedings, Part {I}},
  series    = {Lecture Notes in Computer Science},
  volume    = {10820},
  pages     = {360--384},
  publisher = {Springer},
  year      = {2018}
  }

@inproceedings{CCS19,
  author    = {Hao Chen and
               Ilaria Chillotti and
               Yongsoo Song},
  editor    = {Yuval Ishai and
               Vincent Rijmen},
  title     = {Improved Bootstrapping for Approximate Homomorphic Encryption},
  booktitle = {Advances in Cryptology - {EUROCRYPT} 2019 - 38th Annual International
               Conference on the Theory and Applications of Cryptographic Techniques,
               Darmstadt, Germany, May 19-23, 2019, Proceedings, Part {II}},
  series    = {Lecture Notes in Computer Science},
  volume    = {11477},
  pages     = {34--54},
  publisher = {Springer},
  year      = {2019},
  note = {https://eprint.iacr.org/2018/1043.pdf}
  }

@InProceedings{JM22,
author="Jutla, Charanjit S.
and Manohar, Nathan",
editor="Dunkelman, Orr
and Dziembowski, Stefan",
title="Sine Series Approximation of the Mod Function for Bootstrapping of Approximate HE",
booktitle="Advances in Cryptology -- EUROCRYPT 2022",
year="2022",
publisher="Springer International Publishing",
address="Cham",
pages="491--520"
}

@inproceedings{GHSa12,
  author       = {Craig Gentry and
                  Shai Halevi and
                  Nigel P. Smart},
 
  title        = {Homomorphic Evaluation of the {AES} Circuit},
  booktitle    = {Advances in Cryptology - {CRYPTO} 2012 - 32nd Annual Cryptology Conference,
                  Santa Barbara, CA, USA, August 19-23, 2012. Proceedings},
  series       = {Lecture Notes in Computer Science},
  volume       = {7417},
  pages        = {850--867},
  publisher    = {Springer},
  year         = {2012}
}

@article{fv12,
  author    = {Junfeng Fan and
               Frederik Vercauteren},
  title     = {Somewhat Practical Fully Homomorphic Encryption},
  journal   = {{IACR} Cryptol. ePrint Arch.},
  pages     = {144},
  year      = {2012}
  }

@misc{KDE+21,
      author = {Andrey Kim and Maxim Deryabin and Jieun Eom and Rakyong Choi and Yongwoo Lee and Whan Ghang and Donghoon Yoo},
      title = {General Bootstrapping Approach for {RLWE}-based Homomorphic Encryption},
      howpublished = {Cryptology ePrint Archive, Paper 2021/691},
      year = {2021},
      note = {\url{https://eprint.iacr.org/2021/691}},
      url = {https://eprint.iacr.org/2021/691}
}

@misc{CDKS20,
      author = {Hao Chen and Wei Dai and Miran Kim and Yongsoo Song},
      title = {Efficient Homomorphic Conversion Between (Ring) {LWE} Ciphertexts},
      howpublished = {Cryptology ePrint Archive, Paper 2020/015},
      year = {2020},
      note = {\url{https://eprint.iacr.org/2020/015}},
      url = {https://eprint.iacr.org/2020/015}
}

@inproceedings{DPSZ12,
  author       = {Ivan Damg{\aa}rd and
                  Valerio Pastro and
                  Nigel P. Smart and
                  Sarah Zakarias},
  editor       = {Reihaneh Safavi{-}Naini and
                  Ran Canetti},
  title        = {Multiparty Computation from Somewhat Homomorphic Encryption},
  booktitle    = {Advances in Cryptology - {CRYPTO} 2012 - 32nd Annual Cryptology Conference,
                  Santa Barbara, CA, USA, August 19-23, 2012. Proceedings},
  series       = {Lecture Notes in Computer Science},
  volume       = {7417},
  pages        = {643--662},
  publisher    = {Springer},
  year         = {2012}
  }

@misc{CHH18,
      author = {Jung Hee Cheon and Kyoohyung Han and Minki Hhan},
      title = {Faster Homomorphic Discrete Fourier Transforms and Improved {FHE} Bootstrapping},
      howpublished = {Cryptology {ePrint} Archive, Paper 2018/1073},
      year = {2018},
      url = {https://eprint.iacr.org/2018/1073}
}

@article{APS15,
author = {Albrecht, Martin and Player, Rachel and Scott, Sam},
year = {2015},
month = {10},
pages = {},
title = {On the concrete hardness of Learning with Errors},
volume = {9},
journal = {Journal of Mathematical Cryptology},
doi = {10.1515/jmc-2015-0016}
}

@inproceedings{CCKS23,
author = {Cheon, Jung Hee and Cho, Wonhee and Kim, Jaehyung and Stehl\'{e}, Damien},
title = {Homomorphic Multiple Precision Multiplication for {CKKS} and Reduced Modulus Consumption},
year = {2023},
isbn = {9798400700507},
publisher = {Association for Computing Machinery},
address = {New York, NY, USA},
url = {https://doi.org/10.1145/3576915.3623086},
doi = {10.1145/3576915.3623086},
booktitle = {Proceedings of the 2023 ACM SIGSAC Conference on Computer and Communications Security},
pages = {696–710},
numpages = {15},
location = {Copenhagen, Denmark},
series = {CCS '23}
}

@inproceedings{CHKKS18SAC,
  author       = {Jung Hee Cheon and
                  Kyoohyung Han and
                  Andrey Kim and
                  Miran Kim and
                  Yongsoo Song},
  editor       = {Carlos Cid and
                  Michael J. Jacobson Jr.},
  title        = {A Full {RNS} Variant of Approximate Homomorphic Encryption},
  booktitle    = {Selected Areas in Cryptography - {SAC} 2018 - 25th International Conference,
                  Calgary, AB, Canada, August 15-17, 2018, Revised Selected Papers},
  series       = {Lecture Notes in Computer Science},
  volume       = {11349},
  pages        = {347--368},
  publisher    = {Springer},
  year         = {2018}
  }

@inproceedings{KPK+23,
author = {Kim, Seonghak and Park, Minji and Kim, Jaehyung and Kim, Taekyung and Min, Chohong},
title = {{EvalRound} Algorithm in {CKKS} Bootstrapping},
year = {2023},
isbn = {978-3-031-22965-7},
publisher = {Springer-Verlag},
address = {Berlin, Heidelberg},
url = {https://doi.org/10.1007/978-3-031-22966-4_6},
doi = {10.1007/978-3-031-22966-4_6},
booktitle = {Advances in Cryptology – ASIACRYPT 2022: 28th International Conference on the Theory and Application of Cryptology and Information Security, Taipei, Taiwan, December 5–9, 2022, Proceedings, Part II},
pages = {161–187},
numpages = {27},
location = {Taipei, Taiwan}
}

@inproceedings{BCC+22,
author = {Bae, Youngjin and Cheon, Jung Hee and Cho, Wonhee and Kim, Jaehyung and Kim, Taekyung},
title = {{META-BTS}: Bootstrapping Precision Beyond the Limit},
year = {2022},
isbn = {9781450394505},
publisher = {Association for Computing Machinery},
address = {New York, NY, USA},
url = {https://doi.org/10.1145/3548606.3560696},
doi = {10.1145/3548606.3560696},
booktitle = {Proceedings of the 2022 ACM SIGSAC Conference on Computer and Communications Security},
pages = {223–234},
numpages = {12},
location = {Los Angeles, CA, USA},
series = {CCS '22}
}

@ARTICLE {SYL+23,
author = {S. Shen and H. Yang and Y. Liu and Z. Liu and Y. Zhao},
journal = {IEEE Transactions on Computers},
title = {{CARM}: {CUDA}-Accelerated {RNS} Multiplication in Word-Wise Homomorphic Encryption Schemes for Internet of Things},
year = {2023},
volume = {72},
number = {07},
issn = {1557-9956},
pages = {1999-2010},
doi = {10.1109/TC.2022.3227874},
publisher = {IEEE Computer Society},
address = {Los Alamitos, CA, USA},
month = {jul},
}

@article{JKA+21,
author = {Jung, Wonkyung and Kim, Sangpyo and Ahn, Jung Ho and Cheon, Jung and Lee, Younho},
year = {2021},
month = {08},
pages = {114-148},
title = {Over 100x Faster Bootstrapping in Fully Homomorphic Encryption through Memory-centric Optimization with {GPUs}},
volume = {2021},
journal = {IACR Transactions on Cryptographic Hardware and Embedded Systems},
doi = {10.46586/tches.v2021.i4.114-148}
}

@InProceedings{LLK+22,
author="Lee, Yongwoo
and Lee, Joon-Woo
and Kim, Young-Sik
and Kim, Yongjune
and No, Jong-Seon
and Kang, HyungChul",
editor="Dunkelman, Orr
and Dziembowski, Stefan",
title="High-Precision Bootstrapping for Approximate Homomorphic Encryption by Error Variance Minimization",
booktitle="Advances in Cryptology -- EUROCRYPT 2022",
year="2022",
publisher="Springer International Publishing",
address="Cham",
pages="551--580",
isbn="978-3-031-06944-4"
}

@inproceedings{LLL+21,
author = {Lee, Joon-Woo and Lee, Eunsang and Lee, Yongwoo and Kim, Young-Sik and No, Jong-Seon},
title = {High-Precision Bootstrapping of {RNS-CKKS} Homomorphic Encryption Using Optimal Minimax Polynomial Approximation and Inverse Sine Function},
year = {2021},
isbn = {978-3-030-77869-9},
publisher = {Springer-Verlag},
address = {Berlin, Heidelberg},
url = {https://doi.org/10.1007/978-3-030-77870-5_22},
doi = {10.1007/978-3-030-77870-5_22},
booktitle = {Advances in Cryptology – EUROCRYPT 2021: 40th Annual International Conference on the Theory and Applications of Cryptographic Techniques, Zagreb, Croatia, October 17–21, 2021, Proceedings, Part I},
pages = {618–647},
numpages = {30},
location = {Zagreb, Croatia}
}

@inproceedings{LMSS23,
  author       = {Changmin Lee and
                  Seonhong Min and
                  Jinyeong Seo and
                  Yongsoo Song},
  editor       = {Joseph K. Liu and
                  Yang Xiang and
                  Surya Nepal and
                  Gene Tsudik},
  title        = {Faster {TFHE} Bootstrapping with Block Binary Keys},
  booktitle    = {Proceedings of the 2023 {ACM} Asia Conference on Computer and Communications
                  Security, {ASIA} {CCS} 2023, Melbourne, VIC, Australia, July 10-14,
                  2023},
  pages        = {2--13},
  publisher    = {{ACM}},
  year         = {2023}
  }

@inproceedings{May21,
  author       = {Alexander May},
  editor       = {Tal Malkin and
                  Chris Peikert},
  title        = {How to Meet Ternary {LWE} Keys},
  booktitle    = {Advances in Cryptology - {CRYPTO} 2021 - 41st Annual International
                  Cryptology Conference, {CRYPTO} 2021, Virtual Event, August 16-20,
                  2021, Proceedings, Part {II}},
  series       = {Lecture Notes in Computer Science},
  volume       = {12826},
  pages        = {701--731},
  publisher    = {Springer},
  year         = {2021}
  }

@inproceedings{A17,
  author       = {Martin R. Albrecht},
  editor       = {Jean{-}S{\'{e}}bastien Coron and
                  Jesper Buus Nielsen},
  title        = {On Dual Lattice Attacks Against Small-Secret {LWE} and Parameter Choices
                  in {HElib} and {SEAL}},
  booktitle    = {Advances in Cryptology - {EUROCRYPT} 2017 - 36th Annual International
                  Conference on the Theory and Applications of Cryptographic Techniques,
                  Paris, France, April 30 - May 4, 2017, Proceedings, Part {II}},
  series       = {Lecture Notes in Computer Science},
  volume       = {10211},
  pages        = {103--129},
  year         = {2017}
  }

@inproceedings{Bra12,
  author       = {Zvika Brakerski},
  editor       = {Reihaneh Safavi{-}Naini and
                  Ran Canetti},
  title        = {Fully Homomorphic Encryption without Modulus Switching from Classical
                  {GapSVP}},
  booktitle    = {Advances in Cryptology - {CRYPTO} 2012 - 32nd Annual Cryptology Conference,
                  Santa Barbara, CA, USA, August 19-23, 2012. Proceedings},
  series       = {Lecture Notes in Computer Science},
  volume       = {7417},
  pages        = {868--886},
  publisher    = {Springer},
  year         = {2012}
  }

@inproceedings{SSTX09,
  author       = {Damien Stehl{\'{e}} and
                  Ron Steinfeld and
                  Keisuke Tanaka and
                  Keita Xagawa},
  editor       = {Mitsuru Matsui},
  title        = {Efficient Public Key Encryption Based on Ideal Lattices},
  booktitle    = {Advances in Cryptology - {ASIACRYPT} 2009, 15th International Conference
                  on the Theory and Application of Cryptology and Information Security,
                  Tokyo, Japan, December 6-10, 2009. Proceedings},
  series       = {Lecture Notes in Computer Science},
  volume       = {5912},
  pages        = {617--635},
  publisher    = {Springer},
  year         = {2009}
  }

@inproceedings{LPR10,
  author       = {Vadim Lyubashevsky and
                  Chris Peikert and
                  Oded Regev},
  editor       = {Henri Gilbert},
  title        = {On Ideal Lattices and Learning with Errors over Rings},
  booktitle    = {Advances in Cryptology - {EUROCRYPT} 2010, 29th Annual International
                  Conference on the Theory and Applications of Cryptographic Techniques,
                  Monaco / French Riviera, May 30 - June 3, 2010. Proceedings},
  series       = {Lecture Notes in Computer Science},
  volume       = {6110},
  pages        = {1--23},
  publisher    = {Springer},
  year         = {2010}
  }

@article{DMPS24,
  author       = {Nir Drucker and
                  Guy Moshkowich and
                  Tomer Pelleg and
                  Hayim Shaul},
  title        = {{BLEACH:} Cleaning Errors in Discrete Computations Over {CKKS}},
  journal      = {J. Cryptol.},
  volume       = {37},
  number       = {1},
  pages        = {3},
  year         = {2024}
  }

@inproceedings{BCKS24,
  author       = {Youngjin Bae and
                  Jung Hee Cheon and
                  Jaehyung Kim and
                  Damien Stehl{\'{e}}},
  editor       = {Marc Joye and
                  Gregor Leander},
  title        = {Bootstrapping Bits with {CKKS}},
  booktitle    = {Advances in Cryptology - {EUROCRYPT} 2024 - 43rd Annual International
                  Conference on the Theory and Applications of Cryptographic Techniques,
                  Zurich, Switzerland, May 26-30, 2024, Proceedings, Part {II}},
  series       = {Lecture Notes in Computer Science},
  volume       = {14652},
  pages        = {94--123},
  publisher    = {Springer},
  year         = {2024}
  }

@InProceedings{CHKS25,
author="Cheon, Jung Hee
and Hanrot, Guillaume
and Kim, Jongmin
and Stehl{\'e}, Damien",
editor="Fehr, Serge
and Fouque, Pierre-Alain",
title="SHIP: A Shallow and Highly Parallelizable CKKS Bootstrapping Algorithm",
booktitle="Advances in Cryptology -- EUROCRYPT 2025",
year="2025",
url = {https://eprint.iacr.org/2025/784}
}

@misc{OpenFHE,
      author = {Ahmad Al Badawi and Andreea Alexandru and Jack Bates and Flavio Bergamaschi and David Bruce Cousins and Saroja Erabelli and Nicholas Genise and Shai Halevi and Hamish Hunt and Andrey Kim and Yongwoo Lee and Zeyu Liu and Daniele Micciancio and Carlo Pascoe and Yuriy Polyakov and Ian Quah and Saraswathy R.V. and Kurt Rohloff and Jonathan Saylor and Dmitriy Suponitsky and Matthew Triplett and Vinod Vaikuntanathan and Vincent Zucca},
      title = {{OpenFHE}: Open-Source Fully Homomorphic Encryption Library},
      howpublished = {Cryptology {ePrint} Archive, Paper 2022/915},
      year = {2022},
      url = {https://eprint.iacr.org/2022/915}
}

\appendix 

\iffullver
\else

\newpage

\begin{center}
\Large Supplementary Material
\end{center}

\fi

\section{Proof of lemmas \ref{lemma:noise_encryption},
\ref{lemma:noise_rescaling},
\ref{lemma:noise_multiplication},
\ref{lemma:noise_multiplication_encoding} and \ref{lemma:noise_key_switching}}

\label{a:error_analysis_proofs}

For the following proofs, we will use two facts.
For real numbers $r_1,r_2 \in \mathbb R$, it holds that $\infnorm{\near{r_1\pm r_2}} \le \infnorm{\near{r_1} \pm \near{r_2}} + 1$.
Moreover, for $r \in {\mathbb R}[X]/(X^N+1)$ and $s \in {\mathbb Z}[X]/(X^N+1)$ with hamming weight $h$ and $\infnorm{{s}} = 1$, we have that
$\infnorm{{s} \cdot \near{r(X)} - \near{{s}  \cdot r(X)}} \le h \le N$, 
which can be seen by applying the real number case to polynomial multiplication.

\subsection{Proof of Lemma \ref{lemma:noise_encryption}}

  Let ${\sf pk} = (-a's+e', a')$ be the public key and 
  $\ct$ be the ciphertext, i.e. $\ct = {\sf Enc}_{\sf pk}(m) = v \cdot {\sf pk} + (m + e_0, e_1)=((-a's+e')v+m+e_0,a'v+e_1)$, which gives
  \begin{align*}
 \langle {\sf ct}, {\sf sk} \rangle & =(-a's+e')v + m+e_0 + s (a'v+e_1) \pmod{q_L} \\
 & = m + e_0 +e'v+ se_1 \pmod{q_L}.
 \end{align*}
Writing  $\langle {\sf ct}, {\sf sk} \rangle = m + e_{\sf clean} \pmod {q_L}$, we get
for a signed binary secret key:
  \begin{align*}
  \infnorm{e_{\sf clean}} 
  &\le  \infnorm{e_0} + \infnorm{e' \cdot v} + \infnorm{s \cdot e_1 } \le \kappa + 
  N \cdot \kappa \cdot 1+  N \cdot 1 \cdot \kappa \leq  3N\kappa.
  \end{align*}

\subsection{Proof of Lemma \ref{lemma:noise_rescaling}}

Let $\gamma = q_\ell / q_{\ell'}$. We write $\ct=(b,a)$, which gives $\ct' = (\near{b/\gamma},\near{a/\gamma})$.
We have $\langle \ct, \sk \rangle = b + s \cdot a + \xi \cdot q_\ell$ for some 
$\xi \in {\mathbb Z}$, which gives:
$$  \langle \ct, \sk \rangle / \gamma = b/\gamma + s \cdot a / \gamma + \xi \cdot q_{\ell'},
$$
and therefore we obtain:
\begin{align*}
 \left \lfloor  \langle \ct, \sk \rangle / \gamma \right \rceil  & = \near{b/\gamma}+ s \cdot \near{a/\gamma} - e_{\sf rs} \pmod{q_{\ell'}} \\
 & =  \langle \ct', \sk \rangle - e_{\sf rs} \pmod{q_{\ell'}}, 
 \end{align*}
where $e_{\sf rs}=\near{b/\gamma}-b/\gamma + s \cdot (\near{a/\gamma} -
 a/\gamma )+ \langle \ct, \sk \rangle / \gamma
  -   \lfloor \langle \ct, \sk \rangle / \gamma \rceil$. This implies 
  $\infnorm{e_{\sf rs}} \leq  \|s\|_{1} + 1 \leq 2N$,
  with $\langle \ct',\sk \rangle= \near{\langle \ct,\sk \rangle \cdot q_{\ell'}/q_\ell} 
+ e_{\sf rs} \pmod{q_{\ell'}}$ as required.

\subsection{Proof of Lemma \ref{lemma:noise_multiplication}}

We use as an
input an evaluation key $\evk = (b', a') = (-a's + e' + Ps^2,a')$ with $\infnorm{e'} \le \kappa$ and ciphertexts $\ct_i$ satisfying
$
\langle \ct_i,\sk \rangle = b_i + s \cdot a_i   \pmod{q_\ell}$ for $i \in \{1,2\}$. We have:
$$ \langle \ct_1,\sk \rangle \cdot \langle \ct_2,\sk \rangle
  = b_1b_2+s(b_1a_2 + b_2a_1)+s^2 a_1 a_2 \pmod{q_\ell}.
  $$
Letting $(d_0, d_1, d_2) := (b_1b_2, b_1a_2 + b_2a_1, a_1a_2)$, we rewrite the above as:
$$ 
\langle \ct_1,\sk \rangle \cdot \langle \ct_2,\sk \rangle
  = d_0 + sd_1 + s^2d_2 \pmod{q_\ell}.
$$
We consider $d_2$ over ${\mathbb Z}[X] / (X^N+1)$, with $\infnorm{d_2} \leq q_\ell$. 
 By definition, $\ct_{\sf mult} = (d_0, d_1) + \near{P^{-1} d_2 \cdot \evk}$, where we evaluate $d_2 \cdot \evk$ modulo $Pq_L$ and then divide the latter by $P$ over the reals.
This implies, using $q_\ell | q_L$ and some $\gamma_1, \gamma_2 \in \mathbb Z$:
\begin{align*}
d_2 \cdot \evk & = d_2 \cdot (-a's+e'+Ps^2, a') + Pq_L(\gamma_1, \gamma_2) \qquad \text{over }\mathbb Z \text{, and} \\
\near{P^{-1} d_2 \cdot \evk} & = \left( 
\near{P^{-1}d_2 (-a's + e')} + d_2s^2,\near{P^{-1}a' d_2} \right) \pmod{q_{\ell}}.
\end{align*}
The above now gives:
\begin{align*}
 \langle \ct_{\sf mult},\sk \rangle & = d_0 + d_1 s + \near{P^{-1}d_2 (-a's + e')} + d_2s^2 + s\near{P^{-1}a' d_2} \pmod{q_\ell} \\ 
 & = \langle \ct_1,\sk \rangle \cdot \langle \ct_2,\sk \rangle + e_{\sf mult} \pmod{q_\ell},
 \end{align*}
and now we can estimate the size of $e_{\sf mult}$ as follows:
\begin{align*}
 \infnorm{e_{\sf mult}} 
  &= \infnorm{\near{P^{-1} d_2 (-a's + e')} + s \near{P^{-1} d_2 a'} } \\
  &\le \infnorm{\near{P^{-1} d_2 (-a's + e')} + \near{s \cdot P^{-1} d_2 a'} } + \|s\|_1 \\
  &\le \infnorm{-\near{s P^{-1} d_2 a'}+\near{P^{-1} d_2 e'} + \near{s P^{-1} d_2 a'}  } + \|s\|_1 + 1 \\
  &\le P^{-1} \cdot N q_\ell \kappa + N + 2.
\end{align*}
Finally, if $P \ge Nq_\ell \kappa$, then $\infnorm{e_{\sf mult}} \le 2N$.

\subsection{Proof of Lemma \ref{lemma:noise_multiplication_encoding}}

Letting $\ct_{\sf mult} \leftarrow {\sf Mult}_{\sf pk}(\ct_1,\ct_2)$,
we obtain by applying Lemma \ref{lemma:noise_multiplication}:
\begin{align*}
\langle \ct_{\sf mult},\sk \rangle  & = \langle \ct_1,\sk \rangle \cdot \langle \ct_2,\sk \rangle  +e_{\sf mult} \pmod{q_{\ell}} \\
& = ({\sf Ecd}(x_1)+e_1)({\sf Ecd}(x_2)+e_2) +e_{\sf mult} \pmod{q_{\ell}} \\
& = {\sf Ecd}(x_1){\sf Ecd}(x_2) + e_m +e_{\sf mult} \pmod{q_{\ell}},
\end{align*}
for $e_m=e_1e_2 + {\sf Ecd}(x_1)e_2 + {\sf Ecd}(x_2)e_1$, 
and $\infnorm{e_{\sf mult}} \leq 2N$. 
Letting $\ct_{\sf multR} \leftarrow {\sf RS}(\ct_{\sf mult})$, applying
Lemma \ref{lemma:noise_rescaling}, and using $q_{\ell}/q_{\ell-1}=\Delta$, we get:
\begin{align*}
 \langle \ct_{\sf multR}, \sk \rangle  & = \lfloor \langle \ct_{\sf mult},\sk \rangle/\Delta
\rceil +e_{\sf rs} \pmod{q_{\ell-1}} \\
& = \lfloor ({\sf Ecd}(x_1){\sf Ecd}(x_2) + e_m +e_{\sf mult})/\Delta  \rceil
+ e_{\sf rs} \pmod{q_{\ell-1}},
\end{align*}
where $\infnorm{e_{\sf rs}} \leq 2N$. 
By writing $  {\sf Ecd}(x_1){\sf Ecd}(x_2) / \Delta=
{\sf Ecd} (x_1x_2) + e_{\sf ecd}$, we now reformulate the above as:
\begin{align*}
\langle \ct_{\sf multR}, \sk \rangle 
& = {\sf Ecd}(x_1x_2) + \lfloor e_{\sf ecd} + (e_m +e_{\sf mult})/\Delta \rceil +e_{\sf rs}
\pmod{q_{\ell-1}} \\
& = {\sf Ecd}(x_1x_2)+ e_{\sf multR} \pmod{q_{\ell-1}}
\end{align*}
 Therefore, we have:
$$ \infnorm{e_{\sf multR}} \leq \abs{e_{\sf ecd}}
+ \frac{\infnorm{e_m}}{\Delta} + \frac{\infnorm{e_{\sf mult}}}{\Delta} +  
\infnorm{e_{\sf rs}}+1.
$$
We first bound the error $e_{\sf ecd}$:
\begin{align*}
e_{\sf ecd} &  = \frac{1}{\Delta} \cdot 
\left( {\sf Ecd}(x_1){\sf Ecd}(x_2)-
\Delta \cdot {\sf Ecd} (x_1x_2) \right) = \frac{1}{\Delta} 
\cdot \left( \near{\Delta x_1} \near{\Delta x_2}-\Delta \near{\Delta x_1x_2} \right)  \\
 & = \frac{1}{\Delta} \cdot 
 \left( (\Delta x_1+\varepsilon_1)(\Delta x_2+\varepsilon_2)-\Delta(\Delta x_1 x_2+\varepsilon_{12}) \right) \\
 & = \frac{1}{\Delta} \left( \Delta x_1\varepsilon_2+ \Delta x_2\varepsilon_1 + \varepsilon_1\varepsilon_2 -\Delta \varepsilon_{12} \right)
 \end{align*}
 for some $|\varepsilon_1|,|\varepsilon_2|,|\varepsilon_{12} | \leq 1/2$. 
Using $|x_i| \leq \nu$, this implies 
 $$ |e_{\sf ecd} | \leq \nu + 1.
 $$
 Recall that $e_m=e_1e_2 + {\sf Ecd}(x_1)e_2 + {\sf Ecd}(x_2)e_1$. Using ${\sf Ecd}(x_i) \in {\mathbb Z}$ and  $|{\sf Ecd}(x_i)| =|\near{x_i \cdot \Delta}|\leq \Delta \nu+1$, this gives:
 $$ \infnorm{e_m} \leq NE^2 + 2E \cdot (\Delta\nu +1 ) \leq 2NE^2+2 \nu E \Delta.
 $$
Now, for $e_{\sf multR}$, we have:
$$
\infnorm{e_{\sf multR}} \le \nu + 1 + \frac{2NE^2}{\Delta} + 2 \nu E + \frac{2N}{\Delta} + 2N+1.
$$
% Can we really assume \nu < N ??
Assuming $E^2 \leq \Delta$ and $\nu<N$, we eventually get:
$$\infnorm{e_{\sf multR}} \leq 2\nu E + 8N.
$$

\subsection{Proof of Lemma \ref{lemma:noise_key_switching}}

%Lemma 5: I simplified a bit the proof, since it is similar to the proof of Lemma 3.

 The proof is similar to the proof of Lemma \ref{lemma:noise_multiplication}.
We input the key-switching key $\mathsf{swk} = (b',a')$, for which $b' = -a's + e' + Ps' \pmod{  Pq_L}$.
 Let $\ct = (b,a)$ be the input ciphertext under $\sk'=(1,s')$.
   We consider $a$ over ${\mathbb Z}[X] /(X^N+1)$
 with $\infnorm{a} \leq q_\ell$. The output ciphertext is defined as
$\ct' = (b,0) + \near{P^{-1} a \cdot \mathsf{swk}} \pmod{q_\ell}$,
where we evaluate $a \cdot \mathsf{swk}$ modulo $P q_L$ and then divide by $P$ over the reals.
As in the proof of Lemma \ref{lemma:noise_multiplication}, this gives:
$$ \near{P^{-1} a \cdot \mathsf{swk}} = 
\left( 
\near{P^{-1}a (-a's + e')} + as',\near{P^{-1}a' a} \right) \pmod{q_{\ell}}.
$$
The decryption equation of $\ct'$ now indeed satisfies:
\begin{align*}
\langle \ct',\sk \rangle  & =
b + \near{P^{-1}a (-a's + e')} + as' + s \cdot \near{P^{-1}a' a}  \pmod{q_\ell} \\
& = \langle \ct,\sk' \rangle + \near{P^{-1}a (-a's + e')} + s \cdot \near{P^{-1}a' a} \pmod{q_\ell} \\
& = \langle \ct,\sk' \rangle + e_{\sf ks} \pmod{q_\ell}.
\end{align*}
Finally, we estimate the size of $e_{\sf ks}$ as follows:
  \begin{align*}
    \infnorm{e_{\sf ks}} 
    &= \infnorm{\near{P^{-1}a(-a's + e')} + s \cdot \near{P^{-1} a' a}} \\
    & \leq \infnorm{\near{P^{-1}a(-a's + e')} + \near{s P^{-1} a' a}} + \|s\|_1\\
    & \leq \infnorm{-\near{P^{-1}aa's} + \near{P^{-1}a e'} + \near{s P^{-1} a' a}} + \|s\|_1 + 1 \\
    &\le N+2 + \infnorm{P^{-1} a \cdot e'} \leq N+2+P^{-1} \cdot Nq_\ell \kappa \leq N+3 \leq 2N.
  \end{align*}

\section{The trace and product operators}

\label{a:traceproduct}

\subsection{The trace operator}

\label{a:trace}

\begin{figure}
\centering
\scalebox{.65}{
\begin{tikzpicture}
\pgfmathsetseed{1}

% Define the size of the main rectangle and the number of blocks
\def\blockwidth{2} % Width of each block
\def\blockheight{1}  % Height of the rectangle
\def\nblocks{8}      % Number of blocks
\def\nsubrects{4}    % Number of vertical rectangles in each block

\def\parendistance{0.2}

\draw[very thick] (0,0) rectangle ++(\nblocks*\blockwidth,\blockheight);

% Loop to create the blocks
\foreach \i in {0,...,7} {
    % \draw[very thick] (\i*\blockwidth, 0) rectangle ++(\blockwidth, \blockheight);

    % Loop to create the vertical rectangles within each block
   \foreach \j in {0,...,3} {
        % Calculate x position of the sub-rectangle
       \pgfmathsetmacro{\xpos}{\i*\blockwidth + \j*\blockwidth/\nsubrects}

        % Draw the vertical rectangle with random color
        \draw (\xpos, 0) rectangle ++(\blockwidth/\nsubrects, \blockheight);
    }
    
  }

\draw [thick,decorate,decoration={brace,amplitude=15pt,mirror,raise=.1cm}]
  (1.25,0) -- (1.25+7*\blockwidth,0) ;

\foreach \i in {1,...,6} {

  % Calculate positions for the brace
    \pgfmathsetmacro{\parenleft}{(\i+.37)*\blockwidth + \blockwidth/\nsubrects}
    \pgfmathsetmacro{\parenright}{(\i+1.37)*\blockwidth - \blockwidth/\nsubrects}

    \draw[thick] (\parenleft,-0.1) -- (\parenleft,-0.35);
}

\foreach \i in {0,...,7} {

  % Calculate positions for the brace
    \pgfmathsetmacro{\parenleft}{(\i+.37)*\blockwidth + \blockwidth/\nsubrects}
    \pgfmathsetmacro{\parenright}{(\i+1.37)*\blockwidth - \blockwidth/\nsubrects}

    \draw[thick,arrows = {-Latex[width'=0pt .5, length=6pt]}] (1.25+3.5*\blockwidth,-0.6) -- (\parenleft,-2);
}

\begin{scope}[yshift=-3cm]

\draw[very thick] (0,0) rectangle ++(\nblocks*\blockwidth,\blockheight);

\foreach \i in {0,...,7} {
    % Draw the outer block

    \draw[very thick] (\i*\blockwidth,0) rectangle ++(\blockwidth,\blockheight);

    % Loop to create the vertical rectangles within each block
   \foreach \j in {0,...,3} {
        % Calculate x position of the sub-rectangle
       \pgfmathsetmacro{\xpos}{\i*\blockwidth + \j*\blockwidth/\nsubrects}
        
        % Draw the vertical rectangle with random color
        \draw (\xpos, 0) rectangle ++(\blockwidth/\nsubrects, \blockheight);
    }
    
  }

\end{scope}

\node[circle,draw,thick,minimum size=1.3cm] at (-1.5,-1) (op) {\large ${\sf Tr}_{N/2 \rightarrow a}$};

\draw[thick,->] (0,.5) .. controls (-1,.5) and (-1.5,0) .. (op.north);

\draw[thick,->] (op.south) .. controls (-1.5,-2) and (-1,-2.5) .. (0,-2.5);

\draw [thick,decorate,decoration={brace,amplitude=5pt,raise=4ex}]
  (0,-2.2) -- (\blockwidth,-2.2) node[midway,yshift=3em]{\large $a$};

\draw [thick,decorate,decoration={brace,amplitude=10pt,raise=4ex}]
  (0,1) -- (8*\blockwidth,1) node[midway,yshift=4em]{\large $N/2$};
\end{tikzpicture}
}
\caption{Illustration of the $\tr{N/2}{a}$ 
operator with $N/2=32$ slots and $a=4$. Each slot in the output polynomial
is the sum of all input slots with the same index modulo $a$. Therefore, the output polynomial contains identical blocks of $a$ slots each.}
\label{f:trace}
\end{figure}
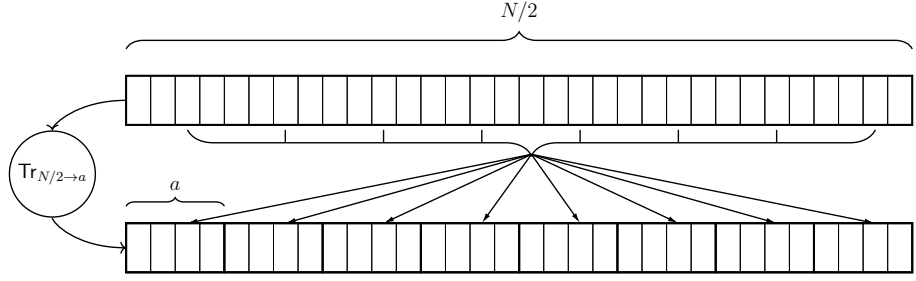

For a power-of-two $a$, we define the operator $\tr{N/2}{a}$ as: 
% : {\mathbb Z}[X] / (X^N+1)  \rightarrow {\mathbb Z}[X^{N/(2a)}] / (X^N+1)$:
$$
\tr{N/2}{a} = ({\sf id} + \Psi_a) \circ 
({\sf id} + \Psi_{2a}) \circ ({\sf id} + \Psi_{4a}) \circ \dots \circ ({\sf id} + \Psi_{N/4}).
$$
One can show recursively that
the $\tr{N/2}{a}$ operator computes partial sums of the slots
sharing the same index modulo $a$, namely for any $\vec{z} \in {\mathbb C}^{N/2}$:
$$\tr{N/2}{a}({\sf iDFT}(z_0,\ldots,z_{N/2-1})) =
{\sf iDFT}(\underbrace{\vec{t},\ldots\ldots,\vec{t}}_{N/(2a)~\textrm{rep.}})$$
where $\vec{t}=(t_k)_{0 \leq k < a}$,
with  $ t_k=z_k + z_{k+a} + \cdots + z_{k+(N/(2a)-1) \cdot a}$. 
Therefore, the  $\tr{N/2}{a}$ operator outputs an encoding of the $a$ slots $t_0,\ldots,t_{a-1}$,
which are repeated $N/(2a)$ times to get $N/2$ slots; see Fig. \ref{f:trace} for
an illustration.

Namely, 
the operator $\tr{N/2}{N/4}={\sf id} + \Psi_{N/4}$ 
computes the sum of pairs of slots sharing the same index modulo $N/4$. 
Assuming the property for $\tr{N/2}{a}$ and using
$\tr{N/2}{a/2}=({\sf id} + \Psi_{a/2}) \circ \tr{N/2}{a}$, the operator 
$\tr{N/2}{a/2}$ computes the sum of slots with the same index modulo $a/2$.

As described in Alg. \ref{alg:hom_trace} below, the $\tr{N/2}{a}$ operator can be homomorphically evaluated, 
requiring $\log_2 (N/(2a))$ automorphisms and key switchings, 
while consuming no multiplicative levels.

% See https://eprint.iacr.org/2020/015.pdf for some reference on the trace field operator.

\begin{algorithm}
  \caption{Homomorphic trace operator}
  \label{alg:hom_trace}
  \begin{algorithmic}[1]
    \Require A ciphertext $\ct$, a power-of-two parameter $a$
    \Ensure Computes $\tr{N/2}{a}(\ct)$
    \smallskip
      \State $r = N/4$
      \While {$r \ge a$}
        \State $\ct = {\sf Add}(\ct, {\sf Rot}(\ct, r))$
        \State $r = r / 2$
      \EndWhile
      \State \Return $\ct$
  \end{algorithmic}
\end{algorithm}

\subsection{The product operator}

\label{a:product}

\begin{algorithm}[H]
  \caption{Homomorphic product operator}
  \label{alg:hom_product}
  \begin{algorithmic}[1]
    \Require A ciphertext $\ct$, two power-of-two parameters $a > b$
    \Ensure Computes $\pr{a}{b}(\ct)$
    \smallskip
      \State $r = a/2$
      \While {$r \ge b$}
        \State $\ct = {\sf MultR}_\evk(\ct, {\sf Rot}(\ct, r))$
        \State $r = r / 2$
      \EndWhile
      \State \Return $\ct$
  \end{algorithmic}
\end{algorithm}

\section{The original \ckks bootstrapping procedure}

\label{app:CKKSboot}

\subsubsection{Bootstrapping a single coefficient.}
We recall the original \ckks bootstrapping approach \cite{CHKKS18}. For simplicity, we 
first consider the bootstrapping of a single coefficient, that is a \ckks ciphertext $\ct$ encrypting a message $m \in {\mathbb Z}$,
with $\langle \ct,\sk \rangle = m + e \pmod q$, where $q$ is
the smallest modulus in the ladder. 

We first perform a modulus lifting to a larger modulus $Q$, which gives:
$$ \langle \ct,\sk \rangle =qI+m + e  \pmod{Q}$$
for some $I \in {\mathbb Z}[X]/(X^N+1)$. The goal of bootstrapping is to get rid
of the $qI$ term. 
 Using the $\tr{N/2}{1}$ operator and also 
${\sf Re2}={\sf id} + {\sf conj}$, we have ${\sf Re2}(\tr{N/2}{1}(qI+m))=
qN \cdot \tilde{I} + N \cdot m$ where $\tilde{I} \in {\mathbb Z}$.
Therefore, by applying these operators 
homomorphically, we can obtain a new ciphertext $\ct'$ such that
$$
\langle \ct',\sk \rangle =qN \cdot \tilde{I}+Nm + e'  \pmod{Q}
$$
for a small error $e'$, 
so we must
remove the term $qN \cdot \tilde{I}$,  where now  $\tilde{I} \in {\mathbb Z}$
instead of $I \in {\mathbb Z}[X]/(X^N+1)$. 

  Such a ciphertext $\ct'$ corresponds to an encoding of $x=(qN \cdot \tilde{I}+Nm)/\Delta \in {\mathbb R}$. Eventually, we want to obtain an encryption of 
$m$, which corresponds to an encoding of $m/\Delta \in {\mathbb R}$. 
Therefore, we homomorphically evaluate on $\ct'$ a polynomial $P(t)$ such that
 $m/\Delta \simeq P(x)$.
 This yields the condition 
 $m/\Delta \simeq P((qN \cdot \tilde{I}+Nm)/\Delta)$, which
 must hold for any small $\tilde{I} \in {\mathbb Z}$. 
We expect the integer  $\tilde{I}$ to be small because the secret key $\sk$ has a small Hamming weight.
 
For $P(t)$, we can therefore  use a polynomial approximation of a periodic function with period $qN/\Delta$, with the appropriate scaling factor:
$$
P(t) \simeq \frac{q}{2\pi  \Delta} \sin \left( \frac{2\pi \Delta \cdot t}{qN}  \right). $$
One could use a Taylor series approximation of the sine function, but the required degree would be too large. Instead, we work over the complex numbers
and first consider a degree-$d$ Taylor series approximation of the complex exponential function,
with a scaling factor $2^r$ as input:
$$
Q(t)=\sum\limits_{k=0}^d \frac{1}{k!} \left( \frac{2i\pi\Delta \cdot t}{qN2^r}
\right)^k \simeq \exp \left( \frac{2i\pi \Delta \cdot t}{qN2^r} \right).
$$
Due to the scaling factor $2^r$, much fewer coefficients are needed 
in the Taylor series approximation. 
 Note that $Q(t)$ is a polynomial with complex coefficients, so when homomorphically 
 evaluating $Q(t)$ on $\ct'$, we consider the encryptions of the real part and the 
 imaginary part separately.
Finally, we can write
\begin{equation}
\label{eq:Pt}
P(t)=\frac{q}{2\pi  \Delta} \cdot {\sf Im} \left( Q(t)^{2^r} \right),
\end{equation}
which we evaluate homomorphically by performing $r$ successive squarings.
Eventually, we only keep the encryption of the imaginary 
 part and apply the scaling factor. Finally, we obtain a new ciphertext 
$\ct''$ such that
\begin{align*}
 \langle \ct'',\sk \rangle &  \simeq {\sf Ecd}(P(x)) \simeq {\sf Ecd}(m/\Delta) \pmod{q'} \\ 
 & = m + e'' \pmod{q'}
 \end{align*}
 for a small error $e''$. 
 Therefore, we have obtained a new encryption of $m$ for the modulus $q'$. The
 previous evaluation procedure has a multiplicative depth of $\ell=d+r+1$. Therefore,
 starting from the largest modulus $Q=q \cdot p \cdot \Delta^{\ell}$, after $\ell$
 rescalings, we end up with a modulus $q'=q \cdot p$. This enables to perform
 one more homomorphic multiplication on $\ct'$, thereby achieving bootstrapping.
We can reduce the depth to $\ell=d+r$ by using a scaling
factor $\delta$ in $Q(t)$ such that $\delta^{2^r}=q/(2\pi\Delta)$.

Moreover, the depth required to evaluate 
a polynomial of degree $d$ can 
 be reduced to ${\cal O}(\log d)$ at the expense 
 of additional multiplications \cite{CKKS17}. 
 For our parameters, where $d=7$ is relatively small, 
 this approach slightly reduces the largest modulus $Q$.
  However, in practice, we did not observe a significant improvement in running time. 
 
 \ignore{
 
\subsubsection{Error analysis.}
The error in the Taylor approximation is upper bounded by $\frac{1}{(d+1)!} \cdot (\frac{2\pi K}{2^r})^{d+1}$. 
Thus, the final
absolute error is then upper-bounded by 
$$
\frac{\Delta \cdot 2^r}{(d+1)!} \cdot \left(\frac{2\pi K}{2^r}\right)^{d+1} 
\simeq \frac{\Delta \cdot 2^r}{\sqrt{2\pi(d+1)}} \cdot \left(\frac{e\pi K}{2^{r-1}(d+1)}\right)^{d+1}.
$$
The authors claim that they can take $d={\cal O}(1)$ and $r={\cal O}(\log(Kq))$.

In practice, one can use a secret vector $\vec{s}$ with small Hamming weight $h$. In that case, if the coefficients of 
$\vec{c}$ are centered around $0$, the value $\tilde{I}$ will behave like a Gaussian with standard deviation $\sqrt{h}$. 
In \cite{CHKKS18}, the authors used $h=64$. Therefore, we can use the bound $K= {\cal O}(\sqrt{h})$. 
 
 The last coefficient of $Q(t)$ is
$ \frac{\Delta}{d_0!} \left( \frac{2\pi i \Delta }{2^r \cdot q} \right)^{d_0}$.
If we want full precision, we may require $\frac{\Delta}{d_0!} \left( \frac{2\pi i \Delta }{2^r \cdot q} \right)^{d_0}= \Delta$, which gives:
$(2 \pi \Delta)/(2^r q)=(d_0!)^{1/d_0}$ and finally:
$$
\Delta=(d_0!)^{1/d_0} \cdot \frac{2^r \cdot q}{2 \pi} \simeq \frac{2^r \cdot q \cdot e}{2 \pi}.
$$

 }
 
\subsubsection{Bootstrapping multiple components.}
The bootstrapping of multiple components proceeds similarly, which
we describe briefly below.
For a power-of-two  $n$ with $2 \leq n \leq N/2$, 
we consider the plaintext space  
${\mathbb Z}[X^{N/n}]/(X^N+1) \simeq {\mathbb Z}[Y] / (Y^n+1)$ for $Y=X^{N/n}$; 
therefore, the  plaintext space
has $n$ non-trivial coefficients.
 Thus, we start from a ciphertext $\ct$ with
$\langle \ct,\sk \rangle = m + e \pmod{q}$ for $m \in {\mathbb Z}[X^{N/n}]/(X^N+1)$.

\begin{enumerate}
\item Modulus raising: As previously, we consider the same $\ct$ modulo a larger modulus $Q$, which gives 
$ \langle \ct,\sk \rangle =qI+m + e  \pmod{Q}$
for a small $I \in {\mathbb Z}[X]/(X^N+1)$,
where $m(X)=\sum_{i=0}^{n-1} m_i \cdot X^{i \cdot N/n}$.

\smallskip

\item Error packing: We homomorphically apply the $\tr{N/2}{n/2}$ operator, which yields a new ciphertext $\ct_1$, for which $\langle \ct_1,\sk \rangle =q\tilde{I} \cdot N/n+m \cdot N/n + e  \pmod{Q}$,
where $\tilde{I} \in {\mathbb Z}[X^{N/n}]/(X^N+1)$.

\smallskip

\item Coefficients to slots: We use the $\cts$ procedure recalled previously
to homomorphically compute a new ciphertext $\ct_2$, such that
$ \langle \ct_2,\sk \rangle = {\sf Ecd}(m_0+q\tilde{I}_0,\ldots,m_{n-1}
+q \tilde{I}_{n-1}) + e_2 \pmod{Q_2}$.

\smallskip

\item Approximate modular reduction: The polynomial approximation of the reduction function modulo $q$ gets independently applied on each of the $n$ slots,
which then yields a new ciphertext $\ct_3$, for which $ \langle \ct_3,\sk \rangle = {\sf Ecd}(m_0,\ldots,m_{n-1}) 
+ e_3 \pmod{Q_3}$.

\smallskip

\item Slots to coefficients: We use the $\stc$ procedure recalled previously
to homomorphically compute a new ciphertext $\ct_4$, such that
$ \langle \ct_4,\sk \rangle = m(X)+ e_4 \pmod{q \cdot p}$, for a modulus $q \cdot p$, which
is larger than the initial modulus $q$. 
\end{enumerate}

Since the complexity of $\cts$ and $\stc$ is ${\cal O}(\log n)$
for $n$ slots (using the homomorphic Fast Fourier Transform; see Appendix \ref{a:slottocoeff}), 
the complexity of the original \ckks bootstrapping for $n$ slots is also ${\cal O}(\log n)$.

\section{Fast ${\sf DFT}$s and the $\stc$ operation}

\label{a:slottocoeff}

% The number of operations in R_q in the homomorphic DFT
% is O(\pho * N^(1/\rho)) where  \rho is the number of Cooley-Tukey recursion steps. The depth
% is linear in \rho.
% More precisely, O(\rho N^(1/2\rho)) rotations and O(rhoN^(1/\rho)) plaintext-ciphertext multiplications

% Plaintext-ciphertext multiplication can be performed without consuming any modulus level,
% as in key switching.

% [BTH22] Bossuat et al. Bootstrapping for approximate homomorphic encryption
%    with negligible failure probability by using spart-secret encapsulation
% They suggest to use two different keys: we use the sparse one before the modraise, and switch
%  to the dense one immediately after. 
%  This applies to both the original CKKS bootstrapping and our new bootstrapping.

%We recall the variant of the Discrete Fourier Transform (DFT) considered
%in \cite{CKKS17} over ${\cal S}_N={\mathbb R}[X] / (X^N+1)$, for a power-of-two $N$. 

\subsection{The {\sf DFT} and {\sf iDFT} isomorphisms}

Recall that the $M$-th cyclotomic polynomial is defined as:
$$ \Phi_M(x)= \prod\limits_{\substack{1 \leq j \leq M \\[.05cm] \gcd(j,M)=1}} \left (x- e^{2I \pi j/M} \right)$$
For a power-of-two $M$, we have $\Phi_M(x)=x^N+1$ where $N=M/2$.
For a power-of-two $N$, the integer $5$ has order $N/2$ modulo $2N$, and generates ${\mathbb Z}^*_{2N}$ together 
with the integer $\{-1\}$.
Let $\zeta=e^{I\pi/N}$ be a primitive $2N$-th root of unity. Then letting $\zeta_j:=\zeta^{5^j}$
for $0 \leq j <N/2$, the set
$\{ \zeta_j,\overline{\zeta_j} : 0 \leq j <N/2 \}$ forms the set of the primitive $2N$-th roots of unity.

The variant of the  Fourier transform ${\sf DFT}_N$ over
${\cal S}_N$ considered in \cite{CKKS17} is defined as
\begin{align*}
{\sf DFT}_N : {\cal S}_N & \rightarrow {\mathbb C}^{N/2} \\
 p & \mapsto \left(p\left(\zeta^{5^0}\right), p\left(\zeta^{5^1}\right), \dots, p\left(\zeta^{5^{N/2-1}}\right) \right).
 \end{align*}
We denote by
${\sf iDFT}_N : {\mathbb C}^{N/2} \rightarrow {\cal S}_N$ the inverse transform.

The ${\sf DFT}_N$ function is a ring
homomorphism from the ring ${\cal S}_N$ (the coefficients space) to
the ring of complex vectors in ${\mathbb C}^{N/2}$ (the slots space).
Its purpose is therefore to map polynomial addition and multiplication in the ring
${\cal S}_N$ to component-wise addition and multiplication in ${\mathbb C}^{N/2}$.

\subsection{${\sf DFT}$ and ${\sf iDFT}$ on sub-rings}

We consider any divisor $n$ of $N$ with $1 \leq n \leq N/2$, 
and a polynomial $p$  in the subring ${\mathbb R}[X^{N/n}] / (X^N+1)$. We can write
$p(X)=\tilde{p}(Y) \in {\mathbb R}[Y] / (Y^n+1)$ where $Y=X^{N/n}$. Then
$${\sf DFT}_N(p)=\left( p(\zeta^{5^i}) \right)_{0 \leq i<N/2}=\left( \tilde{p}(\zeta^{5^i \cdot N/n}) \right)_{0 \leq i<N/2}$$
 Since $5$  has order $n/2$ modulo $2n$, ${\sf DFT}_N(p)$ is periodic with period $n/2$,
 with $N/n$ repetitions of the same pattern:
 $$ {\sf DFT}_N(p)=\rep{{\sf DFT}_n(\tilde{p})}{N/n}.
$$
Conversely, for any $\vec{z} \in {\mathbb C}^{n/2}$, we have:
$$ {\sf iDFT}_N\big(\rep{\vec{z}}{N/n} \big)={\sf iDFT}_n(\vec{z})(X^{N/n}) $$
% ${\sf iDFT}(\vec{z}) \in {\mathbb R}[X^{N/{(2n)}}] / (X^N+1)$.

\subsection{Computing ${\sf DFT}$ and ${\sf iDFT}$}

\label{a:DFTmatrix}

Let $m(X)=\sum\limits_{i=0}^{N-1} m_i X^i \in {\mathbb R}[X] / (X^N+1)$ and let 
$\vec{m}=(m_0,\ldots,m_{N-1}) \in {\mathbb R}^N$
be the vector of its coefficients. Letting 
$\zeta=e^{I\pi/N}$  and   $\zeta_j:=\zeta^{5^j}$ 
for $0 \leq j <N/2$ with $\zeta=e^{I\pi/N}$, we
have $\vec{z}={\sf DFT}_N(m)=U \cdot \vec{m} \in {\mathbb C}^{N/2}$ where
$$ U=\begin{bmatrix}
1 & \zeta_0 & \ldots & \zeta_0^{N-1} \\
\vdots & \vdots & \ddots & \vdots \\
1 & \zeta_{N/2-1} &  \ldots & \zeta_{N/2-1}^{N-1}
\end{bmatrix}
$$
is the $(N/2) \times N$ Vandermonde matrix generated by $\{\zeta_j : 0 \leq j < N/2 \}$.
The decoding algorithm ${\sf DFT}$ is therefore a linear transformation from ${\mathbb R}^N$ to ${\mathbb C}^{N/2}$, given by the matrix $U$.

We consider the full Vandermonde matrix ${\sf CRT}$ generated by the set $\{\zeta_j,\overline{\zeta_j} : 0 \leq j <N/2 \}$, which gives ${\sf CRT}=(U ; \overline{U})$, and $(\vec{z}, \overline{\vec{z}})=(U \vec{m};\overline{U} \vec{m})={\sf CRT} \cdot \vec{m}$. We can then compute $\vec{m}={\sf CRT}^{-1}(\vec{z}, \overline{\vec{z}})$, where
${\sf CRT}^{-1}=\frac{1}{N} \overline{{\sf CRT}}^T$. This enables to write
the coefficients $\vec{m}$ of the polynomial $m(X)={\sf iDFT}_N(\vec{z})$ as the linear transformation from
${\mathbb C}^{N/2}$ to ${\mathbb R}^N$:
$$ \vec{m}=\frac{1}{N} \left( \overline{U}^T \cdot \vec{z} + U^T \cdot \overline{\vec{z}} \right)$$

\subsection{Fast computation of the {\sf DFT} and bit-reversed order}

The drawback of the technique from the previous section to compute the {\sf DFT} is that its complexity
is ${\cal O}(N^2)$ operations. In this section, we recall the fast computation of the {\sf DFT}
function, with complexity ${\cal O}(N \log N)$. 

We work in the ring ${\cal S}= {\mathbb R}[X] / (X^N+1)$. 
Using $X^N+1=(X^{N/2}-I)(X^{N/2}+I)$, we have the isomorphism:
\begin{align*}
\Phi : {\mathbb R}[X] / (X^N+1)  &  \simeq 
{\mathbb C}[X] / (X^{N/2}-I) \\
a & \mapsto (a'=a \bmod (X^{N/2}-I)).
 \end{align*}
The coefficients $a'_j \in {\mathbb C}$ of $a'$ can be computed as 
$a'_j=a_j+ I \cdot a_{j+N/2}$. 
Conversely, we have 
$
a_j  =(a'_j + \bar{a}'_j)/2$ and  $a_{j+N/2} = (a'_j-\bar{a}_j')/(2I)$.
Therefore, the real part of the complex coefficients of $a'$
originates from the first $N/2$ coefficients
of $a \in \cal S$, while the imaginary part comes
from the last $N/2$ coefficients of $a$.

We now consider the ring ${\mathbb C}[X] / (X^{N/2}-I)$. 
We use $\zeta=e^{I \pi/N}$. 
Using
$\zeta^{N/4}=e^{I\pi/4}$ and $\zeta^{5N/4}=e^{5I\pi/4}=-\zeta^{N/4}$, we can further split the ring as follows:
$$
{\mathbb C}[X] / (X^{N/2}-I) \simeq {\mathbb C}[X] / ( X^{N/4}-\zeta^{N/4 \cdot (5^0 \bmod 8)}) \times
{\mathbb C}[X] / ( X^{N/4}-\zeta^{N/4 \cdot (5^1 \bmod 8)}).
$$

We can generalize this approach as follows.
Fix any integer $k \geq 1$ such that $2^k<N$. 
For any $0 \leq j<2^{k-1}$, consider the root $\zeta^{N/2^k \cdot 5^j}$. 
It has two square roots $\zeta^{N/2^{k+1} \cdot 5^j}$
and $-\zeta^{N/2^{k+1} \cdot 5^j}$.
Moreover, we  have $5^{2^{k-1}}=1+2^{k+1} \pmod{2^{k+2}}$.
This gives, for any $0 \leq j<2^{k-1}$:
$$
5^{j+2^{k-1}} =5^j \cdot (1+2^{k+1})=5^j+2^{k+1} \pmod{2^{k+2}}.
$$
This enables to write the  two square roots 
as $\zeta^{N/2^{k+1} \cdot 5^j}$
and 
$ \zeta^{N/2^{k+1}\cdot 5^{j+2^{k-1}}}$, which yields the isomorphism for $0 \leq j<2^{k-1}$:
\begin{align*}
 {\mathbb C}[X] / (X^{N/2^k}-\zeta^{N/2^k\cdot 5^j})
 \simeq  & ~{\mathbb C}[X] / \left(X^{N/2^{k+1}}-\zeta^{N/2^{k+1}\cdot 5^j} \right) \\
 & ~\times {\mathbb C}[X] / \left(X^{N/2^{k+1}}-\zeta^{N/2^{k+1}\cdot 5^{j+2^{k-1}}} \right).
\end{align*}
For the corresponding isomorphism with $u=\zeta^{N/2^{k+1}\cdot 5^j}$
$$ \Phi_k(a)=\left(a'=a \bmod{(X^{N/2^{k+1}}-u)},~a''=a \bmod{(X^{N/2^{k+1}}+u)}\right),$$
The coefficients of $a'$ and $a''$ can be computed efficiently via
\begin{align*}
a'_i & =a_i + u \cdot a_{i+m} \\
a''_i &=a_i - u \cdot a_{i+m}
\end{align*}
for $0 \leq i<m$, for $m=N/2^{k+1}$.
This is known
as the Cooley-Tukey butterfly.

We can then apply this technique recursively.
After step $k$ for $1 \leq k \leq \log_2 (N/2)$,
 we arrive at the isomorphism:
$$
{\mathbb C}[X] / (X^{N/2}-I) \simeq 
\prod_{i=0}^{2^{k}-1} \  {\mathbb C}[X] / \left(X^{N/2^{k+1}}-\zeta^{N/2^{k+1} \cdot 5^{{\sf br}_{k}(i)}} \right).
$$
where ${\sf br}_k(i)$ is the bit-reversal of the $k$-bit integer $i$.
It has the properties ${\sf br}_{k}(2i)={\sf br}_{k-1}(i)$ and ${\sf br}_{k}(2i+1)=2^{k-1}+{\sf br}_{k-1}(i)$.

Eventually, for $N=2^\ell$, we obtain the evaluation of the input polynomial at the roots $\zeta^{5^{{\sf br}_{\ell-1}(i)}}$
for $0 \leq i <N/2$, therefore in bit-reversed order. Recall that 
we have defined ${\sf DFT}_N(p)$ as $\left( p(\zeta^{5^i}) \right)_{0 \leq i<N/2}$.
Let ${\sf Br}_k$ denote the function taking as input a sequence of $2^k$
integers and outputting it in bit-reversed order. 
With this notation, the above procedure yields a sequence $(v_i)_{0 \leq i <N/2}=\left( p(\zeta^{5^{{\sf br}_{\ell-1}(i)}}) \right)_{0 \leq i<N/2}={\sf Br}_{\ell-1}({\sf DFT}_N(p))$.
In total, we obtain the following fast ${\sf DFT}_N$ algorithm, from normal order $({\sf no})$ to bit-reversed order $({\sf bo})$.

\begin{algorithm}[H]
\caption{Fast computation of ${\sf DFT}_N$, ${\sf no} \rightarrow {\sf bo}$}
\label{alg:fastDFTN}
\begin{algorithmic}[1]
\Require A polynomial $p(X) \in {\mathbb R}[X] / (X^N+1)$, 
with $p(X)=\sum_{j=0}^{N-1} p_j X^j$, $\zeta=\exp(I \pi/N)$ and $N=2^\ell$.
\Ensure A sequence $(v_i)_{0 \leq i <N/2}={\sf Br}_{\ell-1}({\sf DFT}_N(p))=\left( p(\zeta^{5^{{\sf br}_{\ell-1}(i)}}) \right)_{0 \leq i<N/2}$. 
\smallskip
\State $(v_j)_{0 \leq j <N/2}=(p_j+ I \cdot p_{j+N/2})_{0 \leq j <N/2}$
\label{l:RnToCn2}
\For{$k=1$ to $\ell-1$}
   \For{$j=0$ to $2^{k-1}-1$}
   \label{l:forj}
   		\State  $u=\zeta^{5^{{\sf br}_{k-1}(j)} \cdot N/2^{k+1}}$
   		\For{$i=0$ to $N/2^{k+1}-1$}
   		   \State  $a,b \leftarrow v_{j \cdot N/2^k+i},~v_{j \cdot N/2^k+i+N/2^{k+1}}$
   		   \label{l:l6}
		   \State $v_{j \cdot N/2^k+i}=a + u \cdot b$ %\Comment{We only compute $u \cdot b$ once.}
		   \label{l:l7}
		   \State $v_{j \cdot N/2^k+i+N/2^{k+1}}=a - u \cdot b$
		   \label{l:l8}
		 \EndFor
	\EndFor
\EndFor
\State \Return  $(v_i)_{0 \leq i <N/2}$
\end{algorithmic}
\end{algorithm}

We describe the corresponding ${\sf iDFT}$ algorithm, that computes the same operations
in reverse order.

\begin{algorithm}[H]
\caption{Fast computation of ${\sf iDFT}_N$, ${\sf bo} \rightarrow {\sf no}$}
\label{alg:fastiDFTN}
\begin{algorithmic}[1]
\Require A sequence $(v_i)_{0 \leq i <N/2}$, $\zeta=\exp(I \pi/N)$ and $N=2^\ell$.
\Ensure A polynomial $p(x) \in {\mathbb R}[X] / (X^N+1)$, such that
$(v_i)_{0 \leq i <N/2}={\sf Br}_{\ell-1}({\sf DFT}_N(p))$
\smallskip
\For{$k=\ell-1$ downto $1$}
  \For{$j=0$ to $2^{k-1}-1$}
 	\State $u=\zeta^{5^{{\sf br}_{k-1}(j)} \cdot N/2^{k+1}}$
	\For{$i=0$ to $N/2^{k+1}-1$}
		\State $a,b \leftarrow v_{j \cdot N/2^k+i},~v_{j \cdot N/2^k+i+N/2^{k+1}}$
		\State $v_{j \cdot N/2^k+i}=(a+b)/2$
		\State $v_{j \cdot N/2^k+i+N/2^{k+1}}=u^{-1} \cdot (a-b)/2$
	\EndFor
 \EndFor
\EndFor
\State \Return $p(x)=\sum\limits_{i=0}^{N/2-1} {\sf Re}(v_i) x^i + {\sf Im}(v_i) x^{i+N/2}$
\end{algorithmic}
\end{algorithm}

In Algorithm \ref{alg:fastDFTN_bo}, we describe the same ${\sf DFT}_N$ algorithm as in
Alg. \ref{alg:fastDFTN}, but with the input
coefficients of the polynomial in bit-reversed order, and the output coefficients in normal order.
In fact, we use a modified bit-reversed indexing, in which the first and second half of the coefficients are encoded separately in bit-reversed order.
This will later facilitate the homomorphic evaluation of the algorithm.

Due to this modified bit-reversed encoding, Line \ref{l:RnToCn2} of Alg. \ref{alg:fastDFTN}
remains the same. At Line \ref{l:forj}, we can
equivalently run $j'$ from $0$ to $2^{k-1}-1$, and let $j={\sf br}_{k-1}(j')$.
At lines \ref{l:l6}, \ref{l:l7}, \ref{l:l8}, we have 
\begin{align*}
{\sf br}_{\ell-1}(j \cdot N/2^k +i) & ={\sf br}_{\ell-k-1}(i) \cdot 2^k+{\sf br}_{k-1}(j) \\
{\sf br}_{\ell-1}(j \cdot N/2^k +i+N/2^{k+1}) & ={\sf br}_{\ell-k-1}(i) \cdot 2^k+2^{k-1}+{\sf br}_{k-1}(j)
\end{align*}
This gives the following algorithm.

\begin{algorithm}[H]
\caption{Fast computation of ${\sf DFT}_N$, ${\sf bo'} \rightarrow {\sf no}$}
\label{alg:fastDFTN_bo}
\begin{algorithmic}[1]
\Require A sequence $(p_i)_{0 \leq i <N/2} \in {\mathbb R}^{N/2}$, $\zeta=\exp(I \pi/N)$ and $N=2^\ell$.
\Ensure A sequence $(v_i)_{0 \leq i <N/2}={\sf DFT}(p)=\left( p(\zeta^{5^i}) \right)_{0 \leq i<N/2}$,
where $p(X)=\sum\limits_{j=0}^{N/2-1} p_{{\sf br}_{\ell-1}(j)} X^{j} + \sum\limits_{j=0}^{N/2-1} p_{N/2+{\sf br}_{\ell-1}(j)} X^{N/2+j} $.
\smallskip
\State $(v_j)_{0 \leq j <N/2}=(p_j+ I \cdot p_{j+N/2})_{0 \leq j <N/2}$
\For{$k=1$ to $\ell-1$}
   \For{$j=0$ to $2^{k-1}-1$}
   		\State  $u=\zeta^{5^j \cdot N/2^{k+1}}$
   		\For{$i=0$ to $N/2^{k+1}-1$}
   		   \State  $a,b \leftarrow v_{i \cdot 2^k+j},~v_{i \cdot 2^k+2^{k-1}+j}$
		   \State $v_{i \cdot 2^k+j}=a + u \cdot b$
		   \State $v_{i \cdot 2^k+2^{k-1}+j}=a - u \cdot b$
		 \EndFor
	\EndFor
\EndFor
\State \Return  $(v_i)_{0 \leq i <N/2}$
\end{algorithmic}
\end{algorithm}

\ignore{
We are now ready to describe the algorithm for the ${\sf EDFT}_{N,n}$ function,
with input in bit-reversed order (using the above modified indexing {\sf Br'}), and output in normal order.
This is the function that will eventually be homomorphically evaluated for the computation
of $\stc$.

\begin{algorithm}[H]
\caption{Fast computation of ${\sf EDFT}_{N,n}$, ${\sf bo} \rightarrow {\sf no}$}
\label{alg:fastEDFTN_bo}
\begin{algorithmic}[1]
\Require A vector $\tilde{\vec t}=
\repl{\vec{t}}{N/(2n)}$, for a $\vec{t} \in {\mathbb R}^n$, $\zeta = \exp(I \pi/n)$ and $n=2^\ell$.
\Ensure A sequence $(v_i)_{0 \leq i <N/2}={\sf EDFT}_{N,n}({\sf Br'}(\vec{t}))$

\smallskip
\State $(v_j)_{0 \leq j <N/2}=\rep{(t_j+ I \cdot t_{j+n/2})_{0 \leq j <n/2}}{N/n}$
\For{$k=1$ to $\ell-1$}
   \For{$i=0$ to $2^{k-1}-1$}
   		\State  $u=\zeta^{5^i \cdot n/2^{k+1}}$
   		\For{$j=0$ to $N/2^{k+1}-1$}
   		   \State  $a,b \leftarrow v_{j \cdot 2^k+i},v_{j \cdot 2^k+2^{k-1}+i}$
		   \State $v_{j \cdot 2^k+i}=a + u \cdot b$
		   \State $v_{j \cdot 2^k+2^{k-1}+i}=a - u \cdot b$
		 \EndFor
	\EndFor
\EndFor
\State \Return  $(v_i)_{0 \leq i <N/2}$
\end{algorithmic}
\end{algorithm}
}

\subsection{Definition of the $\stc$ function}

For convenience, we define the $\stc_N$ operation
over the cyclotomic ring ${\cal S}_N$, 
but the procedure will indeed be the same over the integer ring 
${\cal R}={\mathbb Z}[X] / (X^N+1)$.
The $\stc_N$ function takes as input a polynomial $v(X)$
encoding $N/2$ slots $\vec{t}=(t_0,\ldots,t_{N/2-1}) \in {\mathbb R}^{N/2}$,
that is $ v(X)={\sf iDFT}_N(\vec{t}) $.
It outputs a polynomial $m(X) \in {\mathbb R}[X^2] / (X^N+1)$,
whose $N/2$ coefficients are the $t_j$'s:
$$
m(X)=\stc_N(v)=\sum\limits_{j=0}^{N/2-1}  t_j \cdot X^{2j}.
$$
We can therefore 
write $m(X)=\tilde{m}(X^2)$
for $\tilde{m}(Y)=\sum_{j=0}^{N/2-1} t_j \cdot Y^j$, 
and we let $\vec{\alpha}={\sf DFT}_{N/2}(\tilde{m}) \in {\mathbb C}^{N/4}$. 
This implies
$ {\sf DFT}_N(m)=(\vec{\alpha},\vec{\alpha}) \in {\mathbb C}^{N/2}$.

We generalize the {\sf DFT} function
to take as input the coefficient vector of a polynomial, instead of
the polynomial itself, with the same output.
Therefore, since $\vec{\alpha}={\sf DFT}_{N/2}(\tilde{m})$
and by definition 
$\tilde{m}(Y)=\sum_{j=0}^{n-1} t_j \cdot Y^{j}$, we can write $\vec{\alpha}={\sf DFT}_{N/2}(\vec{t})$. 
In total, we can summarize the above in a commutative diagram:
\[
\begin{tikzcd}
{\rm Coefficient~space} & {\rm Slot~space} \\[-.4cm]
v(X)  \arrow[r, "{\sf DFT}_N"]   \arrow[d, "\stc"'] & \vec{t}
\arrow[d, "{\sf DFT}_{N/2}"] 
 \\
m(X) \arrow[r, "{\sf DFT}_N"] & (\vec{\alpha},\vec{\alpha})
\end{tikzcd}
\]
From the above commutative diagram, 
the goal is therefore to compute the function 
${\mathbb R}^{N/2} \rightarrow {\mathbb C}^{N/2}$,
$\vec{t} \rightarrow ({\sf DFT}_{N/2}(\vec{t}),{\sf DFT}_{N/2}(\vec{t}))$
over the slot space, homomorphically over the coefficient space.

\subsection{The $\stc$ algorithm}

In fact, we cannot compute $\vec{\alpha}={\sf DFT}_{N/2}(\vec{t})$ directly 
with a fast ${\cal O}(\log N)$ algorithm. Either the input or the output
must be indexed in bit-reversed order. We will use the function
${\sf DFT}_{N/2,{\sf bo'} \rightarrow {\sf no}}$
computed by Alg. \ref{alg:fastDFTN_bo}, whose input is bit-reversed and whose output is in normal
order, which gives ${\sf DFT}_{N/2,{\sf bo'} \rightarrow {\sf no}}(\vec{t}')=
{\sf DFT}_{N/2}(\vec{t})$ where $\vec{t'}={\sf Br}'(\vec{t})$. 
 We therefore have the updated commutative diagram:
\[
\begin{tikzcd}
{\rm Coefficient~space} & {\rm Slot~space} \\[-.4cm]
v(X)  \arrow[r, "{\sf DFT}_N"]   \arrow[d, "\stc"'] & {\sf Br}'(\vec{t})
\arrow[d, "{\sf DFT}'_{N}"] 
 \\
m(X) \arrow[r, "{\sf DFT}_N"] & (\vec{\alpha},\vec{\alpha})
\end{tikzcd}
\]
where we define the function
${\sf DFT}'_N(\vec{t})=({\sf DFT}_{N/2,{\sf bo'} \rightarrow {\sf no}}(\vec{t}),
{\sf DFT}_{N/2,{\sf bo'} \rightarrow {\sf no}}(\vec{t}))$. 
The goal is therefore to evaluate the function ${\sf DFT}'_N(\vec{t}')$ homomorphically
over the coefficient space.

\begin{algorithm}[H]
\caption{Fast computation of ${\sf DFT}'_N$}
\label{alg:fastDFTPN}
\begin{algorithmic}[1]
\Require A sequence $(p_i)_{0 \leq i <N/2} \in {\mathbb R}^{N/2}$, $\zeta=\exp(I \pi/N)$ and $N=2^\ell$.
\Ensure A sequence $(v_i)_{0 \leq i <N/2}={\sf DFT}'(\vec{p})$
\smallskip
\State $(v_j)_{0 \leq j <N/2}=(\vec{w},\vec{w})$, where $\vec{w}=(p_j+I \cdot p_{j+N/4})_{0 \leq j<N/4}$
\For{$k=1$ to $\ell-2$}
   \For{$j=0$ to $2^{k-1}-1$}
   		\State  $u=\zeta^{5^j \cdot N/2^{k+1}}$
   		\For{$i=0$ to $N/2^{k+1}-1$}
   		   \State  $a,b \leftarrow v_{i \cdot 2^k+j},~v_{i \cdot 2^k+2^{k-1}+j}$
		   \State $v_{i \cdot 2^k+j}=a + u \cdot b$
		   \State $v_{i \cdot 2^k+2^{k-1}+j}=a - u \cdot b$
		 \EndFor
	\EndFor
\EndFor
\State \Return  $(v_i)_{0 \leq i <N/2}$
\end{algorithmic}
\end{algorithm}

To perform the homomorphic evaluation of the previous algorithm, 
we first rewrite the operations in vector form.
For a vector $\vec{v}=(v_i)_{0 \leq i<N/2}$, 
we write 
$${\sf Rot}_r(\vec{r})=(v_{r+j \bmod N/2})_{0 \leq j < N/2}
= (v_r,\ldots,v_{N/2-1},v_0,\ldots,v_{r-1})$$
the rotation of $\vec{v}$ by $r$ positions to the left.
For a vector $\vec{a}$, we denote by $\vec{a}^n$ the
repetition $n$ times of $\vec{a}$.

For the first line, we can write:
$$ \vec{v} = \vec{p} \odot ((1)^{N/4},(I)^{N/4})+ 
{\sf Rot}_{N/4}(\vec{p}) \odot ((I)^{N/4},(1)^{N/4})$$
We can rewrite the first butterfly equation as:
%$v_{i \cdot 2^k+j} \leftarrow v_{i \cdot 2^k+j} + u \cdot v_{i \cdot 2^k+2^{k-1}+j}$ as:
$$
v_{i \cdot 2^k+j}
 \leftarrow v_{i \cdot 2^k+j} + \zeta^{5^j \cdot N/2^{k+1}}
  \cdot ({\sf Rot}_{2^{k-1}}(\vec{v}))_{i \cdot 2^k + j}
$$
Similarly, we can rewrite the second butterfly equation as:
$$ v_{i \cdot 2^k+2^{k-1}+j}
 \leftarrow 
 ({\sf Rot}_{-2^{k-1}}(\vec{v}))_{i \cdot 2^k+2^{k-1}+j} - \zeta^{5^j \cdot N/2^{k+1}} \cdot 
 v_{i \cdot 2^k+2^{k-1}+j}$$
We can therefore write:
$$ \vec{v} \leftarrow \vec{v} \odot \vec{\omega}_0 
+ {\sf Rot}_{2^{k-1}}(\vec{v}) \odot \vec{\omega}_1 +
{\sf Rot}_{-2^{k-1}}(\vec{v}) \odot \vec{\omega}_2$$
where 
\begin{align*}
\vec{\omega}_0  &= \left( (1)^{2^{k-1}},\left(-\zeta^{5^j \cdot N/2^{k+1}}\right)_{0 \leq j<2^{k-1}} \right)^{N/2^{k+1}} \\
\vec{\omega}_1  &= \left( \left(\zeta^{5^j \cdot 
N/2^{k+1}}\right)_{0 \leq j<2^{k-1}},~(0)^{2^{k-1}}\right)^{N/2^{k+1}} \\
\vec{\omega}_2  &= \left( (0)^{2^{k-1}},~(1)^{2^{k-1}} \right)^{N/2^{k+1}} 
\end{align*}

Finally, we can now define the $\stc$ algorithm, which is a homomorphic
evaluation of the previous algorithm (Alg. \ref{alg:fastDFTPN}).

\begin{algorithm}[H]
\caption{\stc}
\label{alg:stc}
\begin{algorithmic}[1]
\Require A polynomial
$ v={\sf iDFT}_{N}({\sf Br}'(\vec{t}))$ for $\vec{t} \in {\mathbb R}^{N/2}$, $\zeta=\exp(I \pi/N)$ and $N=2^\ell$.
\Ensure A polynomial 
$m(X)=\stc_N(v) \simeq \sum\limits_{j=0}^{N/2-1}  t_j \cdot X^{2j} $
\smallskip
\State $v \leftarrow v \times  {\sf Ecd}((1)^{N/4},(I)^{N/4}))+ \Psi_{N/4}(v) \times
{\sf Ecd}((I)^{N/4},(1)^{N/4})$
\For{$k=1$ to $\ell-2$}
  \State
  $\vec{w}_0=\left( (1)^{2^{k-1}},\left(-\zeta^{5^j \cdot N/2^{k+1}}\right)_{0 \leq j<2^{k-1}} \right) \in {\mathbb C}^{2^k}$
\State $W_0={\sf Ecd}(\vec{w}_0)$
\State
 $\vec{w}_1=\left( \left(\zeta^{5^j \cdot 
N/2^{k+1}}\right)_{0 \leq j<2^{k-1}},~(0)^{2^{k-1}}\right) \in {\mathbb C}^{2^k}$
\State $W_1={\sf Ecd}(\vec{w}_1)$
\State
 $\vec{w}_2=\left( (0)^{2^{k-1}},~(1)^{2^{k-1}} \right) \in {\mathbb C}^{2^k}$
\State $W_2={\sf Ecd}(\vec{w}_2)$

\State
	$v \leftarrow v \times W_0+\Psi_{2^{k-1}}(v) \times W_1 + \Psi_{-2^{k-1}}(v) \times W_2$
\EndFor
\State \Return  $v$
\end{algorithmic}
\end{algorithm}

The above $\stc$ algorithm is essentially the same as in \cite{CHH18}, although in the latter it is described via a matrix approach.
It has multiplicative depth $\log_2 (N/2)$, and
requires $2 \log_2 (N/2) - 1$ homomorphic rotations and
 $3 \log_2 (N/2) -1$ external multiplications. Note that the polynomials $W_0$, $W_1$, and $W_2$
 are independent of the input and can thus be precomputed.
 
\subsection{\stc~ for $n$ slots}

 We  have the updated commutative diagram:
\[
\begin{tikzcd}
{\rm Coefficient~space} & {\rm Coefficient~space} & {\rm Slot~space} \\[-.7cm]
{\mathbb Z}[X^{N/(2n)}] /(X^{N}+1) & {\mathbb Z}[Y] /(Y^{2n}+1) & {\mathbb C}^n \\[-.4cm]
v(X) \arrow[r]  \arrow[d, "\stc_{N,n}"'] & {\tilde v}(Y)  \arrow[r, "{\sf DFT}_{2n}"]   \arrow[d, "\stc_{2n}"'] & {\sf Br}'(\vec{t})
\arrow[d, "{\sf DFT}'_{2n}"] \\
m(X) \arrow[r] & {\tilde m}(Y) \arrow[r, "{\sf DFT}_{2n}"] & (\vec{\alpha},\vec{\alpha})
\end{tikzcd}
\]

\begin{algorithm}[H]
\caption{$\stc_{N,n}$}
\label{alg:stc:slots}
\begin{algorithmic}[1]
\Require A polynomial
$ v={\sf iDFT}_{N}({\sf Br}'(\vec{t})^{N/(2n)})$ for $\vec{t} \in {\mathbb R}^{n}$, $\zeta=\exp(I \pi/(2n))$ and $n=2^\ell$.
\Ensure A polynomial 
$m(X)=\stc_{N,n}(v) \simeq \sum\limits_{j=0}^{n-1}  t_j \cdot X^{j \cdot N/n} $
\smallskip
\State $v \leftarrow v \times  {\sf Ecd}((1)^{n/2},(I)^{n/2}))+ \Psi_{n/2}(v) \times
{\sf Ecd}((I)^{n/2},(1)^{n/2})$
\For{$k=1$ to $\ell-1$}
  \State
  $\vec{w}_0=\left( (1)^{2^{k-1}},\left(-\zeta^{5^j \cdot n/2^{k}}\right)_{0 \leq j<2^{k-1}} \right) \in {\mathbb C}^{2^k}$
\State $W_0={\sf Ecd}(\vec{w}_0)$
\State
 $\vec{w}_1=\left( \left(\zeta^{5^j \cdot 
n/2^{k}}\right)_{0 \leq j<2^{k-1}},~(0)^{2^{k-1}}\right) \in {\mathbb C}^{2^k}$
\State $W_1={\sf Ecd}(\vec{w}_1)$
\State
 $\vec{w}_2=\left( (0)^{2^{k-1}},~(1)^{2^{k-1}} \right) \in {\mathbb C}^{2^k}$
\State $W_2={\sf Ecd}(\vec{w}_2)$
\State
	$v \leftarrow v \times W_0+\Psi_{2^{k-1}}(v) \times W_1 + \Psi_{-2^{k-1}}(v) \times W_2$
\EndFor
\State \Return  $v$
\end{algorithmic}
\end{algorithm}

\subsection{Radix implementation}
\label{a:radix}

As in \cite{CHH18}, one can reduce the multiplicative depth by using radices.
For a power-of-two radix $r$, \cite{CHH18} shows that the homomorphic linear transform 
requires roughly $2 r \log_r n$ homomorphic rotations and $3 r \log_r n$ external multiplications.
The advantage of the radix-based approach is that it yields a smaller multiplicative depth of 
around $\log_r n$ 
compared to the initial depth of $\log_2 n$.
However, since the complexity increases as the radix does, the radix variant can be seen as a trade-off, which becomes disadvantageous for a too large $r$.
As also described in \cite{CHH18}, without affecting the depth,
the number of homomorphic rotations can be even further reduced to $O(\sqrt{r} \log_r n)$ by using a baby-step giant-step approach.

In practice, the authors of \cite{CHH18} suggest to use $r = 2^5$ for a full slots implementation of \ckks, \ie 
with common parameters $n = N/2$ and $N=2^{15}$ or $N=2^{16}$. They show that the running times 
 for radices ranging from $2^1$ to $2^4$ are quite similar.
However, for few slots $n$, we can use a smaller radix since a radix $r$ is only of avail if $r \le \sqrt{n}$.

\ignore{
For our implementation,
we chose to use a radix $r=2^2$, 
which yields a depth of $1+\lfloor (\log_2 n)/2 \rfloor$. 
%This leads to the reasonable assumption that we may take $r = 2^{\lfloor \log_2(n) / 3  \rfloor + 1}$.
%Note that this setting implies that we can bound the depth of the linear transform by $3$.

\subsubsection{Practical experiments.}

Table \ref{t:radix} presents a running time comparison of our new bootstrapping for 
different radix values with $n=64$ slots. 
While increasing the radix reduces the size of the largest modulus $Q$, the optimal radix value for $n=64$ is found to be $r=4$. 

\begin{table}[H]
\centering
\renewcommand{\arraystretch}{1.15}
\resizebox{0.40\columnwidth}{!}{%
\begin{tabular}{|c|c|c|c|c|} \hline
~Parameter~ & ~$r$~ &  ~$\ell$~ & ~$\log_2 Q$~  & ~time~ \\ \hline
\multirow{3}{*}{~Set-I}   & ~2~ & ~13~ & ~497~ & ~47 s~ \\ \cline{2-5}
& ~4~ & ~11~ & ~427~ & ~45 s~ \\ \cline{2-5}
& ~8~ & ~10~ & ~392~ & ~55 s~ \\ \hline
\end{tabular}
}
\medskip
\caption{Running time of our new bootstrapping for Set-I parameters and $n=64$, for various radix values $r$.}
\label{t:radix}
\end{table}
}

\section{Proof of Theorem \ref{theorem:bootstrapping}}

\label{a:proofboot}

  We start with fresh encryptions of the secret-key bits $s_k$.
  By Lemma \ref{lemma:noise_encryption}, their initial noise size is bounded by $E=3N\kappa$.
  By Lemma \ref{lemma:noise_multiplication_encoding},
 since we perform external products with encodings of real values of absolute value
 less than $\nu=1$, the size of the noise of each product is at most $2 \nu E+8N \leq 14N\kappa$.
 
For the product tree, we must consider the error obtained when computing
the product of the encryptions of two complex values, whose real and imaginary 
part are encrypted separately. 
Given the plaintexts $z_1=x_1 + i \cdot y_1$ and $z_2=x_2 + i \cdot y_2$, we have
$z_1  z_2 = x_1 x_2-y_1  y_2 + i \cdot (x_1 y_2 + x_2  y_1)$.
Therefore, for $x=x_1x_2-y_1y_2$,  we must compute the homomorphic difference:
$$ \ct_{x} \leftarrow {\sf MultR}_\evk(\ct_{x_1}, \ct_{x_2})- {\sf MultR}_\evk(\ct_{y_1}, \ct_{y_2}),
$$
and similarly for $\ct_{y}$ for $y=x_1 y_2 + x_2  y_1$.
Since all plaintext values $x_1,y_1,x_2,y_2$ are bounded by $1$ in absolute value,
the errors in ${\sf MultR}(\ct_{x_1}, \ct_{x_2})$ and ${\sf MultR}(\ct_{y_1}, \ct_{y_2})$ 
are upper bounded by $2E+8N$, where $E$ is an upper bound of the errors
in $\ct_{x_1}$, $\ct_{x_2}$, $\ct_{y_1}$ and $\ct_{y_2}$.
 Therefore, the error magnitudes of $\ct_{x}$ and $\ct_{y}$ 
are both upper-bounded by $4E+16N$. 

We now consider the 
% \ell is a bad variable, since it is usually the current level...
 product tree of depth $\ell=\log_2 N$. For $0 \leq i \leq \ell$,
we can upper-bound the noise of the current ciphertext at level $i$ by $u_i N \kappa$, where $u_0=14$, and
the recursive relation
$u_{i+1} = 4u_i+16$ holds.
% I would prefer a proof of this claim about recursions.
 One can show that $u_i \leq 20 \cdot 2^{2i}-6$ by a recursive approach.
For $i=\ell$, the error size after all multiplications is thus upper-bounded by
$20 \cdot 2^{2\ell} N \kappa = 20 N^3 \kappa$. We must therefore
assume that $\Delta\ge (20N^{3}\kappa)^2$ since we require $\Delta \ge E^2$ in each multiplication.

After the final scaling by $q/(2\pi \Delta)$, and by applying
Lemma \ref{lemma:noise_multiplication_encoding} again,
the noise size becomes $20 N^3 \kappa \cdot q/(2\pi \Delta) + 8N$.
Therefore, for $\Delta \geq 4N^3 \kappa q$, the size of the noise is at most $9N$.
This implies that by writing
% Why do you need to write this? Why the <s,c> term??
 $m=\langle \ct,\sk \rangle(0)=\langle \vec{s},\vec{c} \rangle \in {\mathbb Z}$,
 our final ciphertext $\ct'$ satisfies:
 $$
 \langle \ct',\sk \rangle   ={\sf Ecd}(F(m)) + e' \pmod{ p \cdot q} ,
 $$
 for some $e'$ with $\infnorm{e'} \leq 9N$ and $$
F(m)  =\frac{q}{2\pi \Delta} \sin 
\left( \frac{2i\pi m}{q} \right).
$$
% Technically, the Taylor approximation requires an open interval, that's why we must assume that |m| < q^{2/3} and not |m| \leq q^{2/3}.
Given the condition $|m| < q^{2/3}$, we obtain by bounding the residual term in the Taylor expansion:
$$ \left| F(m) - \frac{m}{\Delta} \right| \leq \frac{q}{2\pi\Delta} \cdot 
\frac{1}{3!} \cdot \left( \frac{2 \pi |m|}{q} \right)^3 \leq \frac{7|m|^3}{q^2\Delta} 
\leq \frac{7}{\Delta}.
$$
This implies $|{\sf Ecd}(F(m)) - m | \leq \Delta \cdot |F(m)-m/\Delta|+1 \leq 8$. We can eventually write:
$$ \langle \ct',\sk \rangle =\langle \ct,\sk \rangle(0)+ e_{\sf bt}  \pmod{ p \cdot q},
$$
where $\infnorm{e_{\sf bt}} \leq 10N$ as claimed.

\section{Algorithms for the single slot bootstrapping}

\label{a:algo_single_slot}

\begin{algorithm}[H]
\caption{Bootstrapping key generation, single slot}
\label{alg:bkey_single}
\begin{algorithmic}[1]
	\Require A length $N$ secret key $\vec{s}$ with Hamming weight $h$ and $B = N/h$
	\Ensure A bootstrapping key ${\sf cs} = ({\sf cs}_0,{\sf cs}_1)$
  \smallskip
    \For {{\bf all}  $0 \leq j <B/2$ {\bf and} $0 \leq b <h$}
    \State $\tilde{s}^{0}_{j \cdot h +b} = s_{b \cdot B + j}$ 
    \State $\tilde{s}^{1}_{j \cdot h +b} = s_{b \cdot B + B/2+j}$ 
    \EndFor
\State $S_0(X), S_1(X) \longleftarrow {\sf Ecd}((\tilde{s}^{0}_\iota)_{0 \leq \iota < N/2}), {\sf Ecd}((\tilde{s}^{1}_\iota)_{0 \leq \iota < N/2})$
    \State \Return ${\sf cs} = ({\sf Enc}_\pk(S_0(X)),{\sf Enc}_\pk(S_1(X)))$
\end{algorithmic}
\end{algorithm}

% Maybe we should write down a separate algorithm for parameter selection

\begin{algorithm}[H]
\caption{Bootstrapping, single slot}
\label{alg:boot_single_slot}
\begin{algorithmic}[1]
	\Require A modulus $q$, a bootstrapping key ${\sf cs}$, an \rlwe ciphertext $\ct = (b, a)$, and $\delta = (q/(4\pi \Delta))^{1/h}$
	\Ensure A refreshed ciphertext $\ct'$
  \smallskip
  \State $\vec c = (a_0 + b_0, -a_{N-1}, -a_{N-2}, \dots, -a_1)$
 \For {{\bf all}  $0 \leq j <B/2$ {\bf and} $0 \leq b <h$}
    \State $e^{0}_{j \cdot h +b} = \exp(2 i \pi \cdot c_{b \cdot B + j} / q) \cdot \delta$ 
    \State $e^{1}_{j \cdot h +b} = \exp(2 i \pi \cdot c_{b \cdot B+ B/2+ j} / q) \cdot \delta$ 
    \EndFor
\State $E_0, E_1 \longleftarrow {\sf Ecd} ((e^i_\iota)_{0 \leq \iota <N/2}),  {\sf Ecd} ((e^1_\iota)_{0 \leq \iota <N/2})$
\State $T_0, T_1 \longleftarrow {\sf ExtMultR}({\sf cs}_0, E_0), {\sf ExtMultR}({\sf cs}_1, E_1)$
\State
\Return $\ct'={\sf Im2}(\pr{h}{1}(\tr{N/2}{h}(T_0 + T_1)))$
\end{algorithmic}
\end{algorithm}

\ignore{

\section{Bootstrapping key generation, multiple slots}

\label{a:bootkeygen_multiple}

}

\section{Bootstrapping algorithm for more than $B/2$ slots}

\label{a:boot_max_slots}

The bootstrapping algorithm from Section \ref{s:bootstrapp:nslots}
is limited to bootstrapping up to 
\( n \leq n_{\sf max} = B/2 = N/(2h) \) components. However, it can be extended to support more 
slots, specifically  $n' \leq N$, by shifting the coefficients of the input ciphertext and applying the 
previous bootstrapping procedure as a black box for each group of \( n_{\sf max} \) coefficients.

 More precisely, 
 the previous bootstrapping algorithm  (Alg. \ref{alg:boot_multiple_slots}) inputs a ciphertext 
 $\ct$ for the plaintext
 polynomial $m(X)=\sum_{i=0}^{N-1} m_i \cdot X^i$, and outputs
 a refreshed ciphertext for the
 polynomial $m'(X)=\sum_{i=0}^{n-1} m_{i \cdot N/n} \cdot X^{i \cdot N/n}$.
In other words,  only the $n$ coefficients $m_i$ of $m(X)$ such that $i \equiv 0 \pmod{N/n}$
 are kept and bootstrapped, while the others are lost, because of the initial
 $C \leftarrow {\sf DecMat}(\ct)$ procedure which only considers the coefficients
 of $X^{i \cdot N/n}$ in $m(X)$. 

To bootstrap $n' >n$  coefficients, we can therefore proceed by repeatedly 
shifting the coefficients of the input plaintext (homomorphically on the input ciphertext), by
a shifting polynomial $X^{-j \cdot N/n'}$. We obtain a bootstrapped ciphertext,
which we shift back by $X^{j \cdot N/n'}$.
We then compute the sum for all $0 \leq j <n'/n$ of
all such ciphertexts. At the end, we obtain a refreshed ciphertext for all the $n'$ components.
We describe the concrete algorithm below.
The number of homomorphic operations becomes ${\cal O}(n' + (n'/n) \cdot \log N)$, where $n=n_{\sf max}=B/2$ and $n' \leq N/2$, which gives 
$n'/n \leq h$. Assuming $h={\cal O}(1)$, the complexity
remains ${\cal O}(n' + \log N)$,
while the depth remains unchanged.

\begin{algorithm}[H]
  \caption{Bootstrapping, more than $B/2$ slots}
  \label{alg:boot_more_slots}
  \begin{algorithmic}[1]
    \Require A modulus $q$, a bootstrapping key $({\sf CS}_u)_{0 \le u < 2n}$, an \rlwe ciphertext $\ct=(b,a)$
    containing $n' \geq n$ slots
    \Ensure A refreshed ciphertext $\ct'$
    \smallskip
    \State ${\sf acc} \leftarrow (0,0)$
    \For {$j=0$ to $n'/n-1$}
      \State $\ct_{s} \leftarrow {\sf ExtMult}( \ct , X^{-j \cdot N/n'})$
      \State $\ct'_{s} \leftarrow {\sf Bootstrap}(q,{\sf cs}, \ct_s)$
      \State ${\sf acc} \leftarrow {\sf Add}({\sf acc},{\sf ExtMult}( \ct'_{s} , X^{j \cdot N/n'}))$
      \EndFor
      \State \Return ${\sf acc}$
  \end{algorithmic}
\end{algorithm}

\end{document}